\documentclass[fleqn,11pt]{article}

\usepackage[T1]{fontenc}
\usepackage[utf8]{inputenc}

\usepackage[letterpaper,left=0.8in,top=0.8in,bottom=0.9in,right=0.8in]{geometry} 

\usepackage{cite} 
\usepackage{csquotes}
\usepackage{textgreek}
\usepackage{ulem}
\usepackage{soul}
\usepackage[svgnames]{xcolor}
\usepackage{url}  
\usepackage{mathtools}
\usepackage{amsthm}
\usepackage{amsfonts}
\usepackage{amsopn}
\usepackage{amssymb} 
\usepackage{bm}
\usepackage{centernot}
\usepackage{cancel}
\usepackage{float}
\usepackage{graphicx,graphics,epsf,epsfig}
\usepackage{setspace} 
\usepackage{comment} 
\usepackage{tocloft}
\usepackage{caption}
\usepackage{afterpage}

\usepackage{titlesec}
\usepackage{titletoc}
\usepackage[framemethod=TikZ]{mdframed}
\usepackage[hypertexnames=false]{hyperref}
\hypersetup{colorlinks=true, %
citecolor=blue,%
linkcolor= blue
}

\usepackage{fourier}

\definecolor{midnight}{rgb}{0.,0.,0.5} 

\DeclareCaptionFont{midnight}{\color{midnight}}
\titlespacing{\section} {0pt}{1.5ex plus .1ex minus .2ex}{0pt}
\titlespacing{\subsection} {0pt}{1.5ex plus .1ex minus .2ex}{0pt}
\titlespacing{\subsubsection} {0pt}{1.5ex plus .1ex minus .2ex}{0pt}

\titlecontents{subsection}[2cm]{}{\contentslabel{2.2em}}{\hspace*{3.2em}}{\hspace*{0.15cm}\cftdotfill{1}{.}\contentspage}[]
\titlecontents{subsubsection}[3.2cm]{}{\contentslabel{3em}}{\hspace*{3.2em}}{\hspace*{0.15cm}\cftdotfill{1}{.}\contentspage}[]
\titlecontents{paragraph}[4.7cm]{}{\contentslabel{3.7em}}{\hspace*{3.2em}}{\hspace*{0.15cm}\cftdotfill{1}{.}\contentspage}[]

\renewcommand{\vec}[1]{\bm{#1}}

\def\Nat{\mathbb N}

\newenvironment{raggedy}{\begin{raggedright}\baselineskip20pt\sloppy}{\end{raggedright}\baselineskip20pt\sloppy}

\usepackage{bibunits} 
\defaultbibliography{scaffold}
\defaultbibliographystyle{unsrt}

\begin{document}  

\begin{raggedy}
\normalem
\singlespacing 
\parskip 0.2cm 
\parindent 0cm

\pagenumbering{gobble}

\vspace*{2cm}

\begin{center}
\LARGE
\bf
Combinatorial protein-protein interactions\\
on a polymerizing scaffold
\end{center}

\vspace*{0.5cm}

\begin{center}
Andr\'es Ortiz-Mu\~noz$^{a,1}$, H\'ector F.\@ Medina-Abarca$^{b,1}$, and Walter Fontana$^{b,2}$\\
\vspace*{0.3cm}
{\small $^{a}$ California Institute of Technology, Pasadena, CA 91125}\\
{\small $^{b}$ Systems Biology, Harvard Medical School, Boston, MA 02115}\\[0.4cm]
{\small $^{1}$ both authors contributed equally}\\
{\small $^{2}$ to whom correspondence should be addressed: walter\_fontana@hms.harvard.edu}
\end{center}

\vspace*{1cm}

\begin{abstract}
 Scaffold proteins organize cellular processes by bringing signaling molecules into interaction, sometimes by forming large signalosomes. Several of these scaffolds are known to polymerize. Their assemblies should therefore not be understood as stoichiometric aggregates, but as combinatorial ensembles. We analyze the combinatorial interaction of ligands loaded on polymeric scaffolds, in both a continuum and discrete setting, and compare it with multivalent scaffolds with fixed number of binding sites. The quantity of interest is the abundance of ligand interaction possibilities---the catalytic potential $Q$---in a configurational mixture. Upon increasing scaffold abundance, scaffolding systems are known to first increase opportunities for ligand interaction and then to shut them down as ligands become isolated on distinct scaffolds. The polymerizing system stands out in that the dependency of $Q$ on protomer concentration switches from being dominated by a first order to a second order term within a range determined by the polymerization affinity. This behavior boosts $Q$ beyond that of any multivalent scaffold system. In addition, the subsequent drop-off is considerably mitigated in that $Q$ decreases with half the power in protomer concentration than for any multivalent scaffold. We explain this behavior in terms of how the concentration profile of the polymer length distribution adjusts to changes in protomer concentration and affinity. The discrete case turns out to be similar, but the behavior can be exaggerated at small protomer numbers because of a maximal polymer size, analogous to finite-size effects in bond percolation on a lattice.
\end{abstract}

\vspace*{1cm}

\begin{center}
\end{center}

\newpage
\pagenumbering{roman}
{
\setstretch{1}
\tableofcontents
\newpage
\listoffigures
}

\newpage
\pagenumbering{arabic}


\begin{bibunit}

\addcontentsline{toc}{section}{Introduction}
\subsection*{Introduction}

Protein-protein interactions underlying cellular signaling systems are mediated by a variety of structural elements, such as docking regions, modular recognition domains, and scaffold or adapter proteins \cite{Bhattacharyya2006,Good2011}. These devices facilitate both the evolution and control of connectivity within and among pathways. Since the scaffolding function of a protein can be conditional upon activation and also serve to recruit other scaffolds, the opportunities for plasticity in network architecture and behavior are abundant. 

Scaffolds are involved in the formation of signalosomes --transient aggregations of proteins that process and propagate signals. A case in point is the machinery that tags \textbeta-catenin for degradation in the canonical Wnt pathway. \textbeta-catenin is modified by CK1\textalpha\ and GSK3\textbeta\ without binding any of these kinases directly, but interacting with them through the Axin scaffold \cite{Liu2002,Ikeda1998}. In addition, the DIX domain in Axin allows for oriented Axin polymers \cite{Fiedler2011}, while APC (another scaffold) can bind multiple copies of Axin \cite{Behrens1988}, yielding Axin-APC aggregates to which kinases and their substrates bind. 

By virtue of their polymeric nature, scaffold assemblies like these have no defined stoichiometry and may only exist as statistical ensembles rather than a single stoichiometrically well-defined complex \cite{Deeds2012,Suderman2013}. As a heterogeneous mixture of aggregates with combinatorial state, the \textbeta-catenin destruction system thus appears to be an extreme example of what has been called a \enquote{pleiomorphic ensemble} \cite{r01710}.

Scaffold-mediated interactions are characteristically subject to the prozone or \enquote{hook} effect. At low scaffold concentrations, adding more scaffold facilitates interactions between ligands. Beyond a certain threshold, however, increasing the scaffold concentration further prevents interactions by isolating ligands on different scaffold molecules \cite{Bray97,Ferrell2000,Levchenko2000}. For a scaffold $S$ that binds with affinity $\alpha$ an enzyme $A$ and a substrate $B$, present at concentrations $t_A$ and $t_B$, the threshold is at $1/\alpha+(t_A+t_B)/2$.

In this contribution we define and analyze a simple model of enzyme-substrate interaction mediated by a polymerizing scaffold. The model does not take into account spatial constraints of polymer chains and therefore sits at a level of abstraction that only encapsulates combinatorial aspects of a pleiomorphic ensemble and briefly peeks down the trail of critical phenomena often associated with phase-separation \cite{Li2012,Bergeron-Sandoval2016}. 

\addcontentsline{toc}{section}{The polymerizing scaffold system}
\subsection*{The polymerizing scaffold system}

Let $S$ (the scaffold) be an agent with four distinct binding sites $\{${\tt a},{\tt b},{\tt x},{\tt y}$\}$. At site {\tt y} agent $S$ can reversibly bind site {\tt x} of another $S$ with affinity $\sigma$, forming (oriented) chains. For the time being we exclude the formation of rings. Sites {\tt a} and {\tt b} can reversibly bind an agent of type $A$ (the enzyme) and of type $B$ (the substrate) with affinities $\alpha$ and $\beta$, respectively. All binding interactions are independent. When the system is closed, the total concentrations of $A$, $B$, and $S$ are given by $t_A$, $t_B$, and $t_S$. This setup allows for a variety of configurations as shown on the left of the arrow in Fig.\@ \ref{fig:model}. We posit that each enzyme $A$ can act on each substrate $B$ bound to the same complex. We refer to the number $pq$ of potential interactions enabled by a configuration with sum formula $A_pS_nB_q$ as that configuration's \enquote{catalytic potential} $Q$. By extension we will speak of the catalytic potential $Q$ of a mixture of configurations as the sum of their catalytic potentials weighted by their concentrations.

\begin{figure}[!h]
\centering
\includegraphics[width=0.9\linewidth]{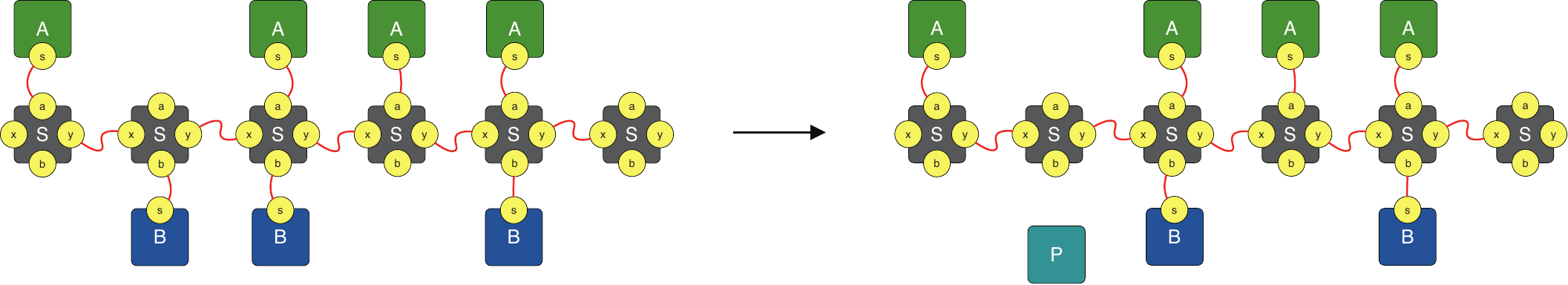}
\caption[Enzyme-substrate interaction on a polymeric scaffold]{Enzyme-substrate interaction on a polymeric scaffold. In the polymerizing model, scaffold protomers $S$ binding each other with affinity $\sigma$ yield a distribution of polymers of varying length to which enzymes $A$ and substrates $B$ bind with affinities $\alpha$ and $\beta$, respectively. For each configuration, the rate of conversion to product is a function of the configuration's catalytic potential $Q$, which is the number of possible interactions between bound $A$ and $B$ agents. Here, each of the four $A$s can interact with each of the three $B$s for a total of $Q=12$ possible interactions.}
\label{fig:model}
\end{figure}

If we assume that the assembly system equilibrates rapidly, the rate of product formation is given by $Q k_{\text{cat}}$ with $k_{\text{cat}}$ the catalytic rate constant and $Q$ the equilibrium abundance of potential interactions between $A$- and $B$-agents. Rapid equilibration is a less realistic assumption than a quasi-steady state but should nonetheless convey the essential behavior of the system. In the following we first provide a continuum description of equilibrium $Q$ in terms of concentrations (which do not imply a maximum polymer length) and then a discrete statistical mechanics treatment for the average equilibrium $Q$ (where $t_S$ is a natural number and implies a maximum length).

In the present context, molecular species $Y_i$ that assemble from $T$ distinct building blocks (\enquote{atoms}) $X_j$ through reversible binding interactions have a graphical (as opposed to geometric) structure that admits two descriptors: $\omega_{i}$, the number of symmetries of $Y_i$ (here $\omega_{i}=1$ because the polymers are oriented), and $\mu_{i,j}$, the number of atoms $X_j$ in $Y_i$. The equilibrium concentration $y_i$ of any $Y_i$ is given by $y_i=\varepsilon_{i}\prod_{j=1}^T (x_j)^{\mu_{i,j}}$, where $\varepsilon_{i}=1/\omega_{i}\prod_{r\in{\cal P}} K_r$ is the exponential of the free energy content of $Y_i$, with $K_r\in\{\alpha, \beta,\sigma\}$ the equilibrium constant of the $r$th reaction along some assembly path $\cal P$. The $x_{j}$ are the equilibrium concentrations of free atoms of type $j$ (here $T=3$). Hence, $\varepsilon_{i}=\alpha^p\beta^q\sigma^r$ for a $Y_i$ that contains $p$ bonds between $A$ and $S$, $q$ bonds between $B$ and $S$, and $r$ bonds between $S$ protomers.

Consider first the polymerization subsystem. From what we just laid out, the equilibrium concentration of a polymer of length $l$ is $\sigma^{l-1}s^l$, where $s$ is the equilibrium concentration of monomers of $S$. Summing over all polymer concentrations yields the total abundance of entities in the system, $W(s)=\sum_{l=1}^{\infty}\sigma^{l-1}s^l=s/(1-\sigma s)$. $W(s)$ gives us a conservation relation, $t_S=s\,dW(s)/ds$, from which we obtain $s$ as:
\begin{align}
\label{eq:free_s}
    s=\frac{1}{4\sigma}\biggl(\sqrt{4+1/(\sigma t_S)}-\sqrt{1/(\sigma t_S)}
    \biggr)^2
\end{align}
Using \eqref{eq:free_s} in $\sigma^{l-1}s^l$ yields the dependence of the polymer size distribution on parameters $t_S$ and $\sigma$. $W(s)$ has a critical point at $s_{\text{cr}}=1/\sigma$, at which the concentrations of all length classes become identical. It is clear from \eqref{eq:free_s} that $s$ can never attain that critical value for finite $\sigma$ and $t_S$. 

\addcontentsline{toc}{section}{The chemostatted case}
\subsection*{The chemostatted case}

In a chemostatted system, $s$ can be clamped at any desired value, including the critical point $1/\sigma$ at which ever more protomers are drawn from the reservoir into the system to feed polymerization. We next include ligands $A$ and $B$ at clamped concentrations $a$ and $b$. Let $A_pS_nB_q$ be the sum formula of a scaffold polymer of length $n$ with $p$ $A$-agents and $q$ $B$-agents. There are $\binom{n}{p}\binom{n}{q}$ such configurations, each with the same catalytic potential $Q=pq$. Summing up the equilibrium abundances of all configurations yields
\begin{align}
\label{eq:Wopen}
    W(s,a,b)=a+b+\frac{s(1+\alpha a) (1+\beta b)}{1-\sigma s(1+\alpha a) (1+\beta b)}.
\end{align}
\eqref{eq:Wopen} corresponds to the $W(s)$ of ligand-free polymerization by a coarse-graining that only sees scaffolds regardless of their ligand-binding state, i.e.\@ by dropping terms not containing $s$ and substituting $s(1+\alpha a)(1+\beta b)\to s$. \eqref{eq:Wopen} indicates that, at constant chemical potential for $A$, $B$ and $S$, the presence of ligands lowers the critical point of polymerization to $s_{\text{cr}}=1/(\sigma(1+\alpha a)(1+\beta b))$ because, in addition to polymerization, free $S$ is also removed through binding with $A$ and $B$.

$Q_{\text{poly}}$, the $Q$ of the system, is obtained by summing up the $Q$ of each configuration weighted by its equilibrium concentration (SI section 1). Using $W$ we compute $Q_{\text{poly}}$ as
\begin{align}
    Q_{\text{poly}}=ab\dfrac{\partial^2}{\partial a\partial b} W =\alpha a\beta b\,s\,\frac{1+\sigma\,s(1+\alpha a) (1+\beta b)}{(1-\sigma s(1+\alpha a) (1+\beta b))^3}. 
    \label{eq:Q}
\end{align}
Note that $Q_{\text{poly}}$ inherits the critical point of $W$. The behavior of the chemostatted continuum model is summarized in Fig.\@ \ref{fig:chemostat}. 
\begin{figure}[!h]
\centering
\includegraphics[width=0.9\linewidth]{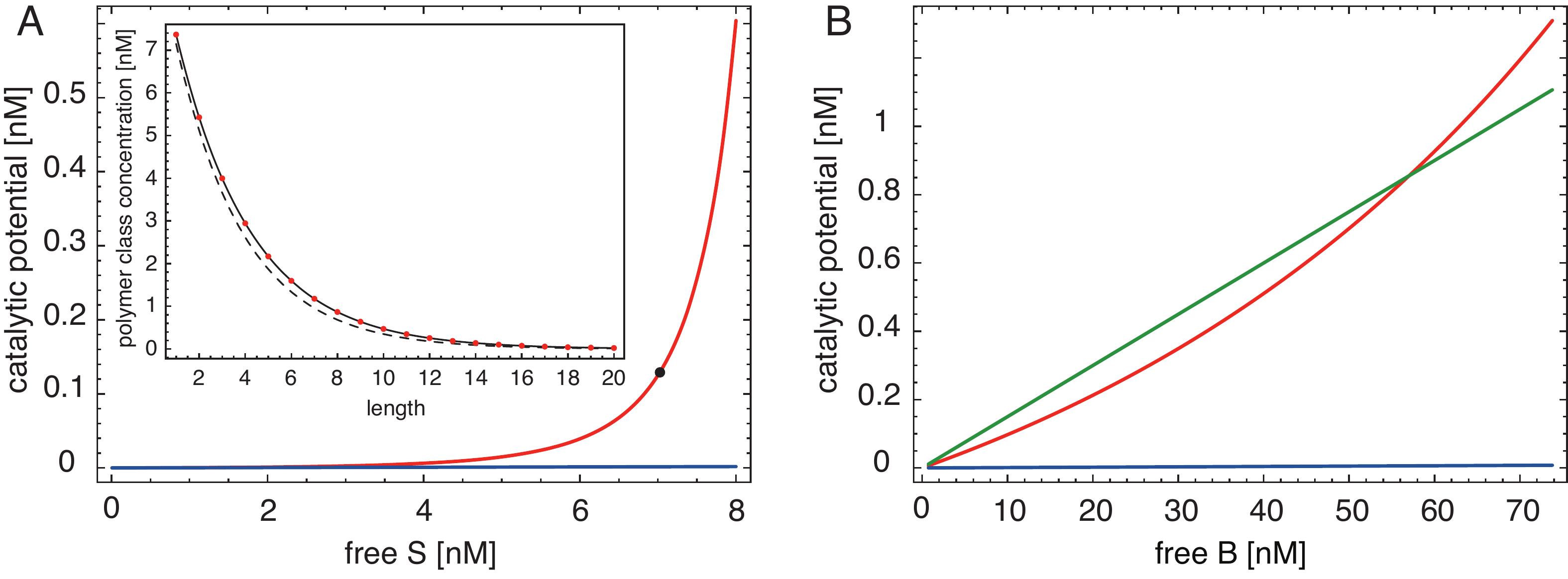}
\caption[Catalysis in a chemostatted polymerizing scaffold system]{Catalysis in a chemostatted polymerizing scaffold system. {\bf A:} The red graph shows the catalytic potential $Q$ as a function of chemostatted $s$ according to \eqref{eq:Q} for $\alpha=\beta=10^6$ M$^{-1}$, $\sigma=10^8$ M$^{-1}$, and $a=b=15\cdot 10^{-9}$ M (about $2\,10^4$ molecules in $10^{-12}$ L). The blue curve is the special case of $\sigma=0$, which is the monovalent scaffold system, $Q=\alpha a\beta b\,s$. The inset shows the scaffold length distribution at $s=7.15$ nM, corresponding to $Q$ at the black filled circle. The critical point in this example is $s_{cr}\sim 9.7$ nM. Panel {\bf B}: The catalytic potential at $s=7.15$ nM as a function of clamped $b$ (the substrate); other parameters as in A. Red: polymerizing scaffold system; blue: monovalent scaffold; green: chemostatted Michaelis-Menten in which $A$ binds directly to $B$ with affinity $\alpha$.}
\label{fig:chemostat}
\end{figure}

$Q_{\text{poly}}$ (red) diverges as the polymerization system approaches the critical point. The inset of Fig.\@ \ref{fig:chemostat}A shows the scaffold length distribution at the black dot on the $Q_{\text{poly}}$-profile. The red dotted curve reports the length distribution in the presence of ligands, $[\{A_*S_kB_*\}]=\sigma^{-1}(\sigma s(1+\alpha a)(1+\beta b))^k$, whereas the black dotted curve reports the length distribution in the absence of ligands, $s_k\equiv[S_k]=\sigma^{k-1}s^k$. The presence of $A$ and $B$ shifts the distribution to longer chains. The blue curve in Fig.\@ \ref{fig:chemostat}A shows the catalytic potential of the monovalent scaffold, $\sigma=0$. It increases linearly with $s$, but at an insignificant slope compared with the polymerizing case, which responds by raising the size (surface) distribution, thus drawing in more $S$ from the reservoir to maintain a given $s$; this, in turn, draws more $A$ and $B$ into the system. In Fig.\@ \ref{fig:chemostat}B, $s$ is fixed and $b$, the substrate concentration, is increased. The green straight line is the Michaelis-Menten case, which consists in the direct formation of an $AB$ complex and whose $Q=\alpha \, a\, b$ is linear in $b$. The red line is the polymerizing scaffold system whose $s_{\text{cr}}$ can be attained by just increasing $b$, \eqref{eq:Q}. All else being equal, there is a $b$ at which more substrate can be processed than through direct interaction with an enzyme. The slope of the monovalent scaffold (blue) is not noticeable on this scale.

\addcontentsline{toc}{section}{The continuum case in equilibrium}
\subsection*{The continuum case in equilibrium}

We turn to the system with fixed resources $t_S$, $t_A$ and $t_B$, expressed as real-valued concentrations. \eqref{eq:Q} for $Q_{\text{poly}}$ is now evaluated at the equilibrium concentrations $s$, $a$ and $b$ of the free atoms. These are obtained by solving the system of conservation equations, $t_S=s\,\partial W/\partial s$, $t_A=a\,\partial W/\partial a$, $t_B=b\,\partial W/\partial b$ (solutions in SI, section 1). The orange curve in Fig.\@ \ref{fig:closed}A depicts the saturation curve of the catalytic potential $Q_{\text{direct}}$ of the Michaelis-Menten mechanism for a fixed concentration $t_A$ of enzyme as a function of substrate $t_B$. The green curves are saturation profiles of the polymerizing scaffold system at varying protomer abundances $t_S$ under the same condition. As in the chemostatted case, beyond some value of $t_S$, the catalytic potential of the polymerizing system exceeds that from direct interaction. 

\begin{figure}[!h]
\centering
\includegraphics[width=0.9\linewidth]{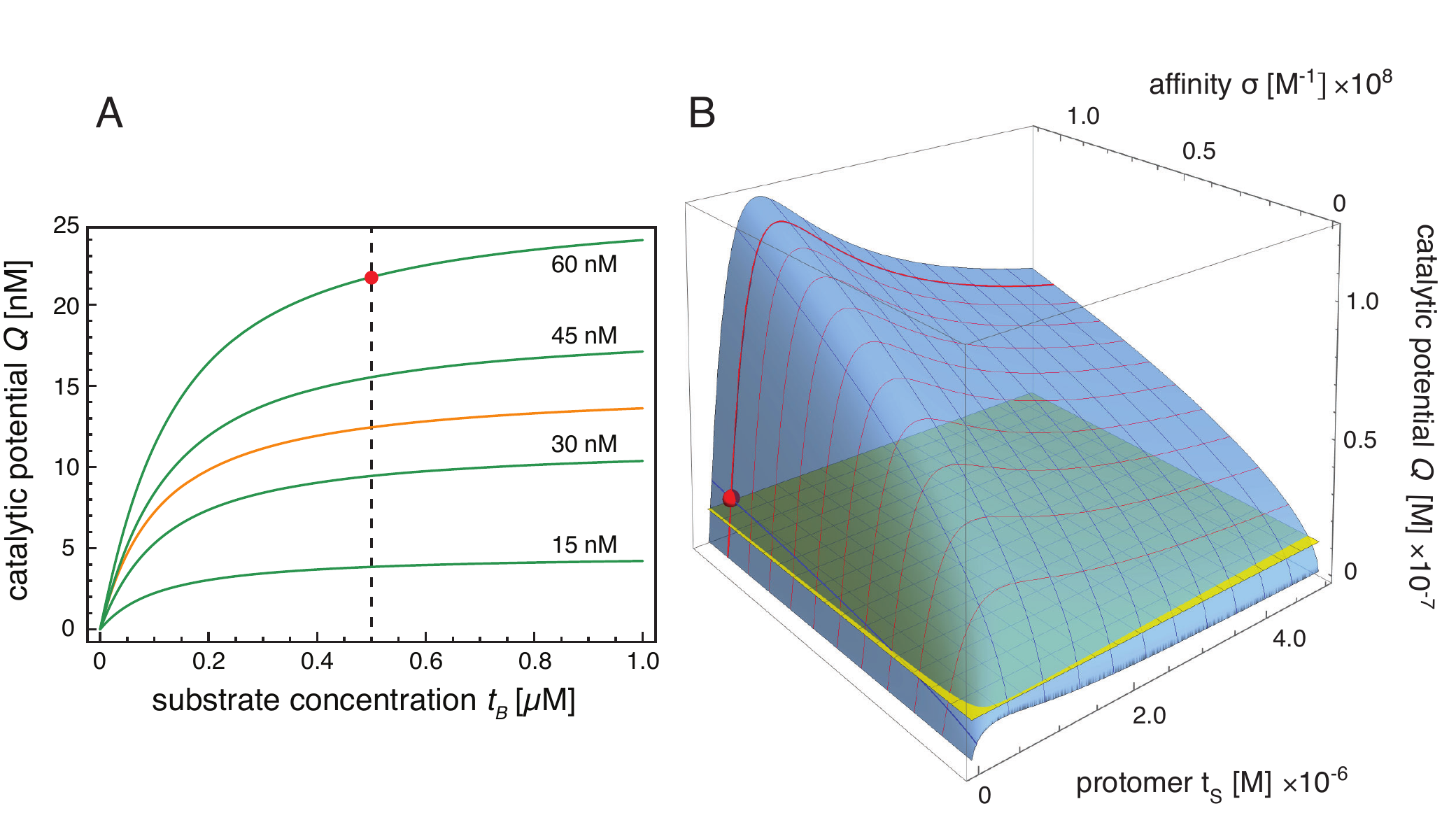}
\caption[Catalysis in a closed polymerizing scaffold system]{Catalysis in a closed polymerizing scaffold system. {\bf A:} The orange curve shows the saturation of catalytic potential $Q$ of the direct enzyme ($A$)-substrate ($B$) interaction, a classic Michaelis-Menten mechanism, as a function of $t_B$ for $\beta=10^7$ M$^{-1}$ and $t_A=15\cdot 10^{-9}$ M. The green curves depict the saturation curves for $Q$ of the poly-scaffold with affinities $\alpha=\beta=10^7$ M$^{-1}$ and $\sigma=10^8$ M$^{-1}$ at various protomer abundances $t_S$. {\bf B}: The catalytic potential surface for the poly-scaffold as a function of $t_S$ and $\sigma$; other parameters as in panel A. The red ball corresponds to the conditions marked by the red dot in panel A ($t_B=5\cdot 10^{-7}$ M). The flat yellow surface is the $Q$ for the direct enzyme-substrate interaction (i.e.\@ the intersection of the vertical dotted line in panel A with the orange curve). See text for discussion.}
\label{fig:closed}
\end{figure}
\ \\

$Q_{\text{poly}}$ can be modulated not only by the protomer concentration $t_S$ but also the protomer affinity $\sigma$ (Fig.\@ \ref{fig:closed}B). Increasing $t_S$ improves $Q_{\text{poly}}$ dramatically at all affinities up to a maximum after which enzyme and substrate become progressively separated due to the prozone effect. At all protomer concentrations, in particular around the maximizing one, $Q_{\text{poly}}$ always increases with increasing affinity $\sigma$. Fig.\@ \ref{fig:closed}B suggests that for the modulation through $\sigma$ to be most effective the protomer concentration should be close to the maximizing $t_S$.

\addcontentsline{toc}{section}{Comparison with multivalent scaffold systems}
\subsubsection*{Comparison with multivalent scaffold systems}

With regard to $Q$, a polymer chain of length $n$ is equivalent to a multivalent scaffold agent $S_{(n)}$ with $n$ binding sites for $A$ and $B$ each. It is therefore illuminating to compare the polymerizing system with multivalent scaffolds and their mixtures. 

It is straightforward to calculate the equilibrium concentration of configurations $A_pS_{(n)}B_q$ for an $n$-valent scaffold by adopting a site-oriented view that exploits the independence of binding interactions. The calculation (SI section 2) yields as a general result that the catalytic potential for an arbitrary scaffolding system, assuming independent binding of $A$ and $B$, consists of two factors:
\begin{align}
    Q=\underbrace{p(t_{\text{sit}},t_A,\alpha) p(t_{\text{sit}},t_B,\beta)}_{I}\underbrace{Q_{\text{max}}(\vec{t}_{S})}_{II}.
    \label{eq:Qn_simple}
\end{align}
The dimensionless function $p(t_{\text{sit}},t_X,\gamma)$ denotes the equilibrium fraction of X-binding \emph{sites}, with total concentration $t_{\text{sit}}$, that are occupied by ligands of type $X$, with total concentration $t_X$:
\begin{align*}
    p(t_{\text{sit}},t_X,\gamma)=\frac{\gamma t_X-\gamma t_{\text{sit}}-1+\sqrt{4\gamma t_X+(\gamma t_X-\gamma t_{\text{sit}} -1)^2}}{\gamma t_X-\gamma t_{\text{sit}}+1+\sqrt{4\gamma t_X+(\gamma t_X-\gamma t_{\text{sit}} -1)^2}}.
\end{align*}
This expression is the well-known dimerization equilibrium, computed at the level of sites rather than scaffolds and taken relative to $t_{\text{sit}}$ (SI section 2).

Factor \emph{I} depends on the total concentration of ligand binding sites (for each type) but not on how these sites are partitioned across the agents providing them. For example, a multivalent scaffold $S_{(n)}$, present at concentration $t_{S_{(n)}}$, provides $t_{\text{sit}}=n t_{S_{(n)}}$ binding sites and the probability that a site of any particular agent is occupied is the same as the probability that a site in a pool of $n t_{S_{(n)}}$ sites is occupied. For a heterogeneous mixture of multivalent scaffold agents we have $t_{\text{sit}}=\sum_{i=1}^n i\, t_{S_{(i)}}$; for a polymerizing system in which each protomer $S$ exposes one binding site we have $t_{\text{sit}}=t_S$.

Factor \emph{II} is the maximal $Q$ attainable in a scaffolding system. This factor depends on how sites are partitioned across scaffold agents with concentrations $\vec{t}_{S}=(t_{S_{(1)}},\ldots,t_{S_{(n)}})$, but does not depend on ligand binding equilibria. For example, a system of multivalent agents at concentrations $\vec{t}_{S}$ has $Q_{\text{max}}=\sum_{i=1}^n i^2 t_{S_{(i)}}$. The polymerizing scaffold system is analogous, but $n=\infty$ and the $t_{S_{(i)}}$ are determined endogenously by aggregation: $t_{S_{(i)}}=s_i=\sigma^{i-1}s^i$. This yields simple expressions for the catalytic potential of a polymerizing scaffold, $Q_{\text{poly}}$, and multivalent scaffold, $Q_{\text{multi}}$:
\begin{align}
    Q_{\text{poly}}&=p(t_S,t_A,\alpha) p(t_S,t_B,\beta)\dfrac{s(1+\sigma s)}{(1-\sigma s)^3} \label{eq:qpoly}\\
    Q_{\text{multi}}&=p(n\,t_{S_{(n)}},t_A,\alpha) p(n\,t_{S_{(n)}},t_B,\beta)n^2 t_{S_{(n)}} \nonumber
\end{align}
with $s$ in \eqref{eq:qpoly} given by \eqref{eq:free_s}. \eqref{eq:qpoly} is equivalent to \eqref{eq:Q}. While \eqref{eq:Q} requires solving a system of mass conservation equations to obtain $a$, $b$, and $s$, $Q_{\text{poly}}$ as given by \eqref{eq:qpoly} does not refer to $a$ and $b$, but only to $s$ as determined by the ligand-free polymerization subsystem. The $Q$ that shapes the Michaelis-Menten rate law under the assumption of rapid equilibration of enzyme-substrate binding has the same structure as \eqref{eq:Qn_simple}: $Q_{\text{direct}}=p(t_A,t_B,\alpha)t_A$, where $t_A$ and $t_B$ are the total enzyme and substrate concentration, respectively. The presence of a second concurrent binding equilibrium in \eqref{eq:Qn_simple} characterizes the prozone effect.

Adding sites, all else being equal, necessarily decreases the fraction $p$ of sites bound. Specifically, factor \emph{I} tends to zero like $1/t_{\text{sit}}^2$ for large $t_{\text{sit}}$. In contrast, $Q_{\text{max}}$ increases monotonically, since adding sites necessarily increases the maximal number of interaction opportunities between $A$ and $B$. For a multivalent scaffold $Q_{\text{max}}$ diverges linearly with $t_{\text{sit}}$. For the polymerizing system $Q_{\text{max}}$ diverges like $t_{\text{sit}}^{3/2}$ (SI section 5). 

Fig.\@ \ref{fig:mix}A provides a wide-range comparison of $Q_{\text{poly}}$ (red) with $Q_{\text{multi}}$ for various valencies (blue) at the same site concentration $t_{\text{sit}}=t_S$. 
\begin{figure}[!h]
\centering
\includegraphics[width=0.75\linewidth]{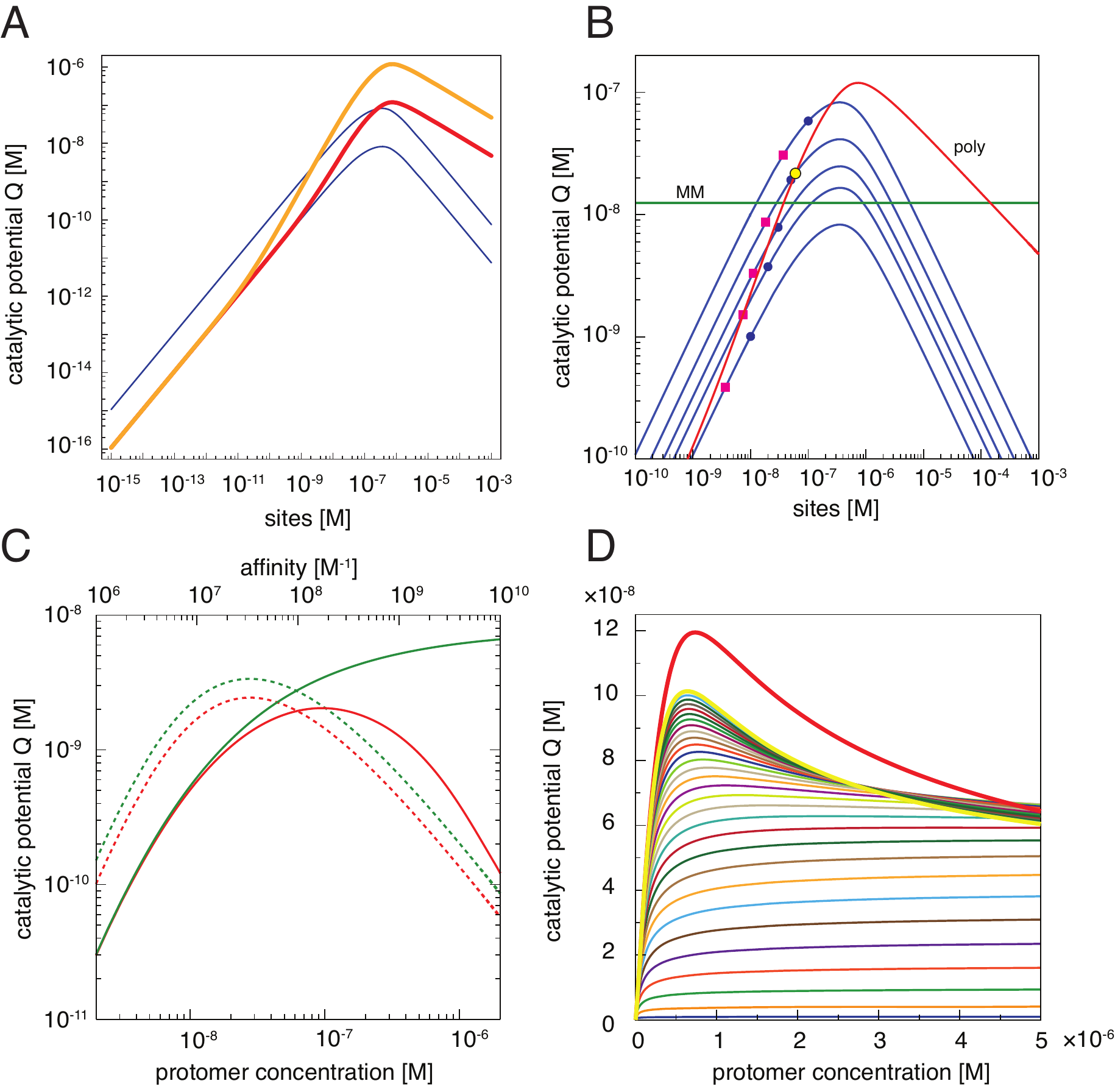}
\caption[Multivalent scaffolds and polymerizing scaffold]{Multivalent scaffolds and polymerizing scaffold. {\bf A:} Large-scale view of the catalytic potential $Q$ as a function of site concentration $t_S$. The blue curves depict $Q_{\text{multi}}$ for $n$-valent scaffolds (lower: $n=1$, higher: $n=10$). The location of the peak of $Q_{\text{multi}}$ is independent of the valency $n$ when expressed as a function of $t_{\text{sit}}=t_S$ (SI section 5, Eq.~38). The red and orange curves depict $Q_{\text{poly}}$ for two affinities (red: $\sigma=10^8$ M$^{-1}$, orange: $\sigma=10^{10}$ M$^{-1}$). Other parameters: $\alpha=\beta=10^7$ molecules$^{-1}$, $t_A=1.5\cdot 10^{-8}$ M, $t_B=5\cdot 10^{-7}$ M. On a log-log scale, the up-slope of $Q_{\text{poly}}$ is $1$ initially---the same as for multivalent scaffolds---and increases to $2$ prior to reaching the prozone peak. The down-slope is $-1/2$, whereas it is $-1$ for multivalent scaffolds (SI section 5). {\bf B:} Close-up of the peak region in panel A for the red curve; multivalent scaffolds were added for $n=2,3,5$. The slight asymmetry in the $Q$ profiles of multivalent scaffolds stems from the differences in ligand concentrations of our running example; see also SI, section 11. The yellow dot on the $Q_{\text{poly}}$ curve corresponds to the red dot in Fig.\@ \ref{fig:closed}. A pink square on a blue curve of valency $n$ marks $Q_{\text{multi}}$ when the scaffold concentration $t_{S(n)}$ is the same as the concentration of polymers of size $n$ ($s_n$) at the $t_S$ at which the length class $n$ dominates the polymerizing system (SI section 3 Fig.\@ S2B). A blue dot indicates the $Q_{\text{multi}}$ when the scaffold concentration $t_{S(n)}=1/\sigma$, which is the asymptotic (and maximal) value of $s_n$, for all $n$, in the limit of infinite $t_S$. These markers serve to show that within the most populated length classes the prozone peak is never reached. MM labels the Michaelis-Menten case of Fig.\@ \ref{fig:closed} for comparison. See text for details. {\bf C:} The solid lines in the graph exemplify the absence of a prozone within an isolated length class $n$, here $n=3$, and the presence of a prozone for the same class in the context of all other classes. Green solid: $Q_{\text{multi}}$ for $n=3$ using $t_{S_{(3)}}=s_3$ and $t_{\text{sit}}=3\,t_{S_{(3)}}$. Red solid: $Q_{\text{multi}}$ for $n=3$ using $t_{S_{(3)}}=s_3$ but $t_{\text{sit}}=t_S$. The dotted lines illustrate the situation for the length class $n=3$ as a function of affinity $\sigma$ (upper abscissa, same ordinate). In this dimension, the bending of the curves is \emph{not} due to a prozone effect, since the number of sites does not increase; see text. {\bf D:} Cumulative sums from $i=1$ to $n=30$ of $Q_{\text{multi}}$ with $t_{S_{(i)}}=s_i$ and $t_{\text{sit}}=\sum_{i=1}^n i\,t_{S_{(i)}}$.}
\label{fig:mix}
\end{figure}
\afterpage{\clearpage}

On a log-log scale, scaffolds of arbitrary valency $n$ exhibit a $Q_{\text{multi}}$ whose slope as a function of $t_{\text{sit}}$ is $1$, with offset proportional to $n$, until close to the peak. For the polymerizing scaffold, the first order term of the series expansion of $Q_{\text{poly}}$ is independent of the affinity $\sigma$ (SI section 5), whereas the second order term is linear in $\sigma$. Hence, for small $t_{\text{sit}}$, the polymerizing system behaves like a monovalent scaffold and any multivalent scaffold offers a better catalytic potential. However, as $t_S$ increases, the equilibrium shifts markedly towards polymerization, resulting in a slope of $2$, which is steeper than that of any multivalent scaffold. The steepening of $Q_{\text{poly}}$ is a consequence of longer chains siphoning off ligands from shorter ones (SI, section 4). All $n$-valent scaffolds reach their maximal $Q_{\text{multi}}$ at the same abundance of sites $t_{\text{sit}}=n\,t_{S_{(n)}}=t_S$ and before $Q_{\text{poly}}$. The superlinear growth in $Q_{\text{max}}$ of the polymerizing system softens the decline of $Q_{\text{poly}}$ to an order $t_S^{-1/2}$ for large $t_S$. In contrast, the decline of $Q_{\text{multi}}$ is of order $t_{\text{sit}}^{-1}$. In sum, the polymerizing scaffold system catches up with any multivalent scaffold, reaches peak-$Q$ later, and declines much slower. 

The mitigation of the prozone effect begs for a mechanistic explanation, since a prozone could occur not only within each length class but also between classes. To assess the within-class prozone, we think of a length class $k$ as if it were an \emph{isolated} $k$-valent scaffold population at concentration $t_{S_{(k)}}=s_k=\sigma^{k-1}s^k$ with $Q_{\text{multi}}=p(k\, s_k,t_A,\alpha) p(k\, s_k,t_B,\beta)k^2 s_k$. For all $k$, $s_k$ approaches monotonically the limiting value $1/\sigma$ as $t_S\to\infty$ (SI section 2, Fig.\@ S1A). Assuming equal affinity $\alpha$ for both ligands $A$ and $B$, peak-$Q_{\text{multi}}$ for a $k$-valent scaffold occurs at $t_{S_{(k)}}^{\text{peak}}=k^{-1} (\alpha^{-1}+\,(t_A+t_B)/2)$. Thus, when established through a polymerization system, $t_{S_{(k)}}$ can never exceed the concentration required for peak-$Q_{\text{multi}}$ for any $k$ up to $k=\sigma/\alpha$ (Fig.\@ \ref{fig:mix}B, blue dots). For the $\alpha$ used in the red curve of Fig.\@ \ref{fig:mix}B this lower bound is $k=10$ and the actual value, given employed values of $t_A$ and $t_B$, is about $k=35$. At the yellow marker and at peak-$Q_{\text{poly}}$ in Fig.\@ \ref{fig:mix}B $98$\% and $68$\%, respectively, of all sites are organized in length classes below $10$. Thus, the most populated lengths avoid the within-class prozone entirely (for example $k=3$ as depicted in Fig.\@ \ref{fig:mix}C, green solid line). Yet, the actual behavior of the $k$th length class occurs in the context of all other classes, i.e.\@ at site concentration $t_S$, not just $k\, s_k$. In this frame, the class indeed exhibits a prozone (Fig.\@ \ref{fig:mix}C, red solid line). The overall prozone of the polymerizing scaffold system is therefore mainly due to the spreading, and ensuing isolation, of ligands \emph{between} length classes. This \enquote{entropic} prozone becomes noticeable only when including all length classes up to relatively high $k$ because the majority of sites are concentrated at low $k$ where they are even jointly insufficient to cause a prozone, Fig.\@ \ref{fig:mix}D.

At constant $t_S$ and in the limit $\sigma\to\infty$, $s_k$ tends toward zero for all $k$ (SI, Fig.\@ S3C). In the $\sigma$-dimension, unlike in the $t_S$-dimension, the class $s_k$ itself has a peak. As $\sigma$ increases, the $k$ of the class that peaks at a given $\sigma$ increases. Consequently, the $Q_{\text{multi}}$ of each length-class in isolation will show a \enquote{fake} prozone with increasing $\sigma$, due entirely to the polymerization wave passing through class $k$ as it moves towards higher $k$ while flattening (Fig.\@ \ref{fig:mix}C, dotted lines). Since there is no site inflation, the overall $Q_{\text{poly}}$ increases monotonically.

Effects of ligand imbalance and unequal ligand binding affinities are discussed in the SI, section 11.

\addcontentsline{toc}{section}{Interaction horizon}
\subsubsection*{Interaction horizon}

The assumption that every $A$ can interact with every $B$ attached to the same scaffold construct is unrealistic. It can, however, be tightened heuristically without leaving the current level of abstraction. We introduce an \enquote{interaction horizon}, $q_{max}(l,h)$, defined as the radius $h$ in terms of scaffold bonds within which a bound $A$ can interact with a bound $B$ on a polymer of size $l$. In this picture, an $A$ can interact with at most $2h+1$ substrate agents $B$: $h$ to its \enquote{left}, $h$ to its \enquote{right} and the one bound to the same protomer. The interaction horizon only modulates the $Q_{\text{max}}$ of a polymer of length $l$, replacing the interaction factor $l^2$ with  (SI section 6):
\begin{align*}
    q_{max}(l,h) =\left\{ 
    \begin{array}{ll} 
    l(2h+1)-h(h+1), & \text{ for } 0\leq h\leq l-1\\ 
    l^2, & \text{ for } h\geq l 
    \end{array} \right.
\end{align*}
The horizon $h$ could be a function of $l$. One case, in which $h$ covers a constant \emph{fraction} of a polymer, is treated in section 6 of the SI. In a more restrictive scenario we assume a fixed horizon independent of length, which could reflect a constant local flexibility of a polymer chain. With the assumption of a constant $h$, \eqref{eq:qpoly} becomes (SI section 6)
\begin{align}
\label{eq:Qhorizon2}
    Q_{\text{poly}}=p(t_S,t_A,\alpha) p(t_S,t_B,\beta)\dfrac{s\left(1+\sigma s-2(\sigma s)^{h+1}\right)}{(1-\sigma s)^3}.
\end{align}
In \eqref{eq:Qhorizon2}, the numerator of the $Q_{\text{max}}$ term of \eqref{eq:qpoly} is corrected by $-2s(\sigma s)^{h+1}$. Since $\sigma s<1$ for all finite $t_S$ and $\sigma$, even moderate values of $h$ yield only a small correction to the base case of a limitless horizon.

\addcontentsline{toc}{section}{The discrete case in equilibrium}
\subsection*{The discrete case in equilibrium}

Replacing concentrations with particle numbers $t_S,t_A,t_B\in\Nat$ in a specified reaction volume yields the discrete case. In this setting, we must convert deterministic equilibrium constants, such as $\sigma$ to corresponding \enquote{stochastic} equilibrium constants $\sigma_s$ through $\sigma_s=\sigma / ({\cal A} V)$, where $\cal A$ is Avogadro's constant and $V$ the reaction volume to which the system is confined. For simplicity we overload notation and use $\sigma$ for $\sigma_s$.

The basic quantity we need to calculate is the average catalytic potential $\langle Q_{\text{poly}}\rangle=\sum_{l,i,j} i\,j\,\langle n_{lij}\rangle$, where $\langle n_{lij}\rangle$ is the average number of occurrences of a polymer of length $l$ with $i$ and $j$ ligands of type $A$ and $B$, respectively. Conceptually, $\langle n_{lij}\rangle$ counts the occurrences of an assembly configuration $A_iS_lB_j$ in every possible state of the system weighted by that state's Boltzmann probability. In the SI (section 7) we show that $\langle n_{lij}\rangle$ is given by the number of ways of building one copy of $A_iS_lB_j$ from given resources ($t_S$, $t_A$, $t_B$) times the ratio of two partition functions---one based on a set of resources reduced by the amounts needed to build configuration $A_iS_lB_j$, the other based on the original resources. The posited independence of all binding processes in our model implies that the partition function is the product of the partition functions of polymerization and dimerization, which are straightforward to calculate (SI section 8). While exact, the expressions we derive for $\langle Q_{\text{poly}}\rangle$ (SI, section 8, Eq.~66) and $\langle Q_{\text{multi}}\rangle$ (SI, section 8, Eq.~69) are sums of combinatorial terms and therefore not particularly revealing. For numerical evaluation of these expressions, we change the size of the system by a factor $\xi$ (typically $\xi=0.01$), i.e.\@ we multiply volume and particle numbers with $\xi$ and affinities with $1/\xi$. Such re-sizing preserves the average behavior. Our numerical examples therefore typically deal with $10$-$1000$ particles and stochastic affinities on the order of $10^{-2}$ to $10$ molecules$^{-1}$.

\begin{figure}[!h]
\centering
\includegraphics[width=0.7\linewidth]{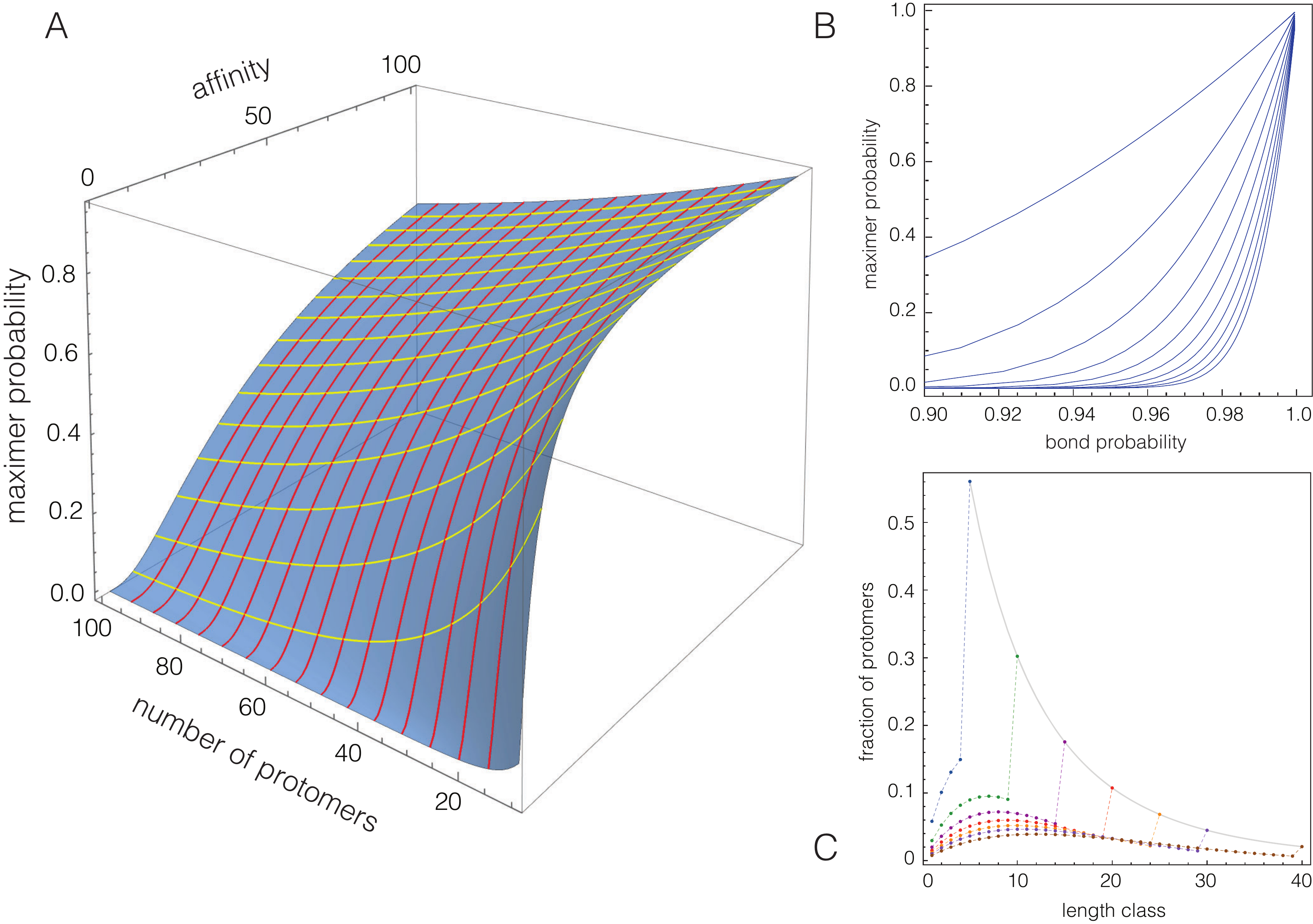}
\caption[Maximer]{Maximer. {\bf A:} The surface depicts the probability of observing the maximer as a function of $t_S$ and $\sigma$. {\bf B:} Here the maximer probability is graphed as function of the probability $p$ that a bond exists between two protomers. $p$ is a function of $t_S$ and $\sigma$ and can be calculated exactly. Each curve corresponds to a particular $t_S$ with varying $\sigma$. $t_S$ ranges from $10$ (topmost curve) to $100$ (bottom curve) in increments of $10$, while $\sigma$ ranges from $1$ to $1000$. {\bf C:} Mass distributions in the polymerizing scaffold model. Any curve depicts the fraction of protomers in all length classes $n$, computed as $n\,\sigma^{n-1}t_S!/(t_S-n)! \,Z^{\text{(poly)}}_{t_S-n}/Z^{\text{(poly)}}_{t_S}$ with $Z^{\text{(poly)}}_{t_S}$ the partition function for polymerization with $t_S$ protomers (SI, section 8). Each curve corresponds to a given number of protomers: $t_S=5$ (blue), $10$ (green), $15$ (plum), $20$ (red), $25$ (orange), $30$ (purple), $40$ (brown); affinity $\sigma=3$ in all cases. When $t_S$ is small, the longest possible polymer---the \enquote{maximer}---is realized with appreciable frequency and dominates the mass distribution. As $t_S$ increases, at fixed $\sigma$, the maximal length class increases too but its dominance fades.}
\label{fig:maximer}
\end{figure}
\ \\

The key aspect of the discrete case is the existence of a largest polymer consisting of all $t_S$ protomers. We refer to it as the \enquote{maximer}; no maximer exists in the continuum case because of the infinite fungibility of concentrations (Fig.\@ S9). Since there is only one maximer for a given $t_S$, its expectation is the probability of observing it: $\langle s_{\text{max}}\rangle=t_S!\, \sigma^{t_S-1}/Z^{\text{(poly)}}_{t_S}$, where $Z^{\text{(poly)}}_{t_S}$ is the partition function of polymerization (SI, sections 8 and 9). This probability is graphed as a function of $t_S$ and $\sigma$ in Fig.\@ \ref{fig:maximer}A. At any fixed $t_S$, the probability of observing the maximer will tend to $1$ in the limit $\sigma\to\infty$. This puts a ceiling to $Q_{\text{max}}$ that is absent from the continuum description. In the $t_S$-dimension, the maximer probability decreases as $t_S$ increases at constant $\sigma$.  

Polymerization as considered here has a natural analogy to bond percolation on a 1-dimensional lattice (SI, section 9). The probability of percolation (in which the entire lattice becomes one connected component) is parametrized by the probability $p$ of a bond between adjacent lattice sites. In the case of polymerization we can compute the probability $p$ that any two protomers are linked by a bond as a function of $t_S$ and $\sigma$. For continuum but not for discrete polymerization the analogy to percolation on an infinite 1D lattice is actually an exact correspondence (SI, section 9). For the present purpose, the percolation perspective is useful in that it combines the two main model parameters $t_S$ and $\sigma$ in the single quantity $p$ (Fig.\@ \ref{fig:maximer}B). As in finite-size percolation, the salient observation is that for small $t_S$ the maximer has a significant probability of already occurring at modest affinities; for example, given $10$ protomers and discrete binding affinity $1$, $p$ is already $0.78$ and the maximer probability a respectable $0.06$. For larger $t_S$, the maximer loses significance unless the affinity is scaled up correspondingly (SI section 10). This is also reflected in the mass distribution, Fig.\@ \ref{fig:maximer}C.

\begin{figure}[!h]
\centering
\includegraphics[width=0.8\linewidth]{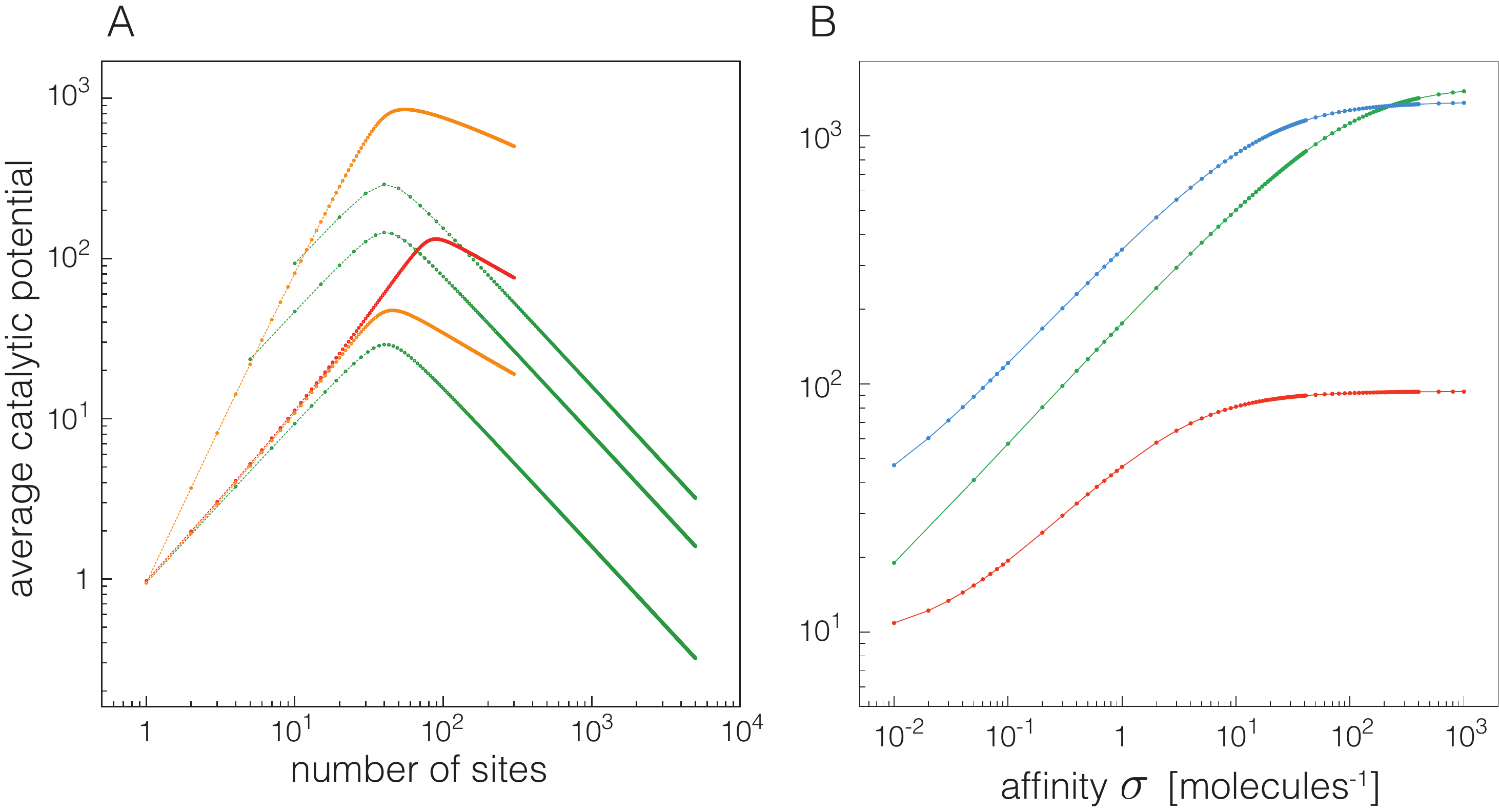}
\caption[Multivalent and polymerizing scaffolds in the discrete case]{Multivalent and polymerizing scaffolds in the discrete case. {\bf A:} Comparison of polymerizing scaffold (orange and red) with multivalent systems of various valencies (green). Orange: $t_A=t_B=40$, $\alpha=\beta=0.9$, $\sigma=10$ (upper) $\sigma=0.01$ (lower). All affinities in units of molecules$^{-1}$. Red: $t_A=t_B=80$, $\alpha=\beta=0.9$, $\sigma=0.01$. Green: $t_A=t_B=40$, $\alpha=\beta=0.9$, valency $n=10$ (top), $n=5$ (middle), $n=1$ (bottom). {\bf B:} $\langle Q_{\text{poly}}\rangle$ as a function of affinity $\sigma$. $t_A=t_B=40$, $\alpha=\beta=0.9$, $t_S=300$ (green), $t_S=10$ (red), $t_S=50$ (blue).}
\label{fig:mvpolydiscrete}
\end{figure}
\ \\

Fig.\@ \ref{fig:mvpolydiscrete}A compares the discrete polymerizing scaffold system with discrete multivalent scaffolds, much like Fig.\@ \ref{fig:mix}A for the continuum case. The behavior of the discrete case is essentially similar to that of the continuum case---with a few nuances that are prominent at low particle numbers and high affinities, such as the topmost orange curve. Its $\langle Q_{\text{poly}}\rangle$-profile does not hug the monovalent profile (bottom green chevron curve) to then increase its slope into the prozone peak as in the continuum case (Fig.\@ \ref{fig:mix}A). A behavior like in the continuum case is observed for the lower orange and red curves, for which $\sigma$ is much weaker. In the continuum case, the affinity does not affect slope---the slope always shifts from $1$ to $2$ within some region of protomer abundance; rather, the affinity determines where that shift occurs (Fig.\@ \ref{fig:mix}A). The higher the affinity, the earlier the shift. The topmost orange curve could be seen as realizing an extreme version of the continuum behavior in which an exceptionally high affinity causes a shift to slope 2 at unphysically low protomer concentrations. That such a scenario can be easily realized in the discrete case is due to the significant probability with which the maximer occurs at low particle numbers, similar to finite-size percolation. It bears emphasis that, as the number $t_S$ of protomers increases, the maximer probability decreases (Fig.\@ \ref{fig:maximer}C), since the length of the maximer is $t_S$. Yet, once the maximer has receded in dominance, the increased number of length classes below it have gained occupancy and control the catalytic potential much like in the continuum case. Likewise, affinity does not appear to affect the slope of the downward leg as $t_S$ increases.

The discrete multivalent scaffold system behaves much like its continuum counterpart.

In the affinity dimension, Fig.\@ \ref{fig:mvpolydiscrete}B, the discrete system shows a behavior similar to the continuum case with the qualification that $\langle Q_{\text{poly}}\rangle$ must level off to a constant, rather than increasing indefinitely. This is because, at constant $t_S$, an ever increasing affinity will eventually drive the system into its maximer ceiling. Because of the volume-dependence of stochastic equilibrium constants, such an increase in affinity at constant protomer number can be achieved by any physical reduction of the effective reaction volume, for example by confinement to a vesicle or localization to a membrane raft.

 We determined standard deviations using stochastic simulations of the cases presented in Fig.\@  \ref{fig:mvpolydiscrete}A (SI, section 12). For a given $\langle Q\rangle$, the standard deviation is larger after the prozone peak than before. Upon adding ligand binding sites, the ratio of standard deviation to mean (noise) increases much slower for the polymerizing system than for multivalent scaffolds.

\addcontentsline{toc}{section}{Main conclusions}
\subsection*{Main conclusions}

Our theoretical analysis of a polymerizing scaffold system shows that, at constant chemical potential, the system can be driven into criticality not only by increasing protomer concentration or affinity, but by just increasing ligand concentrations. 

In equilibrium, the system stands out in how the prozone effect plays out. Compared with multivalent scaffolds, the polymerizing system boosts catalytic potential on the upward leg beyond a certain protomer concentration; delays the prozone peak; and dramatically mitigates the collapse on the downward leg. We explain this behavior by how the polymer length distribution adjusts to changes in protomer concentration and affinity. The discrete case behaves likewise, but, at small protomer numbers, the existence of a maximal polymer manifests itself in behavior only attainable at extreme parameter values in the continuum case. 

A polymerizing scaffold could be viewed as a programmable surface whose extent can be regulated by varying parameters such as protomer concentration, polymerization affinity and, in a discrete setting, reaction volume. The system effectively concentrates interacting ligands, much like a vesicle would, but through a simpler mechanism. Given the pervasive potential for scaffold polymerization through DIX domains and the like, we suspect that many systems of this kind will be discovered. 

Our model is a stylized vignette amenable to analytic treatment and exploitable for insight. Adding a bond distance constraint to the interaction among ligands did not alter the fundamental picture. Taking into account conformational aspects of polymeric chains would be a useful step, as would generalizations in which scaffolding units of distinct types form multiply interconnected aggregates facilitating diverse ligand interactions. We would expect variations in the concentration of scaffold units to have wide ranging effects on the equilibrium mixture of assemblies and the overall catalytic potential. 

\ \\
{\bf Acknowledgements.} We gratefully acknowledge discussions with Tom Kolokotrones, Eric Deeds and Daniel Merkle.

\addcontentsline{toc}{section}{References}
\putbib
\end{bibunit}

\newpage
\begin{bibunit}

\section*{Supplementary Information}
\addcontentsline{toc}{section}{Supplementary Information}
\setcounter{section}{0}
\renewcommand*{\theHsection}{chX.\the\value{section}}
\setcounter{figure}{0}    
\renewcommand\thefigure{S\arabic{figure}}
\setcounter{equation}{0}

\section{$W$ and $Q$ in the polymerizing scaffold model}
\label{sec:W}

In this section we step through the treatment of the polymerizing scaffold model with more granularity.

A polymerizing scaffold protomer $S$ has $1$ binding site for each ligand $A$ and $B$. Let $\{A_pS_nB_q\}$ be the set of complexes (configurations) consisting of a scaffold polymer with $n$ protomers, $p$ agents of type $A$ and $q$ agents of type $B$; let $[\{A_pS_nB_q\}]$ denote their aggregate equilibrium concentration. 
The equilibrium concentration of any particular representative $A_pS_nB_q$ of that class is given by
\begin{align}
\label{eq:pnq}
    [A_pS_nB_q] = \sigma^{n-1}\alpha^p\beta^q s^n a^p b^q = \sigma^{n-1}s^n(\alpha a)^p(\beta b)^q,
\end{align}
where $a$, $b$, $s$ are the equilibrium concentrations of \emph{free} $A$, $B$, and $S$, respectively; $\alpha$ denotes the equilibrium constant of $A$ binding to $S$ and, similarly, $\beta$ and $\sigma$ are the equilibrium constants for $B$ binding to $S$ and for $S$ binding to $S$, respectively. All binding interactions are posited to be mechanistically independent of one another.

In an equilibrium treatment, a system of reactions only serves to define a set of reachable complexes and could be replaced with any other mechanism, no matter how unrealistic, as long as it produces the same set of reachable configurations. Hence we could posit that a polymer of length $n$ is generated by a reversible \enquote{reaction} in which all constituent protomers come together at once. The equilibrium constant of such an imaginary reaction must be the exponential of the energy content of a polymer of length $n$, which in our case is simply $(n-1)$ times the energy content of a single bond, i.e.\@ $\ln{\sigma}$. Thus, the equilibrium constant of the fictitious one-step assembly reaction is $\sigma^{n-1}$ and \eqref{eq:pnq} follows.

To aggregate the equilibrium concentrations of all molecular configurations in the class $\{A_pS_nB_q\}$ we note that the set $\{A_pS_nB_q\}$ includes $\binom{n}{p}\binom{n}{q}$ configurations with the same energy content $\sigma^{n-1}\alpha^p\beta^q$. Summing over all $p$ and $q$, yields the contribution of the polymer length class $n$, $\{A_{\ast}S_nB_{\ast}\}$
\begin{align}
\label{eq:sump}
    [\{A_{\ast}S_nB_{\ast}\}]&=\sigma^{n-1}s^n\left[\sum_{p=1}^{n}\binom{n}{p}\alpha^p\,a^p\right]\left[\sum_{q=1}^{n}\binom{n}{q}\beta^q\,b^q\right]=\sigma^{n-1}s^n(1+\alpha\,a)^n(1+\beta\,b)^n\nonumber\\
    &=\frac{1}{\sigma}\left(\sigma\,s\,(1+\alpha\,a)(1+\beta\,b)\right)^n
\end{align}
Summing over all equilibrium concentrations defines a function $W$:
\begin{align}
    W=a+b+\frac{1}{\sigma}\sum_{n=1}^\infty\left(\sigma\,s\,(1+\alpha a)\, (1+\beta b)\right)^n
    =a+b+s(1+\alpha a) (1+\beta b)\sum_{n=0}^\infty\left(\sigma s (1+\alpha a) (1+\beta b)\right)^n \label{eq:W_m0_}
\end{align}
When viewing $a$, $b$ and $s$ as formal variables, $W$ acts as a generating function of energy-weighted configurational counts. By differentiating $W$ with respect to $s$, each $s$-containing term gets multiplied with the exponent of $s$, which is the $S$-content of the respective configuration. Multiplying by $s$ then restores the exponent and recovers the equilibrium concentration of the respective configuration. Summing over all configurations so treated, yields the total amount of $S$ protomers in the system and thus a conservation relation. This holds for all formal variables representing the \enquote{atoms}, or building blocks, of the system:
\begin{align}
\label{eq:free_equations}
t_A=a\frac{\partial W(a,b,s)}{\partial a},\quad t_B=b\frac{\partial W(a,b,s)}{\partial b}\quad t_S=s\frac{\partial W(a,b,s)}{\partial s}.
\end{align}
By solving the equations \eqref{eq:free_equations}, we obtain the equilibrium concentrations of free $A$, $B$, and $S$ needed to compute the equilibrium concentration of any configuration:

\begin{align}
a&=\dfrac{\alpha t_A - \alpha t_S - 1 + \sqrt{(\alpha t_A+\alpha t_S+1)^2-4 \alpha t_A \alpha t_S}}{2 \alpha} \label{eq:am0closed}\\
b&=\dfrac{\beta t_B - \beta t_S - 1 + \sqrt{(\beta t_B+\beta t_S+1)^2-4 \beta t_B \beta t_S}}{2 \beta} \label{eq:bm0closed}\\
s&=\dfrac{2}{\sigma^2 t_S} \dfrac{2 \sigma  t_S+1 - \sqrt{4 \sigma  t_S+1}}{\left(\alpha  t_A-\alpha  t_S+1+\sqrt{(\alpha  t_A+\alpha  t_S+1)^2-4 \alpha t_A  \alpha  t_S}\right) \left(\beta  t_B-\beta  t_S+1+\sqrt{(\beta  t_B+\beta  t_S+1)^2-4 \beta  t_B \beta  t_S}\right)}
\label{eq:sm0closed}
\end{align}

Carrying out the geometric sum in \eqref{eq:W_m0_} yields equation (2) in the main text:
\begin{align}
    W(a,b,s)=a+b+\frac{s(1+\alpha a) (1+\beta b)}{1-\sigma s(1+\alpha a) (1+\beta b)}.
    \label{eq:W_m0}
\end{align}
The same manipulation of $W$ used to obtain \eqref{eq:free_equations} can be carried out twice, once for $a$ and once for $b$, to yield the catalytic potential of the system:
\begin{align}
    Q=a\,b\,\frac{\partial^2}{\partial a\partial b} W(a,b,s),
\end{align}
given as equation (3) in the main text.

By setting $a=b=0$, we recover the standalone polymerization system with 
\begin{align}
    W(s)=\frac{s}{1-\sigma s}
    \label{eq:W_poly}
\end{align}
and $s$ obtained from solving $t_S=dW(s)/ds$:
\begin{align}
\label{eq:free_s_poly}
    s=\frac{1}{4\sigma}\left(\sqrt{4+\frac{1}{\sigma t_S}}-\sqrt{\frac{1}{\sigma t_S}}
    \right)^2,
\end{align}
as in equation (1) of the main text. We discuss the main properties of the standalone polymerization system in section \ref{sec:poly} of this Appendix. In an equilibrium setting, the critical point of the model with ligands $A$ and $B$ should be the same as that of the polymerization system without ligands, namely $t_S\to\infty$ or $\sigma\to\infty$. This is not obvious from $W$ (whose critical point $Q$ inherits) as given in \eqref{eq:W_m0} with solutions \eqref{eq:am0closed}-\eqref{eq:sm0closed}. However, it is made explicit in an alternative, more insightful derivation of the equilibrium catalytic potential $Q$ given in section \ref{sec:general} of this Appendix.

\section{Derivation of the general expression for the catalytic potential}
\label{sec:general}

In this section we derive expression (4) of the main text.

We consider a \emph{multivalent} scaffold agent $S$ with $n_A$ binding sites for $A$ and $n_B$ binding sites for $B$. Our goal is to calculate the catalytic potential $Q_{\text{multi}}$ of a system consisting of $A$-agents at concentration $t_A$, $B$-agents at concentration $t_B$, and $S$-agents at concentration $t_S$. 

The function $W(a,b,s)$, introduced in the main text for the polymerizing scaffold system, sums up the equilibrium concentrations of all possible entities in the system. The same concept applies to a multivalent scaffold:
\begin{align}
    W_{\text{multi}}(a,b,s)=a+b+s(1+\alpha a)^{n_A} (1+\beta b)^{n_B}
\end{align}
with $a$, $b$, and $s$ the equilibrium concentrations of the free $A$, $B$, and $S$, respectively. The catalytic potential $Q_{\text{multi}}$ of the multivalent scaffold system is
\begin{align}
\label{eq:iprozone_Q}
    Q_{\text{multi}}=a\,b\,\dfrac{\partial^2}{\partial a \partial b} W_{\text{multi}}(a,b,s)=s\,\alpha\,\beta\,a\,b\,n_A\,n_B\,(1+\alpha a)^{n_A-1} (1+\beta b)^{n_B-1}.
\end{align}
The equilibrium concentrations $a$, $b$, and $s$ are determined by the system of conservation equations
\begin{align}
\label{eq:indiv_prozone}
    a\dfrac{\partial}{\partial a}W=t_A, \quad b\dfrac{\partial}{\partial b}W=t_B, \quad s\dfrac{\partial}{\partial s}W=t_S.
\end{align}
However, we can bypass solving these equations by calculating the concentrations directly, which serendipitously gives us an intelligible expression for the catalytic potential $Q$ in general.

We first calculate the equilibrium concentration of the fully occupied scaffold configuration, $[A_{n_A}SB_{n_B}]$ by reasoning at the level of binding \emph{sites}. The concentration of \emph{sites} \emph{available} for binding to $S$ are denoted by $a$, which is also the concentration of free $A$-agents. Since each $A$-binding site on $S$ is independent, the equilibrium fraction of $S$-agents that are fully occupied with $A$-agents is simply
\begin{align}
\label{eq:equia}
    \dfrac{[\{A_{n_A}S\}]}{t_S}=\left(\dfrac{\alpha a}{1+\alpha a}\right)^{n_A}
\end{align}
The expression in parentheses is the single-site binding equilibrium. Likewise, let $[s]$ be the concentration of free $A$-binding sites on $S$-agents and $[as]$ the concentration of bonds between $A$- and $S$-agents. In equilibrium we have that
\begin{align}
    \alpha a\,[s]=[as], \quad n_A t_S=[s]+[as],\quad t_A=a+[as].
\end{align}
Hence, $a=[as]/(\alpha[s])$ or $a=(t_A-a)/(\alpha[s])=(t_A-a)/(\alpha(n_A t_S-t_A+a))$, which yields a quadratic in $a$ whose solution is
\begin{align}
\label{eq:sites-on-a}
    a=\dfrac{1}{2\alpha}\left(\alpha t_A-n_A\alpha t_S-1+\sqrt{(\alpha t_A-n_A\alpha t_S-1)^2+4\alpha t_A}\right).
\end{align}
We plug \eqref{eq:sites-on-a} into \eqref{eq:equia} to obtain
\begin{align}
\label{eq:AS}
    \dfrac{[\{A_{n_A}S\}]}{t_S}=\left(\dfrac{\alpha t_A-n_A\alpha t_S-1+\sqrt{(\alpha t_A-n_A\alpha t_S-1)^2+4\alpha t_A}}{\alpha t_A-n_A\alpha t_S+1+\sqrt{(\alpha t_A-n_A\alpha t_S-1)^2+4\alpha t_A}}\right)^{n_A}.
\end{align}
The same reasoning holds for the (independent) binding of $B$ to $S$:
\begin{align}
\label{eq:SB}
    \dfrac{[\{SB_{n_B}\}]}{t_S}=\left(\dfrac{\beta t_B-n_B\beta t_S-1+\sqrt{(\beta t_B-n_B\beta t_S-1)^2+4\beta t_B}}{\beta t_B-n_B\beta t_S+1+\sqrt{(\beta t_B-n_B\beta t_S-1)^2+4\beta t_B}}\right)^{n_B}.
\end{align}
At this point it is useful to abbreviate

\begin{equation}
\label{eq:abbrev}
\begin{aligned}
    a_{\pm}&\equiv a_{\pm}(t_A,t_S,\alpha,n_A)=\alpha t_A-n_A\alpha t_S\pm 1+\sqrt{(\alpha t_A-n_A\alpha t_S-1)^2+4\alpha t_A}\\
    b_{\pm}&\equiv b_{\pm}(t_B,t_S,\beta,n_B)=\beta t_B-n_B\beta t_S\pm 1+\sqrt{(\beta t_B-n_B\beta t_S-1)^2+4\beta t_B}\\
\end{aligned}
\end{equation}
Note that these abbreviations are dimensionless functions of the parameters $t_A$, $t_S$, $\alpha$ and $n_{A/B}$. Because $A$ and $B$ bind independently, we can combine \eqref{eq:AS} and \eqref{eq:SB} to obtain:
\begin{align}
\label{eq:fullyoccupied}
    [A_{n_A}SB_{n_B}]=t_S
    \dfrac{a_{-}^{n_A}}{a_{+}^{n_A}}\dfrac{b_{-}^{n_B}}{b_{+}^{n_B}}=(\alpha a)^{n_A}(\beta b)^{n_B}s,
\end{align}
where the last equation is the equilibrium concentration in terms of free $A$, free $B$, and free $S$, as mentioned in the Introduction of the main text (and section \ref{sec:W} of this Appendix). The expression $a$ for free $A$ is given by \eqref{eq:sites-on-a}, or $a=a_{-}/(2\alpha)$. The expression $b$ for free $B$ is analogous, $b=b_{-}/(2\beta)$. Equation \eqref{eq:fullyoccupied} now yields $s$:
\begin{align}
\label{eq:iprozone_s}
    s = t_S\dfrac{1}{(\alpha a)^{n_A}(\beta b)^{n_B}}\dfrac{a_{-}^{n_A}}{a_{+}^{n_A}}\dfrac{b_{-}^{n_B}}{b_{+}^{n_B}}
    =t_S\dfrac{2^{n_A}2^{n_B}}{a_{+}^{n_A}b_{+}^{n_B}}
\end{align}
To summarize, using abbreviations \eqref{eq:abbrev}:
\begin{align}
\label{eq:free}
    a =\dfrac{a_{-}}{2\alpha},\quad 
    b =\dfrac{b_{-}}{2\beta}, \quad
    s =t_S\left(\dfrac{2}{a_{+}}\right)^{n_A}\left(\dfrac{2}{b_{+}}\right)^{n_B}.
\end{align}
Keep in mind that $a_{+/-}$ and $b_{+/-}$ are not constants, but functions of the system parameters. We now insert \eqref{eq:free} into \eqref{eq:iprozone_Q} to obtain
\begin{align}
     Q_{\text{multi}}&=n_A\,n_B\,s\,\left(\dfrac{\alpha a}{1+\alpha a}\right)\left(\dfrac{\beta b}{1+\beta b}\right)(1+\alpha a)^{n_A} (1+\beta b)^{n_B} \nonumber \\
     &=n_A\,n_B\,t_S\left(\dfrac{2}{a_{+}}\right)^{n_A}\left(\dfrac{2}{b_{+}}\right)^{n_B}\left(\dfrac{\alpha a}{1+\alpha a}\right)\left(\dfrac{\beta b}{1+\beta b}\right)(1+\alpha a)^{n_A} (1+\beta b)^{n_B} \nonumber \\
     &=n_A\,n_B\,t_S\left(\dfrac{\alpha a}{1+\alpha a}\right)\left(\dfrac{\beta b}{1+\beta b}\right)\left(\cancel{\dfrac{2+2\alpha a}{a_{+}}}\right)^{n_A}\left(\cancel{\dfrac{2+2\beta b}{b_{+}}}\right)^{n_B} \nonumber \\
     &=n_A\,n_B\,t_S\left(\dfrac{\alpha a}{1+\alpha a}\right)\left(\dfrac{\beta b}{1+\beta b}\right) \nonumber \\
     &=n_A\,n_B\,t_S\dfrac{a_{-}}{a_{+}}\dfrac{b_{-}}{b_{+}}.\label{eq:Qn}
\end{align}
The cancellations are due to $2\alpha a=a_{-}$ (from \eqref{eq:free}) and $a_{+}=a_{-}+2$ (from \eqref{eq:abbrev}). 

Return to equation \eqref{eq:AS} and set $n_A=1$. This gives the fraction of $A$-binding sites (of monovalent scaffold agents) that are occupied, that is, the probability that an $A$ is bound:
\begin{align}
\label{eq:reallyimportant}
    p(t_S,t_A,\alpha) =\dfrac{a_{-}(t_A,t_S,\alpha,1)}{a_{+}(t_A,t_S,\alpha,1)}
    =\frac{\alpha t_A-\alpha t_S-1+\sqrt{(\alpha t_A-\alpha t_S-1)^2+4\alpha t_A}}{\alpha t_A-\alpha t_S+1+\sqrt{(\alpha t_A-\alpha t_S-1)^2+4\alpha t_A}}
\end{align}
In the site-oriented view it does not matter whether an $A$-binding site belongs to a monovalent scaffold agent or to an $n$-valent scaffold agent. At the same agent concentration $t_S$, the $n$-valent agent simply provides $n$ times more sites. Thus, the probability that an $A$ is bound if the scaffolds are $n$-valent is
\begin{align}
\label{eq:pas}
    p(n t_S,t_A,\alpha) =\dfrac{a_{-}(t_A,t_S,\alpha,n)}{a_{+}(t_A,t_S,\alpha,n)}
    =\dfrac{a_{-}(t_A,n t_S,\alpha,1)}{a_{+}(t_A,n t_S,\alpha,1)},
\end{align}
since the number of binding sites only scales $t_S$ in \eqref{eq:abbrev}. With these observations, we can rephrase \eqref{eq:Qn} as the product of two terms:
\begin{align}
    Q_{\text{multi}}=\underbrace{p(n_A t_S,t_A,\alpha) p(n_B t_S,t_B,\beta)}_{I}\underbrace{n_A\,n_B\,t_S}_{II}.
    \label{eq:Qn_simpleSI}
\end{align}

Term (I) is the probability that a site of \emph{some} $S$ is occupied by $A$ and a site of \emph{some} $S$ is occupied by $B$. Term (II) counts the maximal number of possible interactions between $A$ and $B$ agents in the system. 

Let $S_{(i)}$ denote an agent of valency $i$ for both ligands and let $t_{S_{(i)}}$ denote its concentration. In a mixture of multivalent scaffold types of distinct valencies $i=1,\ldots,n$ present at concentrations $t_{S_{(i)}}$, the catalytic potentials of each type add up to that of the mixture, $Q_{\text{mix}}$:
\begin{align}
    Q_{\text{mix}}=p\left({\scriptstyle \sum_{i=1}^n} i\, t_{S_{(i)}},t_A,\alpha\right) p\left({\scriptstyle \sum_{i=1}^n} i\, t_{S_{(i)}},t_B,\beta\right)\sum_{i=1}^n i^2 t_{S_{(i)}}.
    \label{eq:Qn_mix}
\end{align}
Generally, we can write $Q_{\text{mix}}$ as
\begin{align}
    Q_{\text{mix}}=p(t_{\text{sit}},t_A,\alpha) p(t_{\text{sit}},t_B,\beta)\,Q_{\text{max}}(\vec{t}_S).
    \label{eq:Qn_general}
\end{align}
 In \eqref{eq:Qn_general}, $t_{\text{sit}}$ is the total concentration of binding sites, regardless of how they are partitioned across scaffold agents, $\vec{t}_S=(t_{S(i)},\dots,t_{S(n)})$ is a partition of sites across scaffold molecules of different valencies, and $Q_{\text{max}}$ is the maximal attainable number of enzyme-substrate interactions in the system, which depends on the concentration of scaffolds and their valency.

If the mixture results from a polymerization process between monovalent scaffolds $S\equiv S_{(1)}$, we identify a polymer of length $l$ with an $l$-valent scaffold agent (Figure \ref{fig:s-agent}).

\begin{figure}[!h]
\centering
\includegraphics[width=0.7\linewidth]{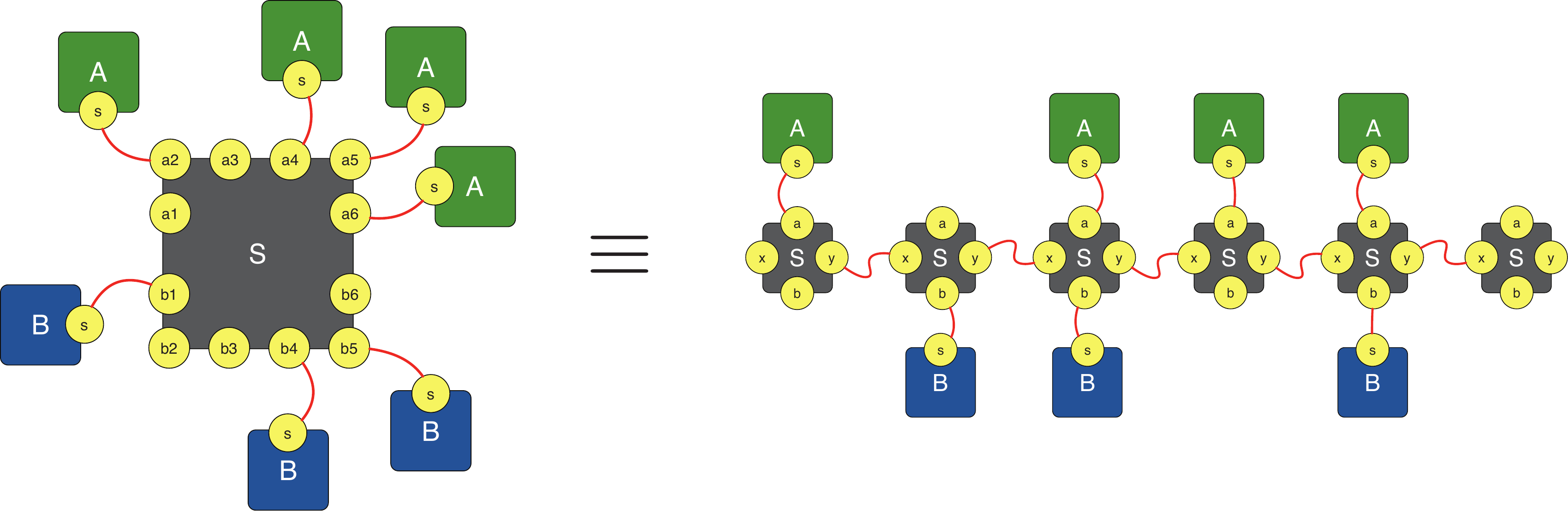}
\caption[Scaffold polymers and multivalent agents]{A multivalent scaffold agent can be thought as representing a particular scaffold polymer configuration.}
\label{fig:s-agent}
\end{figure}
\ \\

The concentrations $t_{S_{(l)}}$ are endogenously determined by polymerization at equilibrium:
\begin{align}
t_{S_{(l)}}=\sigma^{l-1}s^l, \nonumber
\end{align}
where the expression for $s$ is given by the expression for the equilibrium concentration of free monomer in the polymerization system \emph{absent ligands}, expression \eqref{eq:free_s_poly} in section \ref{sec:W} (equation (1) in the main text). Using these $t_{S_{(l)}}$ in the sum \eqref{eq:Qn_mix}, which in the continuum case runs to $n=\infty$, yields the expression (5) for $Q_{\text{poly}}$ in the main text:
\begin{align}
Q_{\text{poly}}=p(t_S,t_A,\alpha) p(t_S,t_B,\beta)\sum_{n=1}^{\infty} n^2 \sigma^{n-1}s^n=p(t_S,t_A,\alpha) p(t_S,t_B,\beta)\dfrac{s(1+\sigma s)}{(1-\sigma s)^3}, \label{eq:qpolySI}
\end{align}
with $p(\cdots)$ given by \eqref{eq:reallyimportant}.

\section{Overview of the polymerization system}
\label{sec:poly}

In this section we summarize some combinatorial properties of the polymerization subsystem. Understanding the concentration profile of the polymer length distribution is useful for rationalizing the overall behavior with respect to catalytic potential, because we can view the polymerizing scaffold system as a mixture of multivalent scaffolds whose concentration is set by polymerization. Since this is the simplest conceivable polymerization system, it would surprise us if anything being said here isn't already known in some form or another. Some of the features described can be found in Flory \cite{Flory1936}.

Let $S_n$ be a polymer of length $n$ and let $s_n$ denote the equilibrium concentration of polymers in length class $n$. To conform with our previous notation, we shall refer to the equilibrium concentration of the monomer as $s\equiv s_1$ and to the monomer species as $S\equiv S_1$. As stated repeatedly,
\begin{align}
    s_n=\sigma^{n-1}s^n \quad \text{ with }\quad s=\frac{1}{4\sigma}\left(\sqrt{4+\frac{1}{\sigma t_S}}-\sqrt{\frac{1}{\sigma t_S}}
    \right)^2 
    \label{eq:free_s_again}
\end{align}

Figure \ref{fig:sigma_ts} shows the dependency of $s_n$ on the total protomer concentration $t_S$ (panels A and B) and the affinity $\sigma$ (panels C and D). Obviously, $s_n$ is a geometric progression, thus linear in a lin-log plot for all parameter values (insets of panel A and C). 

\begin{figure}[!h]
\centering
\includegraphics[width=0.6\linewidth]{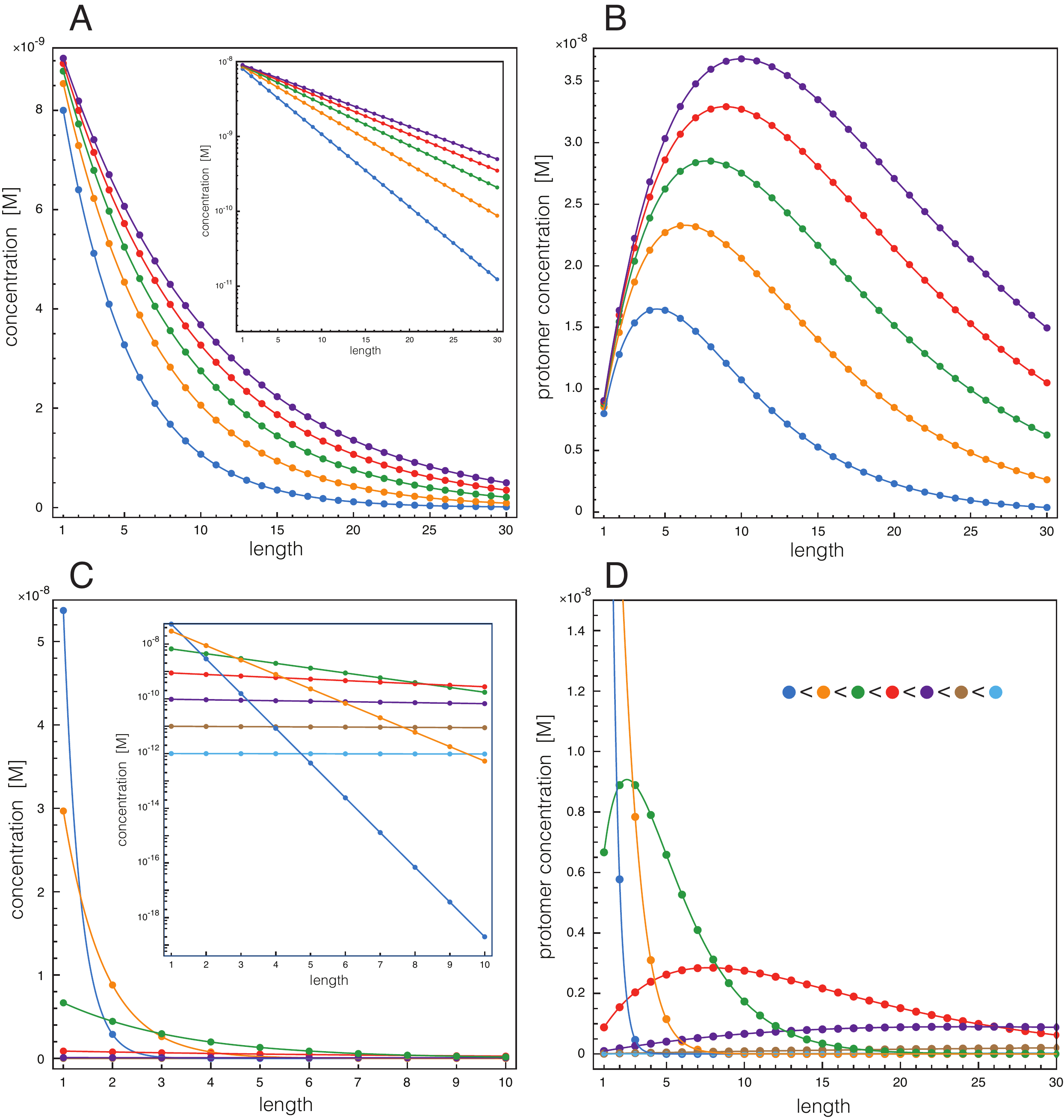}
\caption[The dependence of the length distribution on the protomer concentration $t_S$ and the affinity $\sigma$]{The dependence of the length distribution on the protomer concentration $t_S$ and the affinity $\sigma$. {\bf A:} The curves depict the length distribution $s_i$ of the linear polymerization subsystem with varying $t_S$ at $\sigma=10^8$ M$^{-1}$. Blue: $t_S=2\cdot 10^{-7}$ M, orange: $t_S=4\cdot 10^{-7}$ M, green: $t_S=6\cdot 10^{-7}$ M, red: $t_S=8\cdot 10^{-7}$ M, purple: $t_S=1\cdot 10^{-6}$ M. The inset plots the same curves in lin-log. {\bf B:} The curves depict the concentrations of protomers in each length class, that is, the \enquote{mass} distribution $i\,s_i$ under the same conditions as in panel {\bf A}. {\bf C:} The curves depict the length distribution $s_i$ with varying polymerization affinity $\sigma$ at $t_S=6\cdot 10^{-8}$ M. Blue: $\sigma=10^{6}$ M$^{-1}$, orange: $\sigma=10^{7}$ M$^{-1}$, green: $\sigma=10^{8}$ M$^{-1}$, red: $\sigma=10^{9}$ M$^{-1}$, purple: $\sigma=10^{10}$ M$^{-1}$, brown: $\sigma=10^{11}$ M$^{-1}$, light blue: $\sigma=10^{12}$ M$^{-1}$. {\bf D:} As in panel B, but with varying affinity $\sigma$ (as in panel C) at $t_S=6\cdot 10^{-8}$. For all panels $\alpha=\beta=10^7$ M$^{-1}$, $t_A=15\cdot 10^{-9}$ M and $t_B=5\cdot 10^{-7}$ M.}
\label{fig:sigma_ts}
\end{figure}
\ \\

In the $t_S$ dimension, $s_n$ approaches $1/\sigma$ from below for each $n$ and there is no value of $t_S$ that maximizes $s_n$. In the $\sigma$ dimension, $s_n$ approaches $0$ like $1/\sigma$ (in the lin-log plot, inset of panel C, the straight lines become less tilted and sink toward $0$); see also expansions \eqref{eq:snts} and \eqref{eq:snsigma} below. However, for any given length class $n$, there is a $\sigma$ that maximizes the concentration of that class:
\begin{align}
    \sigma=\frac{n^2-1}{4t_S}.
\end{align} 
At that $\sigma$, the respective $s_n$ is the most frequent, i.e.\@ the most dominant, length class. It does not mean that $s_n$ is at its most frequent, for $s_n$ rises to $1/\sigma$ as $t_S\to\infty$. In the continuum description, the most frequent polymer class is always the monomer, for any $t_S$ or $\sigma$. This is much more pronounced in the $t_S$ dimension than the $\sigma$ dimension.

\begin{figure}[!h]
\centering
\includegraphics[width=0.6\linewidth]{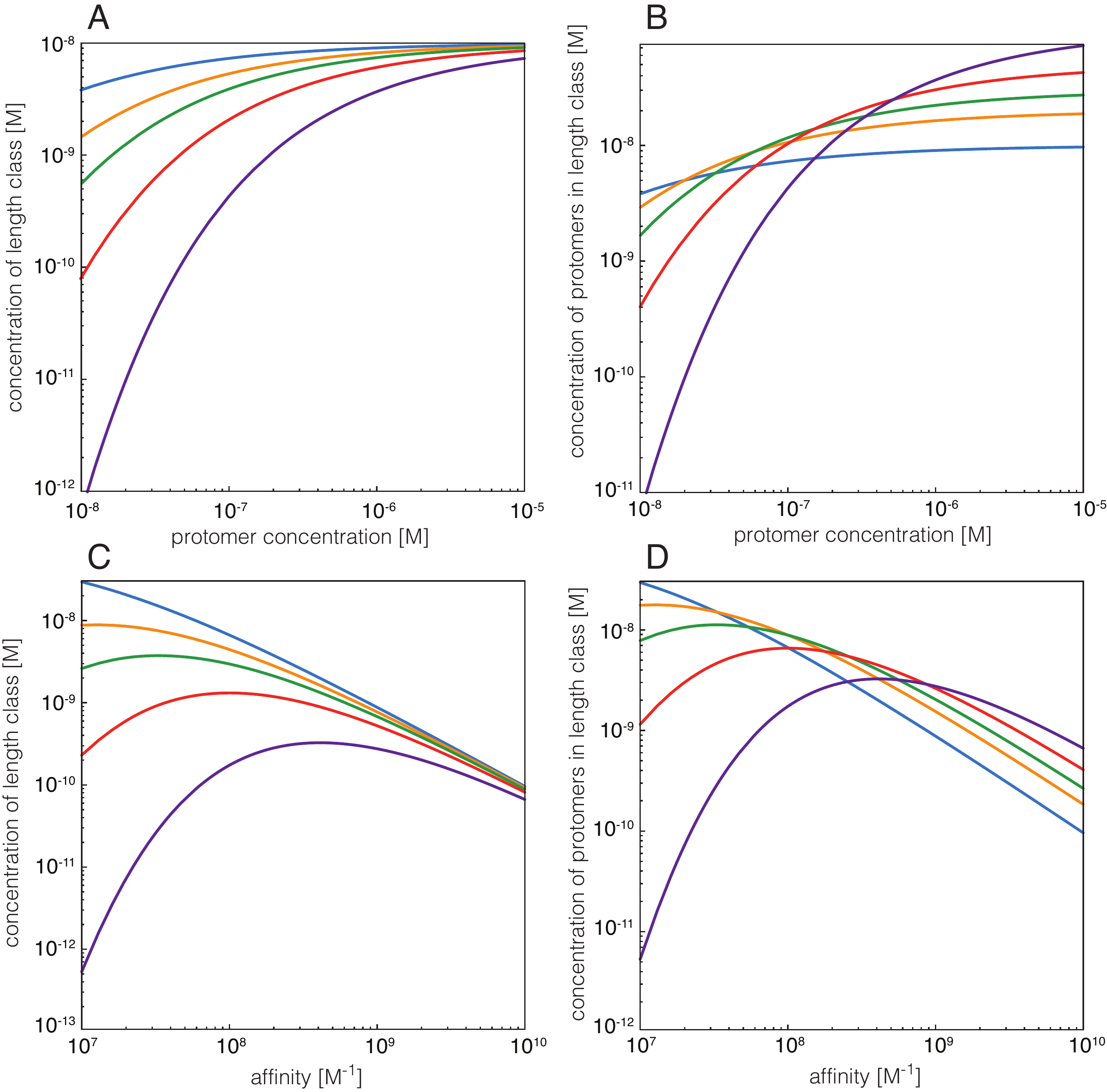}
\caption[Concentrations within length classes]{Concentrations within length classes. These panels are complementary to those in Figure \ref{fig:sigma_ts}. Each curve tracks the concentration of a particular length class $n$ as protomer concentration $t_S$ and affinity $\sigma$ are varied, effectively following the changes along a vertical cut across the curves in Figure \ref{fig:sigma_ts}. Blue: $n=1$, orange: $n=2$, green: $n=3$, red: $n=5$, purple: $n=10$. All other parameters as in Figure \ref{fig:sigma_ts}. {\bf A:} Concentration $s_n$ of length class $n$ with varying $t_S$. {\bf B:} Concentration $n s_n$ of the mass in length class $n$ with varying $t_S$. Panel {\bf C:} Concentration $s_n$ of length class $n$ with varying $\sigma$. Panel {\bf D:} Concentration $n s_n$ of the mass in length class $n$ with varying $\sigma$.}
\label{fig:oligo}
\end{figure}
\ \\

Panels B and D of Figure \ref{fig:sigma_ts} show the \enquote{mass} distribution, $n s_n$, i.e.\@ the concentration of protomers in each length class. For all values of $t_S$ and $\sigma$ the mass exhibits a maximum at some class length. This maximum wanders towards ever larger $n$ with increasing $t_S$ and $\sigma$, while its value steadily increases with $t_S$, whereas it decreases with increasing $\sigma$. The length class $n$ whose mass is maximized at a given $t_S$ and $\sigma$ is 
\begin{align}
\label{eq:nmax}
    n_{\text{max}}=\left[\log\left(\dfrac{4t_S\sigma}{\left(\sqrt{1+4t_S\sigma}-
    1\right)^2}\right)\right]^{-1},
\end{align}
and, for given $\sigma$ and $n$, the $t_S$ at which the class $n$ becomes the most massive of all classes is given by
\begin{align}
    t_S=\frac{\exp(1/n)}{\sigma(1-2\exp(1/n)+\exp(2/n))}.
\end{align}
The pink squares on the blue multivalent scaffold curves in Figure 4B of the main text correspond to the catalytic potential $Q$ that obtains at this concentration of sites. The same expression obtains for $\sigma$ by swapping $t_S$ and $\sigma$. At the $t_S$ at which the mass in class $n$ peaks, the concentration of the class is
\begin{align}
\label{eq:snmax}
    s_{n_{\text{max}}}=\dfrac{1}{e\sigma},
\end{align}
independent of $n_{\text{max}}$. Equation \eqref{eq:nmax} assumes a continuous $n$; thus, to account for the discrete nature of polymer length, the actual $n_{\text{max}}$ should be the nearest integer to the $n_{\text{max}}$ given in \eqref{eq:nmax}. Accordingly, the actual value of $s_{n_{\text{max}}}$ in expression \eqref{eq:snmax} will wobble slightly.

Switching perspective from the length distribution to the behavior within a length class yields Figure \ref{fig:oligo}. The expansion of $s_n$ shows how each length class approaches its limit as $t_S\to\infty$ or $\sigma\to\infty$ (multiply by $n$ for the mass distribution):
\begin{align}
    &\text{As } t_S\to \infty,\, s_n\to\frac{1}{\sigma} \text{ with } \frac{1}{\sigma}-\frac{n}{\sigma^{3/2}}\frac{1}{t_S^{1/2}}+O\left(\frac{1}{t_S}\right) \label{eq:snts}\\
    &\text{As } \sigma\to \infty,\, s_n\to 0 \text{ with } \frac{1}{\sigma}-\frac{n}{t_S^{1/2}}\frac{1}{\sigma^{3/2}}+O\left(\frac{1}{\sigma^2}\right) \label{eq:snsigma}
\end{align}

\section{Mixtures of multivalent scaffolds}
\label{sec:mixtures}

\begin{figure}[!h]
\centering
\includegraphics[width=0.8\linewidth]{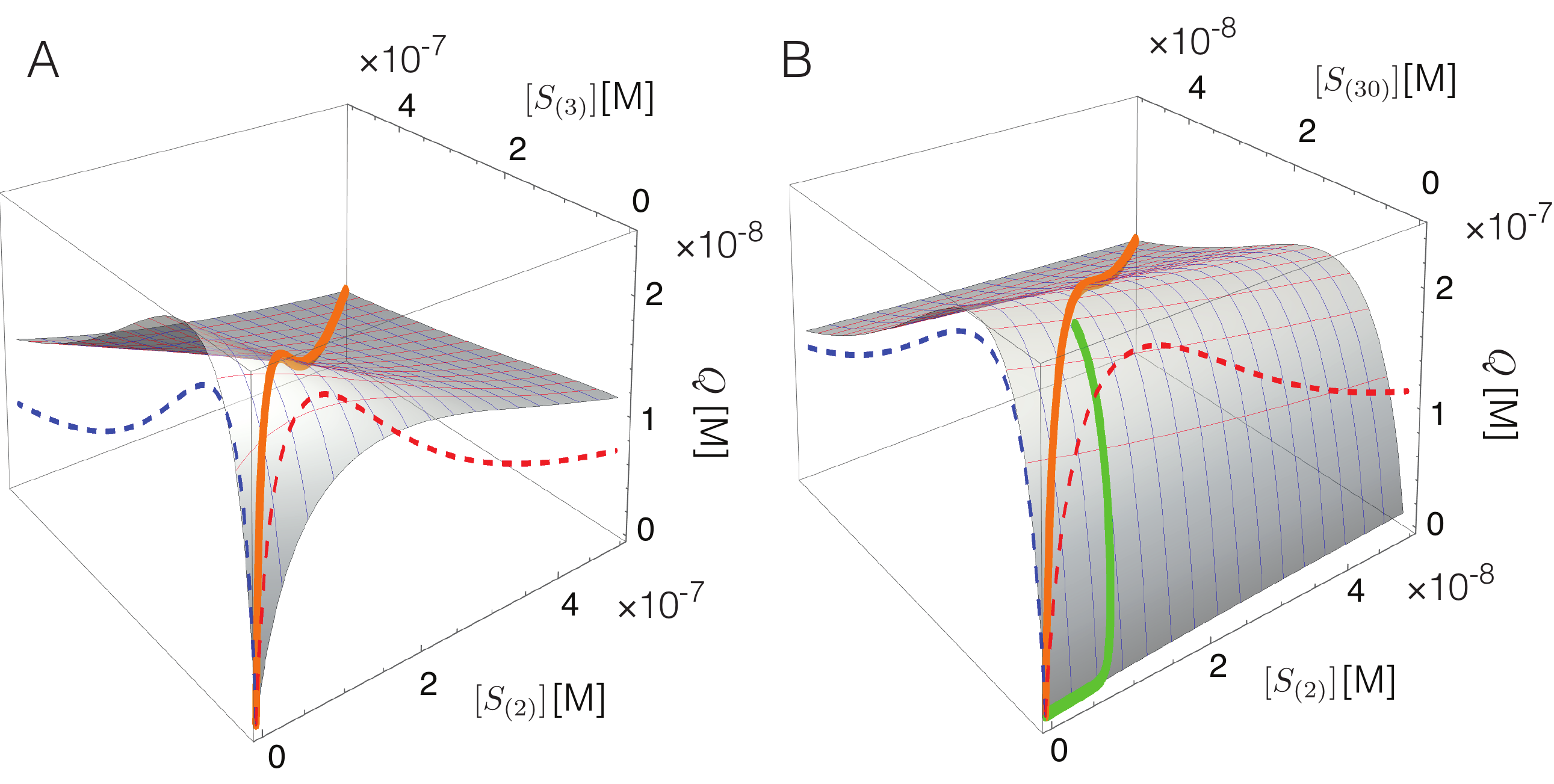}
\caption[Mixtures of multivalent scaffolds]{Mixtures of multivalent scaffolds. {\bf A:} The graphics renders the $Q_{\text{multi}}$-surface of a mixture of a bivalent and trivalent scaffold. The orange line is the $Q$-profile when both agents are added in equal amounts to the mix. The dotted lines are projections of the orange line for comparison with the homogeneous scaffold systems. {\bf B:} Same as in panel {\bf A} but for a mixture of $S_{(2)}$ and $S_{(30)}$; only the portion of the surface at low scaffold concentrations is shown. The green curve shows the $Q$-trajectory for the binary mixture that would obtain when $[S_{(2)}]$ and $[S_{(30)}]$ are set by the polymerizing scaffold system with increasing $t_S$. The green curve is the \emph{whole} trajectory, because both $[S_{(2)}]$ and $[S_{(30)}]$ converge to $1/\sigma=10^{-8}$ M (Figure \ref{fig:oligo}). Other parameters: $\alpha=\beta=10^7$ M$^{-1}$, $t_A=15\cdot 10^{-9}$ M, $t_B=5\cdot 10^{-7}$ M.}
\label{fig:mixSI}
\end{figure}
\ \\

Figure \ref{fig:mixSI}A shows the $Q_{\text{mix}}$-surface \eqref{eq:Qn_mix} of a bivalent and trivalent scaffold mixture. The main observation is the asymmetry in the effect on $Q$ upon adding $S_{(3)}$ to a fixed amount of $S_{(2)}$ compared to the other way around---blue versus red mesh lines in Figure \ref{fig:mixSI}. Upon adding $S_{(3)}$, the ligands $A$ and $B$ re-equilibrate over the available binding sites. Over a range of $[S_{(2)}]$, this equilibration is more likely to result in $A$ and $B$ agents ending up on the same $S_{(3)}$ scaffold than on the same $S_{(2)}$ scaffold. This is most pronounced at small $[S_{(2)}]$ and disappears gradually as the addition of binding sites drives the system past the prozone peak due to the $p^2$ term in \eqref{eq:Qn_mix}. The orange curve shows the $Q$-profile of a mixture in which $S_{(3)}$ and $S_{(2)}$ are increased in equal amounts. The dotted curves are the projections of the mixture curve on each component axis for the purpose of comparison with the $Q$-curves of each component in isolation. This behavior is more dramatic in binary mixtures of multivalent scaffolds with large valency differences (Figure \ref{fig:mixSI}B). 

In a polymerizing scaffold system, the concentrations $s_i\equiv [S_{(i)}]$ and $s_j\equiv [S_{(j)}]$ do not increase in equal amounts when $t_S$ is increased, but are related by a factor $(\sigma s)^{i-j}$. Since $\sigma s <1$ for $t_S<\infty$, there is a lag between the rise of $S_{(i)}$ and $S_{(j)}$, where $S_{(i)}$ increases before $S_{(j)}$ for $i<j$; this lag is more dramatic the bigger the difference $|i-j|$ (Figure \ref{fig:mixSI}B, green curve). In the polymerizing system, as $t_S$ increases, the ratio of $S_{(i)}$ and $S_{(j)}$ will tend to $1$, but by then the between-class prozone is taking its toll. In sum, the \enquote{stealing} of ligands by higher length classes from lower ones is the reason for the turn towards a steeper slope of $Q_{\text{poly}}$ at $t_S$ values at which polymerization becomes effective (Figure 4A in the main text). Incidentally, the shift of ligands from lower towards higher valency classes also tends to flatten the intrinsic slope of the downward leg of lower valency classes after the prozone peak, contributing further to prozone mitigation in the overall system.

\section{Comparison between polymerizing and multivalent scaffold systems}
\label{sec:multivspoly}

In the main text, Figure 4A and 4B, we compare multivalent scaffolds with the polymerizing scaffold system. Figure \ref{fig:multivspoly} places that comparison in the context of the full $Q_{\text{poly}}$ surface to show the effectiveness of regulating the affinity $\sigma$.

While even for $n_A=n_B=n$ and $\alpha=\beta$, $Q_{\text{multi}}$ is a cumbersome expression, determining the concentration of scaffold agents $t_S$ for which $dQ_{\text{multi}}/dt_S=0$ yields a simple solution
\begin{align}
\label{eq:prozonepeak}
    t_S=\frac{1}{n}\left(\frac{1}{\alpha}+\frac{t_A+t_B}{2}\right).
\end{align}
Equation \eqref{eq:prozonepeak} shows that when plotting $Q_{\text{multi}}$ against the concentration of sites $t_{\text{sit}}=n t_S$, as in Figure \ref{fig:multivspoly} and Figure 4A of the main text, the prozone peaks line up for all valencies $n$.

\begin{figure}[!h]
\centering
\includegraphics[width=0.4\linewidth]{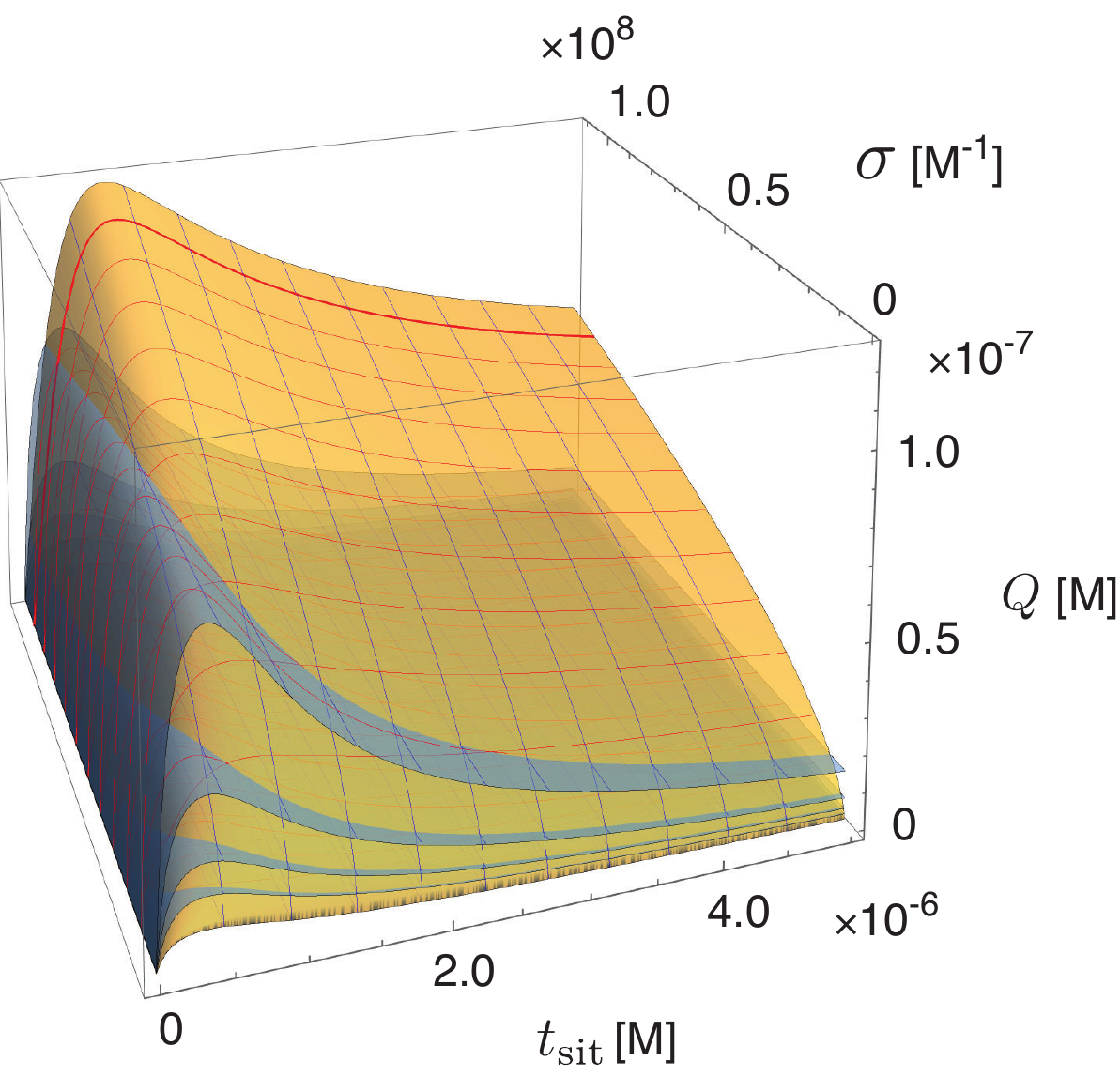}
\caption[Polymerizing scaffold and multivalent scaffolds]{Polymerizing scaffold and multivalent scaffolds. The surface shows $Q_{\text{poly}}$ as function of $t_S$ and $\sigma$, giving more context to Figure 4B in the main text. The emphasized mesh line (red) at $\sigma=10^8$ corresponds to the $Q$-function of the polymerizing scaffold system shown in Figure 4B of the main text. $\alpha=\beta=10^7$ M$^{-1}$, $t_A=15\cdot 10^{-9}$ M, $t_B=5\cdot 10^{-7}$ M.}
\label{fig:multivspoly}
\end{figure}

Expanding $Q_{\text{multi}}$ (assuming $n_A=n_B=n$) in $t_S$ near zero, yields
\begin{align}
    Q_{\text{multi}}=\frac{\alpha t_A \beta t_B}{1 +\alpha t_A+\beta t_B+\alpha\beta t_A t_B}n^2 t_S + O(t_S^2).
\end{align}
Hence in a log-log plot, the up-leg of $Q_{\text{multi}}(n)$ has, to leading order, slope $1$ and offset $n$ when plotted against sites $t_{\text{sit}}=n t_S$ as in Figure 4A of the main text. Similarly, expanding $Q_{\text{multi}}$ in $t_S$ near infinity, yields
\begin{align}
    Q_{\text{multi}}=t_A t_B\frac{1}{t_S} + O(1/t_S^2),
\end{align}
and hence, to leading order, a slope of $-1$ in a log-log plot in the down-leg after the prozone peak and an offset of $n$ when plotted against $t_{\text{sit}}$ as in Figure 4A of the main text.

The expansion of $Q_{\text{poly}}$ in $t_S$ ($=t_{\text{sit}}$) around zero yields
\begin{align}
    Q_{\text{poly}}=\frac{\alpha t_A \beta t_B}{1 +\alpha t_A+\beta t_B+\alpha\beta t_A t_B} t_S + \left[f(\alpha,\beta,t_A,t_B) + g(\alpha,\beta,t_A,t_B)\sigma\right] t_S^2 + O(t_S^3)
    \label{eq:initialslope_cont}
\end{align}
with $f()$ and $g()$ functions of the indicated parameters. The leading-order term is the same as the $Q_{\text{multi}}$ of the monovalent scaffold, \emph{and is independent of} $\sigma$, which enters the second-order term. Accordingly, for small $t_S$, $Q_{\text{poly}}$ hugs the $Q$ of the monovalent scaffold as if there was no polymerization; as $t_S$ increases, $\sigma$ (i.e.\@ polymerization) becomes effective and $Q_{\text{poly}}$ doubles its slope upward. This is clearly seen in Figure 4A of the main text. Some microscopic consequences from building up a length distribution as $t_S$ increases are discussed in section \ref{sec:mixtures}.

Expanding $Q_{\text{poly}}$ in $t_S$ at infinity yields
\begin{align}
    Q_{\text{poly}}=2 t_A t_B \sqrt{\sigma}\sqrt{\frac{1}{t_S}} +O(1/t_S^{3/2}),
\end{align}
where the $p(t_S,t_A,\alpha) p(t_S,t_B,\beta)$ component scales with $t_At_B/t_S^2$ and the $Q_{\text{max}}$ component with $2t_S^{3/2}\sqrt{\sigma}$ to leading order. As a result, the slope of the down-leg of $Q_{\text{poly}}$ after the prozone peak in a log-log plot is $-1/2$.

\section{Catalytic horizon}

\begin{figure}[!h]
\centering
\includegraphics[width=0.8\linewidth]{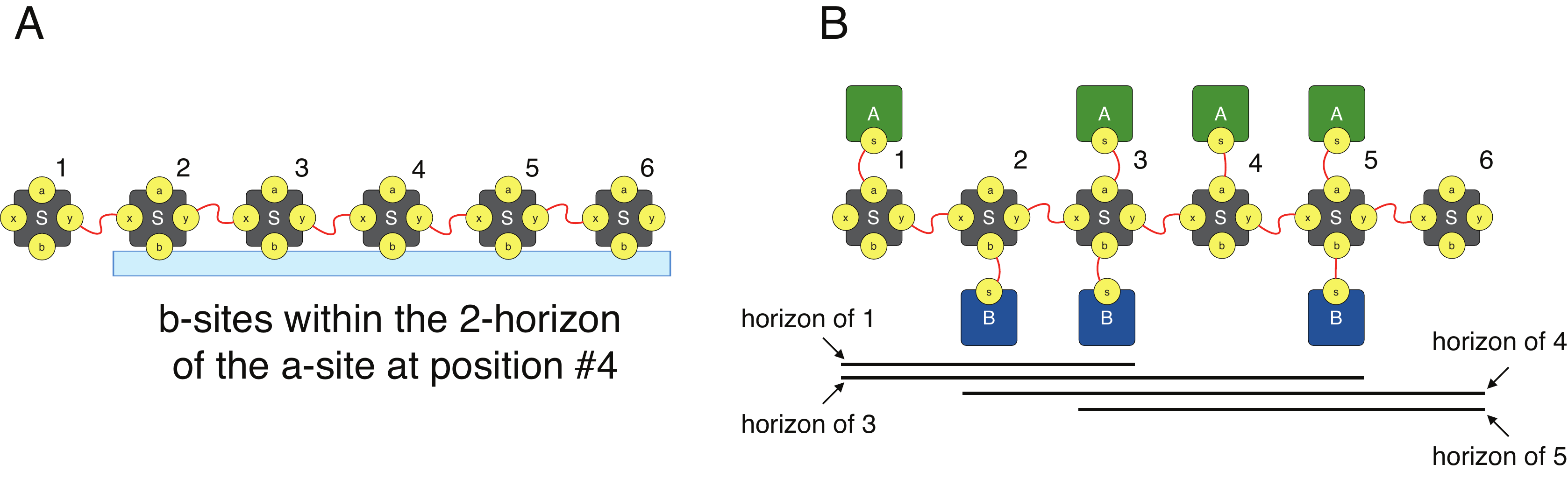}
\caption[Catalytic horizon]{Catalytic horizon. The schematic illustrates the case in which the horizon $h$ is less than the polymer length $n$. In this case, each $A$-binding position can interact with at most $h$ $B$-binding positions on its \enquote{left} or \enquote{right} side. When $h\geq n$, every $A$-position can interact with every $B$-position.}
\label{fig:horizon}
\end{figure}
\ \\

Structural constraints might prevent every catalyst $A$ on a polymeric scaffold from interacting with all substrates $B$ bound to the same polymer. To obtain a rough sense of how such constraints could impact the catalytic potential $Q$, we define a \enquote{catalytic horizon}, $h$, Figure \ref{fig:horizon}. The horizon $h$ is the farthest distance in terms of scaffold bonds that a bound $A$ can \enquote{reach}. This means that a given bound enzyme $A$ can interact with at most $2h+1$ substrate agents $B$: $h$ to its \enquote{left}, $h$ to its \enquote{right} and the one bound to the same protomer, Figure \ref{fig:horizon}A. For example, in Figure \ref{fig:horizon}B, the $2$-horizon of the $A$ at position $1$ includes the $B$s at positions $2$ and $3$, but not at position $5$. Likewise, the $B$ at position $2$ is outside the $2$-horizon of the $A$ at position $5$, whereas all $B$s are within reach of the $A$ at position $3$. Clearly, the catalytic horizon only modulates the $Q_{max}$ in equation \eqref{eq:Qn_general} of a polymer of length $n$; more precisely, it modulates the interaction factor---the $n^2$ in the first equation of \eqref{eq:qpolySI}. We now write this factor as $q_{max}(n,h)$; it replaces the $n^2$ in \eqref{eq:qpolySI}.

To reason about the catalytic combinations, we first consider the case $0\le h\le\lfloor n/2\rfloor$:
\begin{align}
\label{eq:horizon1}
    q_{max}(n,h) = \underbrace{\vphantom{\sum_{k=1}^{h-1}} (n-2h)(2h+1)}_{\text{I}} + \underbrace{\vphantom{\sum_{k=1}^{h-1}} 2h(h+1)}_{\text{II}} + \underbrace{2\sum_{k=1}^{h-1}(h-k)}_{\text{III}}=\quad n(2h+1)-h(h+1)
\end{align}
Term I refers to the $n-2h$ positions in the middle region of the chain that can interact with the full complement of $2h+1$ sites within its horizon. Term II refers to the $h$ positions at each end of the chain and accounts for all $h+1$ sites reachable towards the interior of the chain. Term III accounts for the remaining $h-k$ locations towards the end of the chain that can be reached from a position considered in term II; these locations depend on that position's distance $k$ from the end of the chain. 
\noindent
For $\lfloor n/2\rfloor < h \le n-1$ we obtain 
\begin{align}
\label{eq:horizon2}
    q_{max}(n,h) = \underbrace{\vphantom{\sum_{k=1}^{h-1}} (2h-n)n}_{\text{I'}} + \quad {\underbrace{\vphantom{\sum_{k=1}^{2h-n}} 2(n-h)(h+1)}_{\text{II'}}} \quad + {\underbrace{\vphantom{\sum_{k=1}^{2h-n}} 2 \sum_{k=1}^{n-h}(k-1)}_{\text{III'}}} = \quad n(2h+1)-h(h+1)
\end{align}
In analogy to \eqref{eq:horizon1}, Term I' refers to the $2h-n$ positions that can access the whole chain; term II' accounts for the $h+1$ locations spanned by the inward-facing side of the remaining $n-h$ positions at each end of the chain. Finally, term III' accounts for the locations covered by the outward facing side of these $n-h$ positions.

If the horizon $h$ is larger than the polymer length $n$, then every $A$-position can interact with every $B$-position on the polymeric scaffold and $q_{max}(n,h)=n^2$. Merging this with \eqref{eq:horizon1} and \eqref{eq:horizon2} yields
\begin{align}
\label{eq:horizon}
    q_{max}(n,h) =\left\{ 
    \begin{array}{ll} 
    n(2h+1)-h(h+1), & \text{ for } 0\leq h\leq n-1\\ 
    n^2, & \text{ for } h\geq n 
    \end{array} \right.
\end{align}
which appears in the main text. The corner cases are covered correctly: $q_{max}(n,0)=n$ and $q_{max}(n,n-1)=n^2$. (Note that $h=n$ yields the same result as $h=n-1$, which is useful below.)

We use \eqref{eq:horizon} to calculate two scenarios. In scenario 1, $h$ is a simple linear function of the length $n$: $h=\xi n$ with $0\le\xi\le 1$. In other words, every $A$ can monitor the same fraction $\xi$ of $B$-binding sites on a polymer of any size. This seems rather unrealistic (and makes $h$ a continuous variable, although that appears to work just fine). However, scenario 1 may serve as a comparison with the subsequent, more realistic scenario 2.

When $h=\xi n$, $h$ is always less or equal than $n$ and the first case of \eqref{eq:horizon} applies. Using $q_{max}(n,h)$ with $h=\xi n$ instead of $n^2$ in the first equation of \eqref{eq:qpolySI} yields
\begin{align}
    Q_{max}(\xi)&=\sum_{n=1}^{\infty} \left[n(2h+1)-h(h+1)\right] \sigma^{n-1}s^n 
    =\sum_{n=1}^{\infty} \left[n(2\xi n+1)-\xi n(\xi n+1)\right] \sigma^{n-1}s^n \nonumber\\
    &=\dfrac{1}{\sigma}\left[\xi(2-\xi)\sum_{n=1}^{\infty} n^2 \sigma^{n}s^n + (1-\xi)\sum_{n=1}^{\infty} n \sigma^{n}s^n \right] 
    =\xi(2-\xi)\dfrac{s(1+\sigma s)}{(1-\sigma s)^3}+(1-\xi)\dfrac{s}{(1-\sigma s)^2},
\end{align}
which leads to 
\begin{align}
    \label{eq:Qhorizon}
    Q=p(t_S,t_A,\alpha) p(t_S,t_B,\beta)\left(\xi(2-\xi)\dfrac{s(1+\sigma s)}{(1-\sigma s)^3}+(1-\xi)\dfrac{s}{(1-\sigma s)^2}\right)
\end{align}

For $\xi=1$, the expression \eqref{eq:Qhorizon} becomes \eqref{eq:qpolySI}, as a horizon that equals the length of any polymer does not affect $Q_{max}$. For $\xi=0$ we get
\begin{align}
    Q=p(t_S,t_A,\alpha) p(t_S,t_B,\beta)\dfrac{s}{(1-\sigma s)^2} =
    p(t_S,t_A,\alpha) p(t_S,t_B,\beta)t_S,
    \label{eq:Qmono}
\end{align}
because of $t_S=s\, dW/ds$ for the polymer-only system. Thus, for $\xi=0$, we recover the $Q$ of the simple monovalent scaffold, since in this case the organization of protomers into polymers doesn't affect catalytic potential. Scenario 1 is shown in Figure \ref{fig:hzcompare}, panels A and B.

In scenario 2, $h=\text{const}$ for all lengths $n$, which means a \enquote{hard} horizon independent of polymer size. This scenario is more realistic. $Q_{max}(h)$ becomes
\begin{align}
    Q_{max}(h)&=\sum_{n=1}^{\infty}q_{max}(n,h)\sigma^{n-1}s^n=\sum_{n=1}^h n^2\sigma^{n-1}s^n+\sum_{n=h+1}^{\infty} [n(2h+1)-h(h+1)]\sigma^{n-1}s^n \nonumber\\
    &=\dfrac{1}{\sigma}\left\{\sum_{n=1}^h n^2(\sigma s)^n+(2h+1)\sum_{n=h+1}^{\infty}n(\sigma s)^n-h(h+1)\sum_{n=h+1}^{\infty}(\sigma s)^n
    \right\} \nonumber \\
    &=\dfrac{1}{\sigma}\left\{\dfrac{\sigma s(1+\sigma s)-(\sigma s)^{h+1}[(h+1)^2-(2h^2+2h-1)\sigma s+h^2(\sigma s)^2]}{(1-\sigma s)^3}\right. \nonumber\\
    &\left.\phantom{=}+(2h+1)\dfrac{(\sigma s)^{h+1}(h+1-h\sigma s)}{(1-\sigma s)^2}- h(h+1)\dfrac{(\sigma s)^{h+1}}{1-\sigma s} \right\} \nonumber\\
    &=\dfrac{s\left(1+\sigma s-2(\sigma s)^{h+1}\right)}{(1-\sigma s)^3},
    \label{eq:consthorizon}
\end{align}
yielding
\begin{align}
\label{eq:Qhorizon2SI}
    Q=p(t_S,t_A,\alpha) p(t_S,t_B,\beta)\dfrac{s\left(1+\sigma s-2(\sigma s)^{h+1}\right)}{(1-\sigma s)^3},
\end{align}
which is equation (6) of the main text. Expression \eqref{eq:Qhorizon2SI} becomes \eqref{eq:Qmono} for $h=0$, as we would expect. As $h$ increases, \eqref{eq:Qhorizon2SI} quickly converges to the infinite horizon case \eqref{eq:qpolySI}, since $\sigma s < 1$ raised to the power of $h$ becomes negligible. Scenario 2 is shown in Figure \ref{fig:hzcompare}, panels B and D. As suggested in Figure \ref{fig:MMcomparison}, even restrictive structural constraints (small $h$) make only a relatively modest dent in the catalytic potential of the polymerizing scaffold when compared to that of the plain Michaelis-Menten scenario.

\begin{figure}[!h]
\centering
\includegraphics[width=0.55\linewidth]{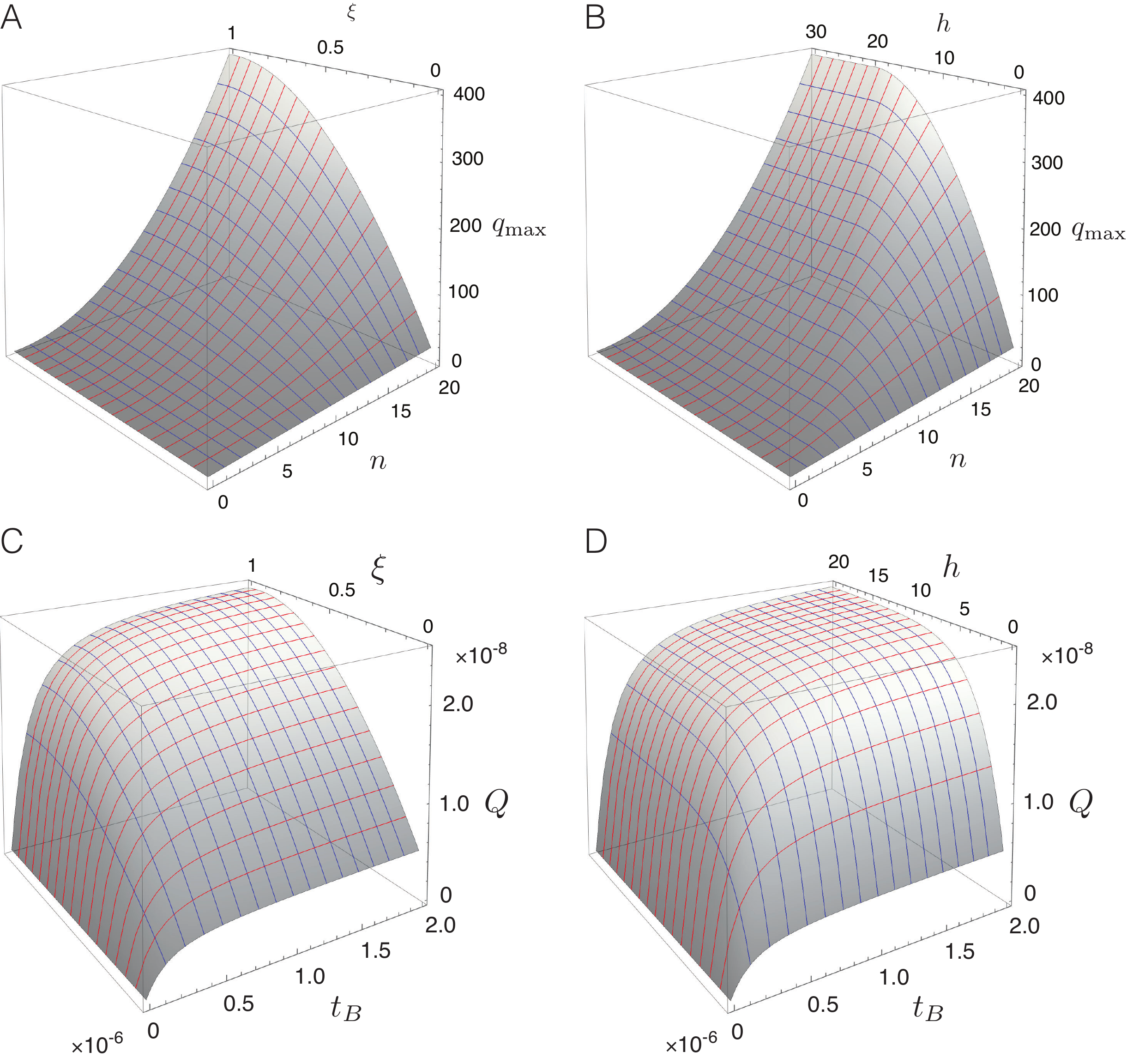}
\caption[Catalytic horizon scenarios]{Catalytic horizon scenarios. {\bf A:} $q_{max}(n,h)$, equation \eqref{eq:horizon}, for scenario 1 when $h=\xi n$ ($0\le \xi\le 1$). {\bf B:} $q_{max}(n,h)$, equation \eqref{eq:horizon}, for scenario 2 when $h$ is a constant independent of $n$. The difference to panel {\bf A} is that the surface of scenario 2, once $h$ exceeds $n$, is a quadratic extension of the surface of scenario 1 in panel {\bf A} at $\xi=1$. {\bf C:} The $Q$-surface \eqref{eq:Qhorizon} for scenario 1 as a function of substrate concentration $t_B$. {\bf D:} The $Q$-surface \eqref{eq:Qhorizon2SI} for scenario 2 as a function of substrate concentration $t_B$. In Figure \ref{fig:MMcomparison}, this surface is compared against the Michaelis-Menten case. The parameter values in {\bf C} and {\bf D} are: $\alpha=\beta=10^7$ M and $\sigma=10^8$ M, $t_A=15\cdot 10^{-9}$ M, and $t_S=60\cdot 10^{-9}$ M.}
\label{fig:hzcompare}
\end{figure}

\ \\

\begin{figure}[!h]

\vspace*{0.4cm}

\centering
\includegraphics[width=0.4\linewidth]{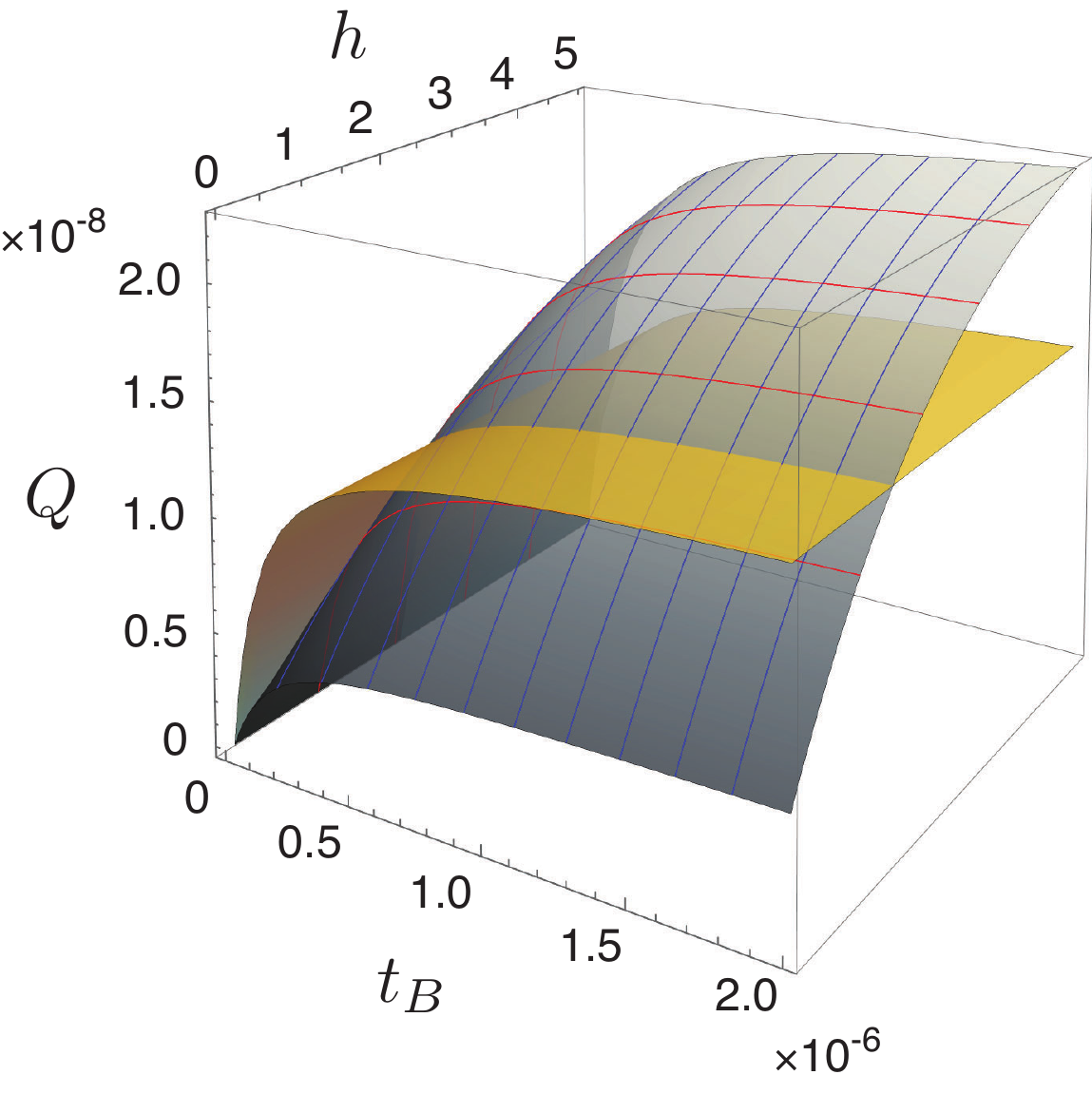}
\caption[The impact of catalytic horizon]{The impact of catalytic horizon. The $Q$-surface \eqref{eq:Qhorizon2SI} with hard horizon $h$, gray, and the plain dimerization (Michaelis-Menten) surface, orange, for the parameter settings corresponding to Figure 3A in the main text ($\alpha=\beta=10^7$ M$^{-1}$ and $\sigma=10^8$ M$^{-1}$, $t_A=15\cdot 10^{-9}$ M, $t_S=60\cdot 10^{-9}$ M). At $t_S=60$ nM (the curve with the red dot in Figure 3A of the main text) a horizon $h=2$ is already sufficient to achieve a higher catalytic potential than the direct binding of enzyme to substrate. This suggests that structural constraints forcing a small catalytic horizon might not undermine the efficacy of a polymerizing scaffold.}
\label{fig:MMcomparison}
\end{figure}

\section{The discrete case}
\label{sec:statmech}

While we strive for a reasonably self-contained exposition, some details are only asserted for brevity and are developed in a forthcoming manuscript providing a more general treatment of equilibrium assembly. 

In the following, we use the same symbols for the binding affinities $\alpha$, $\beta$, and $\sigma$ as in the continuum case, but they must now be understood as \enquote{stochastic affinities}. Specifically, if $\gamma'$ is a binding affinity in the continuum case, the stochastic affinity $\gamma$ (in units of molecules$^{-1}$) is related as $\gamma=\gamma'/({\cal A}V)$, where $V$ is the effective volume hosting the system and ${\cal A}$ is Avogadro's constant. Thus a polymerization affinity of $3$ molecules$^{-1}$ in the discrete case corresponds to about $1.8\cdot 10^{12}$ M$^{-1}$ in a cell volume of $10^{-12}$ L in the continuum setting.

\subsection{Average catalytic potential} 

Our objective is to calculate the average catalytic potential $\langle Q\rangle$ of a scaffold mixture, defined as
\begin{align}
\label{eq:averageQdiscrete}
    \langle Q \rangle =\sum_{i=0}^{\min(t_A,n)}\sum_{j=0}^{\min(t_B,n)} i\, j\,\langle S_{ij}\rangle,
\end{align}
where $S_{ij}$ is any scaffold (polymer or multivalent) with $n$ $A$-binding sites, of which $i$ are occupied, and $n$ $B$-binding sites, of which $j$ are occupied. More precisely, $S_{ij}$ is the set of all configurations, or molecular species, with $i$ and $j$ agents of type $A$ and $B$ bound, respectively. $\langle S_{ij}\rangle$ is the average or expected total number of such configurations in an equilibrium system with resource vector $\vec{t}=(t_A,t_B,t_S)'\in\mathbb{N}_0^3$. The $'$ means a transpose. ($t_S$ is typically the number of scaffolds of a given valency $n$ or the number of protomers in a polymerizing system. When considering mixtures of scaffolds of different valencies $i$, $t_S$ is generalized accordingly.)

This raises the need to compute $\langle S_{ij}\rangle$, which requires a little detour. We start by defining a few well-known quantities.

Assume a system of molecular interactions with a set of atomic building blocks, or atoms for short, $\{X_1,\ldots,X_T\}$ (in the main text typically $T=3$, namely $A$, $B$, and $S$) that give rise to a set of configurations $\{Y_1,\ldots,Y_C\}$. Since we are interested in equilibrium, the precise nature of the interactions is irrelevant as long as the resulting systems have the same set of reachable molecular species. The assembly scenarios considered in the main text only require binding and unbinding interactions.

\subsection{Boltzmann factor of a molecular species} 

Each molecular species $Y_i$ has a Boltzmann factor given by
\begin{align}
\varepsilon_{i}=\prod_{r} \gamma_r,
\label{eq:energy}
\end{align}
where $\gamma_r=\exp(-\frac{\Delta G_r^0}{kT})$ is the binding constant of the $r$-th reaction and the product runs over a series of reactions $r$ that constitute an assembly path from atomic components ($A$, $B$, and $S$). Note that, in the discrete case, $\varepsilon_{i}$ is not divided by the number of symmetries $\omega_{i}$ as in the continuum case (main text leading up to Eq.\@ [1]). The effect of symmetries is accounted for in the state degeneracy, Eq.\@ \eqref{eq:degeneracy} below, which considers all instances of $Y_i$ in a given state. As a consequence, $-kT\log\varepsilon_{i}$ is not the free energy of formation, but just the internal energy due to bond formation. 

\subsection{Boltzmann factor of a state} 

By extension, the Boltzmann factor of a system \emph{state} $\vec{n}=(n_1,n_2,\ldots,n_C)'$, where $n_i$ is the number of particles of species $Y_{i}$, is given by
\begin{align}
\varepsilon(\vec{n})=\prod_{i=1}^{C}(\varepsilon_i)^{n_i}. \label{eq:state_energy}
\end{align}
More precisely, \eqref{eq:state_energy} is the Boltzmann factor associated with a particular \emph{realization} of the state $\vec{n}$, as all atoms are labelled (distinguishable).

\subsection{Degeneracy of a state}

\ A state $\vec{n}$ is the specification of a multiset of species in which atom labels are ignored. The degeneracy $d(\vec{t},\vec{n})$ of a state $\vec{n}$ with resource vector $\vec{t}=(t_1,\ldots,t_T)$ is the number of distinct ways of realizing it by taking into account atom labels. Let $\mu_{i,j}$ denote the number of atoms of type $X_j$ contained in one instance of $Y_{i}$. For a given resource vector $\vec{t}$ the set $\Sigma(\vec{t})$ of states $\vec{n}$ that are compatible with it satisfy $t_j=\sum_{i=1}^C\mu_{i,j}n_i$ for every atom type $X_j$. Hence, the degeneracy of a state $\vec{n}\in\Sigma(\vec{t})$ is given by
\begin{align}
d(\vec{t},\vec{n})=\dfrac{\prod_{i=1}^{T}\limits t_i!}{\prod_{i=1}^{C}\limits n_i! \prod_{i=1}^{C}\limits (\omega_{i})^{n_i}}. \label{eq:degeneracy}
\end{align}
The numerator counts all permutations of the atoms that constitute the system, the first product in the denominator corrects for all orderings among the $n_i$ copies of species $Y_{i}$ and the second product corrects for all symmetries associated with $Y_{i}$. 

\subsection{The partition function for a given resource vector} 

\ As usual,
\begin{align}
Z(\vec{t})=\sum_{\vec{n}\in\Sigma(\vec{t})}d(\vec{t},\vec{n})\varepsilon(\vec{n}), \label{eq:partition}
\end{align}
where the sum runs over all admissible states given resource vector $\vec{t}$. The equilibrium probability of a state $\vec{n}$ is given by
\begin{align}
p(\vec{t},\vec{n})=\dfrac{d(\vec{t},\vec{n})\varepsilon(\vec{n})}{Z(\vec{t})}. \label{eq:prob}
\end{align}

\subsection{The average number of instances of a specific configuration in equilibrium}

\ For a given resource vector $\vec{t}$ a species $Y_{i}$ occurs in various numbers $n_i$ across the states $\vec{n}$ in the admissible set $\Sigma(\vec{t})$. The average abundance of $Y_{i}$, $\langle n_i\rangle$ then is
\begin{equation}
\langle n_i\rangle = \sum_{\vec{n}\in\Sigma(\vec{t})} n_ip(\vec{t},\vec{n})=\dfrac{1}{Z(\vec{t})}\sum_{\vec{n}\in\Sigma(\vec{t})} n_id(\vec{t},\vec{n})\varepsilon(\vec{n}). \label{eq:average}
\end{equation}
The workhorse for the discrete treatment of the scaffolding systems discussed in the main text is the following Theorem.

\noindent
{\bf Theorem:}\\
\emph{The average equilibrium abundance $\langle n_i\rangle$ of species $Y_{i}$ in an assembly system with resource vector $\vec{t}$ is given by}
\begin{equation}
\langle n_i\rangle = \varrho(\vec{t},Y_{i})\varepsilon_i\dfrac{Z(\vec{t}-\vec{\mu_{i}})}{Z(\vec{t})}, \label{eq:Ortiz}
\end{equation}
\emph{where $\vec{\mu_{i}}=(\mu_{i,1},\ldots,\mu_{i,T})'$ is the atomic content vector of species $Y_{i}$; $\varrho(\vec{t},Y_{i})$ is the number of distinct realizations of a single instance of $Y_{i}$ given resources $\vec{t}$; and $Z(\vec{t}-\vec{\mu_{i}})$ is the partition function of a system in which the atomic resources have been decreased by the amount needed to build one instance of $Y_{i}$.}

It is immediate from \eqref{eq:degeneracy} that 
\begin{align}
    \varrho(\vec{t},Y_{i})=d(\vec{t},\vec{Y_i})=\frac{\prod_{j=1}^{T}\limits t_j!}{\prod_{j=1}^{T}\limits (t_j-\mu_{i,j})!\,\omega_{i}},
\end{align}

where $\vec{Y_i}$ denotes a unit vector in the $Y_i$ direction. We provide a proof of the theorem using generating functions elsewhere. However, to see why the claim holds, we reason as follows. The subset of $\Sigma(\vec{t})$ in which we restrict ourselves to states $\vec{n}$ that contain at least one copy of $Y_{i}$ stands in a 1-1 correspondence to the unrestricted state space $\Sigma(\vec{t}-\vec{\mu_{i}})$, because any realization of $Y_{i}$ in $\Sigma(\vec{t})$ occurs in all possible contexts and these contexts are precisely the states of $\Sigma(\vec{t}-\vec{\mu_{i}})$. The question then is how the degeneracy and the energy content of a state $\vec{n}\in\Sigma(\vec{t}-\vec{\mu_{i}})$ change by adding $\vec{\mu_{i}}$ atoms to realize one instance of $Y_{i}$. The degeneracy of state $\vec{n}\in\Sigma(\vec{t}-\vec{\mu_{i}})$ is amplified (multiplied) by $\varrho(\vec{t},Y_{i})$ realizations of $Y_{i}$, but one instance of $Y_{i}$ is added to those the state already had and so we also need to divide by $n_i+1$ to compensate for indistinguishable permutations within the instances of $Y_{i}$, see \eqref{eq:degeneracy}. Thus, $d(\vec{t},\vec{n}+\vec{Y_{i}})= (\varrho(\vec{t},Y_{i})/(n_i+1)) d(\vec{t}-\vec{\mu_{i}},\vec{n})$ and the Theorem follows as summarized symbolically:
\begin{equation}
\dfrac{1}{Z(\vec{t})}\sum_{\substack{\vec{n}\in\Sigma(\vec{t}) \\ n_i\ge 1}}n_id(\vec{t},\vec{n})\varepsilon(\vec{n})= 
\dfrac{1}{Z(\vec{t})}\sum_{\vec{n}\in\Sigma(\vec{t}-\vec{\mu_{i}})} (n_i+1)\dfrac{\varrho(\vec{t},Y_{i})}{n_{i}+1} d(\vec{t}-\vec{\mu_{i}},\vec{n})\varepsilon_i\varepsilon(\vec{n})=\varrho(\vec{t},Y_{i})\varepsilon_i\dfrac{Z(\vec{t}-\vec{\mu_{i}})}{Z(\vec{t})}.
\end{equation}

It remains to compute the partition function of the assembly systems discussed in the main text, which is not too difficult and provided in the subsequent section \ref{sec:partition_functions}.

\section{Partition functions and average catalytic potential}
\label{sec:partition_functions}

\subsection{Polymerizing scaffold without ligands}

\ Let a state contain $i$ bonds (not necessarily in the same polymer). Any such state has a Boltzmann factor $\sigma^i$, where $\sigma$ is the binding affinity between two scaffold protomers. We count the number of ways to realize $i$ bonds as follows. Line up the $t_S$ (labelled) protomers and observe that there are $t_S-1$ slots between protomers where a bond could be inserted. Thus there are $\binom{t_S-1}{i}$ ways of inserting $i$ bonds and the insertion of $i$ bonds always creates $t_S-i$ molecules. For each choice of $i$ slots there are $t_S!$ permutations of the protomers. Since the order in which a choice of bond locations creates the $t_S-i$ molecules is irrelevant, we must reduce the label permutations by $(t_S-i)!$ object permutations to obtain the degeneracy $d_i$ of a state with $i$ bonds.
The partition function is therefore
\begin{align}
Z^{\text{poly}}_{t_S}=\sum_{i=0}^{t_S-1}\limits \sigma^{i}\begin{pmatrix} t_S-1 \\ i \end{pmatrix} \dfrac{t_S!}{(t_S-i)!}
\label{eq:partition_lin}
\end{align}
The number of possible realizations of a single polymer $s_n$ of length $n$ is $t_S!/(t_S-n)!$, which yields with \eqref{eq:Ortiz} for the average number of polymers of length $n$, $\langle s_n\rangle$:
\begin{align}
\langle s_n\rangle=\dfrac{t_S!}{(t_S-n)!} \sigma^{n-1} \dfrac{Z^{\text{poly}}_{t_S-n}}{Z^{\text{poly}}_{t_S}}.
\label{eq:homopoly_s}
\end{align}

Figure \ref{fig:lendist} compares the length distributions of equivalent continuum and discrete polymerization systems
\begin{figure}[!h]
\centering
\includegraphics[width=0.4\linewidth]{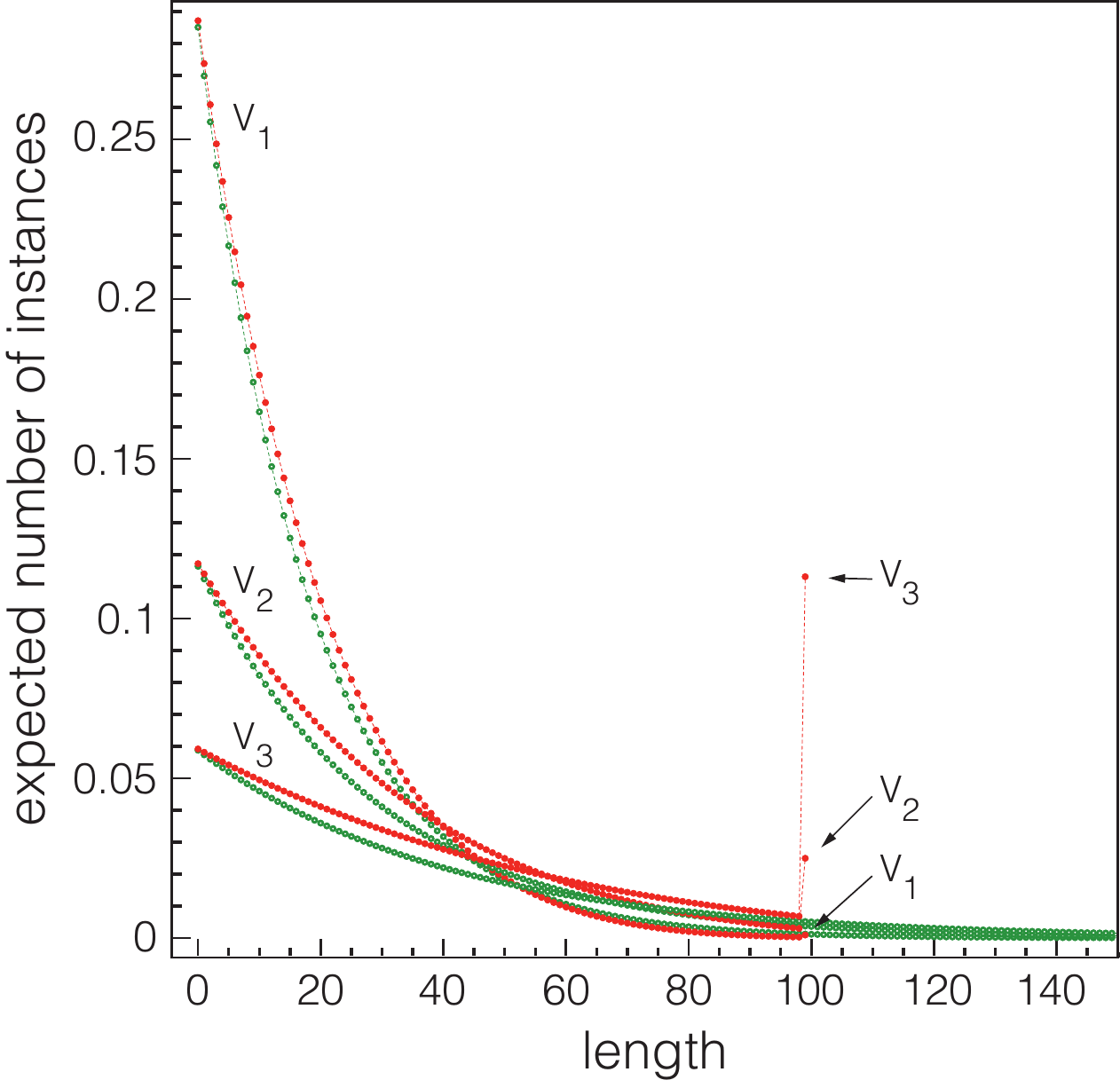}
\caption[Length distribution in continuum and discrete polymerization]{Length distribution in continuum and discrete polymerization. A continuum and discrete polymerization system are set up with equivalent parameters assuming a base volume $V=10^{-15}$ L (the order of magnitude of a bacterial cell). Their length distributions are compared for three volumes: $V_1=0.05 V$, $V_2=0.02 V$, $V_3=0.01 V$. A change in volume means a change in affinity for the discrete system and a change in protomer concentration for the continuum system, i.e.\@ $t_S=100$ protomers or $t_S=100 / ({\cal A} V_i)$ M; discrete affinity $\sigma_s=10^8/ ({\cal A} V_i)$ molecules$^{-1}$ or continuum affinity $\sigma_d=10^8$ M$^{-1}$. The green curves are associated with the continuum system (equation \ref{eq:free_s_again} and the red ones with the discrete case (equation \ref{eq:homopoly_s}. Associated volumes are as indicated in the graph. Since the curves cross, the maximer is also marked with the corresponding volume. The continuum distribution is cut off at $150$.}
\label{fig:lendist}
\end{figure}

\subsection{Average catalytic potential of the polymerizing scaffold with ligands}
\label{sec:polyQ}

\ Because of binding independence, the partition function of this system is the product of three partition functions: $Z^{\text{poly}}_{t_S} Z^{\text{dimer}}_{t_S,t_A} Z^{\text{dimer}}_{t_S,t_B},$ with $Z^{\text{dimer}}_{t_S,t_X}$ the partition function of a system in which $S$-agents and $X$-agents can dimerize with affinity $\gamma$. $Z^{\text{dimer}}_{t_S,t_X}$ is simple to obtain: choose $i$ agents of type $A$, $i$ agents of type $S$, and pair them:
\begin{align}
Z^{\text{dimer}}_{t_S,t_X}=\sum_{i=0}^{\min(t_S,t_X)}\gamma^i\,\binom{t_S}{i}\binom{t_X}{i}\, i!.
\label{eq:dimer}
\end{align}
Putting this together yields the partition function for resource vector $\vec{t}=(t_A,t_B,t_S)$
\begin{align}
\label{partition2}
    Z(\vec{t})&=
    \left[\sum_{k=0}^{t_S-1} \sigma^k\,\binom{t_S-1}{k}\dfrac{t_S!}{(t_S-k)!}\right]\,
    \left[\sum_{i=0}^{\min(t_S,t_A)}\alpha^i\,\binom{t_A}{i}\binom{t_S}{i}\, i!\right]\,
    \left[\sum_{j=0}^{\min(t_S,t_B)}\beta^j\,\binom{t_B}{j}\binom{t_S}{j}\, j!\right] \nonumber\\
    &=Z_{t_S}^{\text{poly}}Z_{t_S,t_A}^{\text{dimer}}Z_{t_S,t_B}^{\text{dimer}}
\end{align}

The total number of realizations, $\varrho(\vec{t},\{A_iS_lB_j\})$ of polymers of length $l$ with $i$ $A$-agents and $j$ $B$-agents attached, and thus each with Boltzmann factor $\sigma^{l-1}\alpha^i\beta^j$, is given by
\begin{align}
    \varrho(\vec{t},\{A_iS_lB_j\})&=\dfrac{t_S!}{(t_S-l)!}\binom{l}{i}\binom{t_A}{i}i!\,\binom{l}{j}\binom{t_B}{j}j!
    =\binom{l}{i}\binom{l}{j}\dfrac{t_S!}{(t_S-l)!}\dfrac{t_A!}{(t_A-i)!}\dfrac{t_B!}{(t_B-i)!} \nonumber\\
    &=\binom{l}{i}\binom{l}{j}\dfrac{\vec{t}!}{(\vec{t}-\vec{v})!}
\end{align}
where $\vec{v}=(i,j,l)$ is the composition vector of the configuration and we define for brevity the factorial of a vector as the product of the factorials of its components.
Putting all this together yields the average catalytic potential $\langle Q\rangle$
\begin{align}
\label{eq:Qpolydiscrete}
    \langle Q_{\text{poly}}\rangle=\sum_{l=1}^{t_S}\sum_{i=0}^{\min\{l,t_A\}}\sum_{j=0}^{\min\{l,t_B\}}
    \underbrace{i\, j \vphantom{\binom{l}{i}}}_{\substack{\text{\# of}\\ \text{interactions}}}\,\underbrace{\underbrace{\binom{l}{i}\binom{l}{j}\dfrac{\vec{t}!}{(\vec{t}-\vec{v})!}}_{\substack{\text{total realizations of} \\ \text{configurations with $\vec{v}$}}}\sigma^{l-1}\alpha^i\beta^j\dfrac{Z(\vec{t}-\vec{v})}{Z(\vec{t})}}_{\text{average total counts}}
\end{align}

\subsection{Average catalytic potential of the multivalent scaffold with ligands}

\ The case of a multivalent scaffold with $m$ binding sites for $A$ and $n$ binding sites for $B$ follows the lines of section \ref{sec:polyQ}. For each type of binding sites one can formulate a partition function in full analogy to $Z^{\text{dimer}}_{t_S,t_X}$, but with $m\,t_S$ (or $n\,t_S$) sites available to bind $i$ agents of type $A$ (or $j$ agents of type $B$) to yield a state with Boltzmann factor $\alpha^i\beta^j$. Thus, the partition function for a multivalent scaffold system is 
\begin{align}
\label{eq:partition_multi}
    Z_{t_A,t_B,t_S}^{\text{multi}}&= \sum_{i=0}^{\min(m\,t_S,t_A)}\sum_{j=0}^{\min(n\,t_S,t_B)}\alpha^i\,\beta^j\binom{t_A}{i}\binom{m\,t_S}{i}\, i!\binom{t_B}{j}\binom{n\,t_S}{j}\,j!
\end{align}
The average number of scaffolds loaded with $i$ ligands of type $A$ and $j$ ligands of type $B$ \emph{in a particular configuration} then becomes
\begin{align}
    \langle n_{ij}\rangle=\frac{t_A!}{(t_A-i)!}\frac{t_B!}{(t_B-j)}t_S\,\alpha^i\,\beta^j
    \frac{Z^{\text{multi}}_{t_A-i,t_B-j,t_S-1}}{Z^{\text{multi}}_{t_A,t_B,t_S}}.
\end{align}
Finally, for the average catalytic potential we have
\begin{align}
    \label{eq:Qmultidiscrete}
    \langle Q_{\text{multi}}\rangle=\sum_{i=0}^{\min(t_A,m)}\sum_{j=0}^{\min(t_B,n)} i\,j\,\binom{m}{i}\binom{n}{j} \langle n_{ij}\rangle.
\end{align}

\subsection{Remarks on numerical evaluation}

\ While expressions \eqref{eq:Qpolydiscrete} and \eqref{eq:Qmultidiscrete} are explicit, their use with large particle numbers---$t_S$, $t_A$ and $t_B$---is limited by numerical instabilities (even after efficiency rearrangements). In a separate paper we connect assembly systems with the theory of analytic combinatorics \cite{Flajolet2009}, which provides direct approximations based on viewing generating functions as analytic functions over the complex numbers. In our hands, these approximations are not accurate enough over the entire parameter range for the present context. Our figures were therefore generated using the exact expressions \eqref{eq:Qpolydiscrete} and \eqref{eq:Qmultidiscrete}, using arbitrary-precision calculations (to $100$ significant digits) in {\tt Mathematica} \cite{Mathematica}, and employing relatively modest particle numbers to keep computation times reasonable.

\section{The maximer probability and 1D percolation}

The probability of observing the longest possible polymer, given protomer resources, is obtained from \eqref{eq:homopoly_s} by setting $n=t_S$:
\begin{align}
\langle s_{\text{max}}\rangle=\dfrac{t_S!\, \sigma^{t_S-1}}{Z^{\text{poly}}_{t_S}}.
\label{eq:maximer}
\end{align}
This probability is graphed as a function of $t_S$ and $\sigma$ in Figure 5A of the main text. 

There is an analogy between 1D bond percolation and polymerization at our level of abstraction. The analogy is an exact correspondence in the case of continuum polymerization and bond percolation on an infinite 1D lattice. 

A basic quantity in 1D percolation is the mean number of chains (clusters) of size $n$ normalized per lattice site, which is given by $p^{n-1}(1-p)^2$, where $p$ is the probability of a bond between adjacent lattice sites and functions as a parameter. The same expression obtains in terms of the concentration of polymers of length $n$ normalized per protomer \cite{Flory1936,Reynolds1977}:
\begin{align}
\dfrac{s_n}{t_S}= p^{n-1}(1-p)^2.
\label{eq:flory}
\end{align}
In the context of polymers, the bond probability is not the primary parameter, but a function of the basic parameters $t_S$ and $\sigma$. Following Flory \cite{Flory1936}, we can express $p$ as
\begin{align}
p=\dfrac{t_S-W}{t_S}=1-\dfrac{1}{t_S}\dfrac{s}{1-\sigma s},
\label{eq:floryp}
\end{align}
with $W$ the concentration of all polymers as defined in \eqref{eq:W_m0_} for $a=b=0$ and given more compactly by \eqref{eq:W_poly}. The first equality defines $p$ in terms of the difference between the maximal possible concentration of objects in the system ($t_S$) and the actual concentration of objects; this difference is the concentration of bonds. Using \eqref{eq:free_s_again} for $s$ yields
\begin{align}
p=1-\dfrac{2}{1+\sqrt{1+4\sigma t_S}}.
\label{eq:p}
\end{align}
Together, expressions \eqref{eq:flory} and \eqref{eq:p} are equivalent to \eqref{eq:free_s_again} and connect simple polymerization to percolation. As well-known, in the infinite/continuum case, percolation can only occur at $p=1$, which is to say in the limit of $t_S\to\infty$ or $\sigma\to\infty$.

\begin{figure}[!h]
\centering
\includegraphics[width=0.6\linewidth]{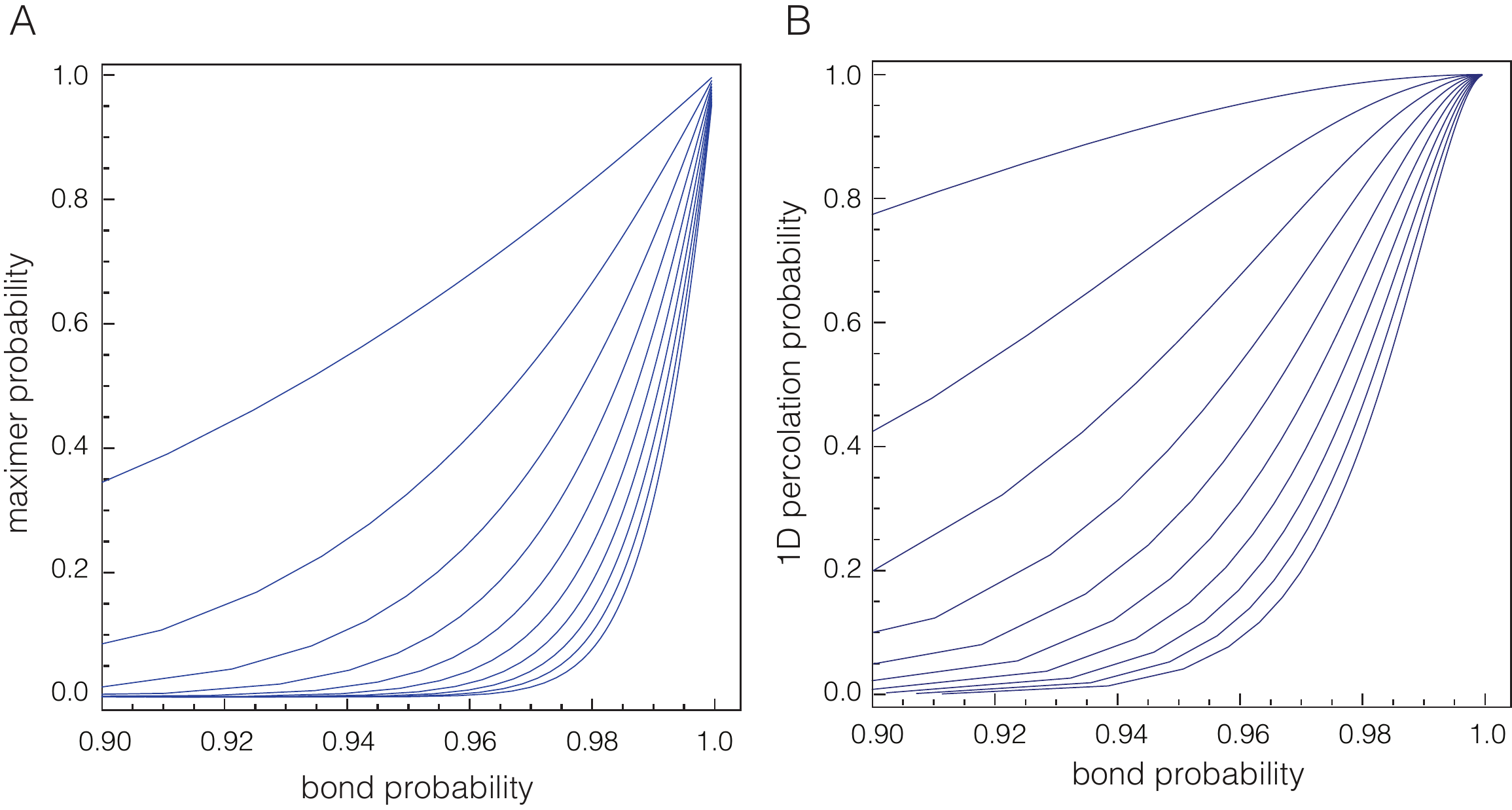}
\caption[Finite size 1D bond percolation and polymerization]{Finite size 1D bond percolation and polymerization. {\bf A:} This panel is panel B of Figure 5 in the main text. It depicts the probability of the maximer \eqref{eq:maximer} as a function of $p_{\text{bond}}$ as given by \eqref{eq:pBond}. Each curve represents a particular $t_S$-value for which $\sigma$ sweeps from $1$ to $1000$ molecules $^{-1}$. $t_S$ ranges from $10$ (topmost curve) to $100$ (lowest curve) in increments of $10$. {\bf B:} The plot depicts the 1D bond percolation probability \eqref{eq:finiteperc} as a function of the same bond probabilities used in panel A. The comparison serves to illustrate the difference between 1D bond percolation and polymerization while also emphasizing the analogy. On the other hand, bond percolation on an infinite 1D lattice is equivalent to polymerization described in terms of continuous concentrations.}
\label{fig:perc}
\end{figure}
\ \\

The analogy persists but the exact correspondence breaks down in the finite, i.e.\@ discrete, case. The percolation probability in the polymerization case is $\langle s_{\text{max}}\rangle$ as given by \eqref{eq:maximer}. The bond probability, $p_{\text{bond}}$, is the expected fraction of bonds and can be computed following the arguments that led to \eqref{eq:partition_lin}. We obtain
\begin{align}
p_{\text{bond}} = \dfrac{1}{t_S-1}\dfrac{\sum\limits_{i=1}^{t_S-1} i  \sigma^i \begin{pmatrix} t_S-1 \\ i \end{pmatrix} \dfrac{t_S!}{(t_S-i)!}}{Z^{\text{poly}}_{t_S}}.
\label{eq:pBond}
\end{align}
In 1D bond percolation, the percolation probability is 
\begin{align}
p_{\text{perc}} = 1-(1-p)^2\sum_{i=0}^{t_S-2}ip^{i-1}=p^{t_S-2}(t_S-p(t_S-2)-1),
\label{eq:finiteperc}
\end{align}
with $t_S$ the size of the lattice and $p$ the bond probability. 

In Figure 5B of the main text we sweep across a range for $t_S$ and $\sigma$. For each $(t_S, \sigma)$ pair we calculate the corresponding $p_{\text{bond}}$ via \eqref{eq:pBond} as the abscissa and $\langle s_{\text{max}}\rangle$ via \eqref{eq:maximer} as the ordinate. This graph is reproduced as Figure \ref{fig:perc}B for comparison with finite-size bond percolation, Figure \ref{fig:perc}A. Clearly in \eqref{eq:finiteperc} $p$ is just a parameter, but in Figure \ref{fig:perc}A we compute it via \eqref{eq:pBond} using the same sweep over $t_S$ and $\sigma$ as for Figure \ref{fig:perc}B to make comparison meaningful. The view from percolation is useful because it packages the dependency on $t_S$ and $\sigma$ into the single quantity $p$ (or $p_{\text{bond}}$). 

\section{Scaling behavior}

\begin{figure}[!h]
\centering
\includegraphics[width=0.8\linewidth]{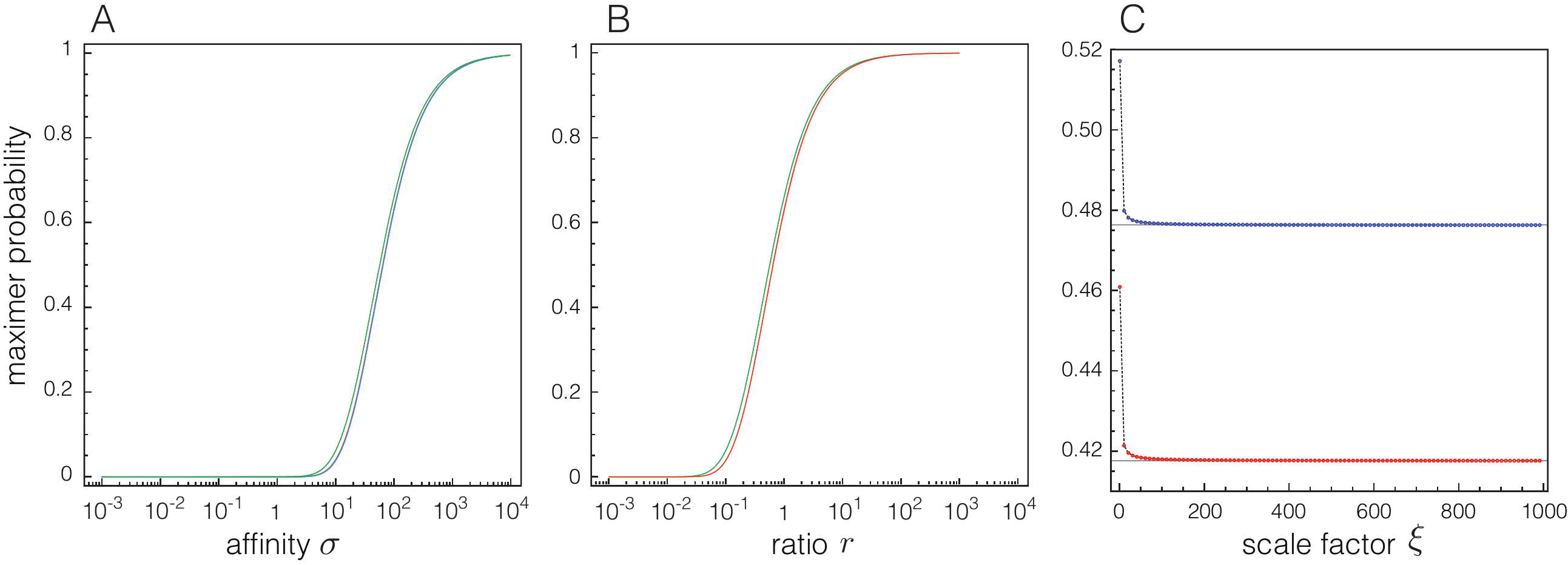}
\caption[Scaling behavior of the maximer distribution]{Scaling behavior of the maximer distribution. The panels illustrate the approximate scaling behavior of $\langle s_{\text{max}}\rangle$ from different perspectives implied by \eqref{eq:maximer_scaling}. In all three panels, the ordinate is the maximer probability as given by \eqref{eq:maximer}. {\bf A:} The graph exemplifies the relation \eqref{eq:maximer_scaling} by plotting three curves, blue: $\langle s_{\text{max}}\rangle[10,0.1 \sigma]$, red: $\langle s_{\text{max}}\rangle[100,\sigma]$, and green: $\langle s_{\text{max}}\rangle[1000,10 \sigma]$ as a function of the affinity $\sigma$. The blue and green graphs are related to the (arbitrary) red baseline graph by scale factors $\xi=0.1$ and $\xi=10$, respectively. The red and blue graphs sit on top of each other, while green has a slight (and slightly $\sigma$-dependent) shift to the left. {\bf B:} This panel illustrates the scaling version \eqref{eq:maximer_scaling2}, comparing red: $\langle s_{\text{max}}\rangle[1000,r\, 1000]$ with green: $\langle s_{\text{max}}\rangle[10,r\,10]$, sweeping along $r$. {\bf C:} The graph in this panel shows an integer sweep of the scale factor $\xi$, as per \eqref{eq:maximer_scaling}, for two pairs, $[t_S,\sigma]=[10,5]$ (red), $[t_S,\sigma]=[10,6]$ (blue). The scaling relation is well fulfilled except for very small particle numbers.}
\label{fig:scaling}
\end{figure}
\ \\

We refine the notation for the maximer probability \eqref{eq:maximer} to emphasize the dependence on the parameters $t_S$ and $\sigma$,
\begin{align}
\langle s_{\text{max}}\rangle [t_S,\sigma]\equiv \langle s_{\text{max}}\rangle,
\label{eq:maximer_notation}
\end{align}
in order to note an approximate scaling relation that we observe numerically:
\begin{align}
\langle s_{\text{max}}\rangle[t_S,\sigma]\approx \langle s_{\text{max}}\rangle [\xi t_S,\xi \sigma],
\label{eq:maximer_scaling}
\end{align}
with $\xi>0$ a dimensionless scale factor. Two systems are approximately equivalent if their protomer numbers and affinities are related by the same scale factor: $t_S^{(1)}=\xi t_S^{(2)}$ and $\sigma^{(1)}=\xi \sigma^{(2)}$. This implies that $t_S^{(1)}/t_S^{(2)}=\sigma^{(1)}/\sigma^{(2)}$ or $r=\sigma^{(1)}/t_S^{(1)}=\sigma^{(2)}/t_S^{(2)}$. The latter says that two systems behave approximately the same if the ratio $r$ of their respective affinity to protomer number is the same, which yields another way of expressing the scaling observation as
\begin{align}
\langle s_{\text{max}}\rangle[t^{(1)}_S, r\, t^{(1)}_S]\approx \langle s_{\text{max}}\rangle[t^{(2)}_S,r\, t^{(2)}_S].
\label{eq:maximer_scaling2}
\end{align}
These relations are depicted in Figure \ref{fig:scaling}.

\section{Unequal ligand concentrations and ligand binding affinities}

\subsection{Polymerizing scaffold system}

As in Figure 6 of the main text, Figure \ref{fig:diffs}A evidences the $\sigma$-dependence of the initial slope in the discrete system and illustrates the effect of ligand imbalance: Once the scarcer ligand, here $A$, is mostly bound up and the number of scaffold protomers increases further, $A$-ligands must spread across an increasingly wider range of length classes, thereby reducing the likelihood of multiple occupancy on the same polymer. As a result, although the binding opportunities for the more abundant ligand, here $B$, increase (up to the overall prozone peak), $B$-particles bound to a particular polymer are less likely to encounter any $A$s bound to it. The result is a slope reduction compared to a situation in which both ligands are present in equal numbers. A substantive difference between ligand binding constants causes not only a slope reduction prior to the prozone but has, in particular, the effect of delaying the prozone peak considerably beyond what one would expect based on particle numbers alone. It is worth noting that in the Wnt signaling cascade, ligand affinities------enzyme-scaffold, i.e.\@ GSK3\textbeta--Axin, and substrate-scaffold, i.e.\@ \textbeta-catenin--Axin---are regulated by the signaling process \cite{Luo2007,Willert1999}.

\begin{figure}[!h]
\centering
\includegraphics[width=0.6\linewidth]{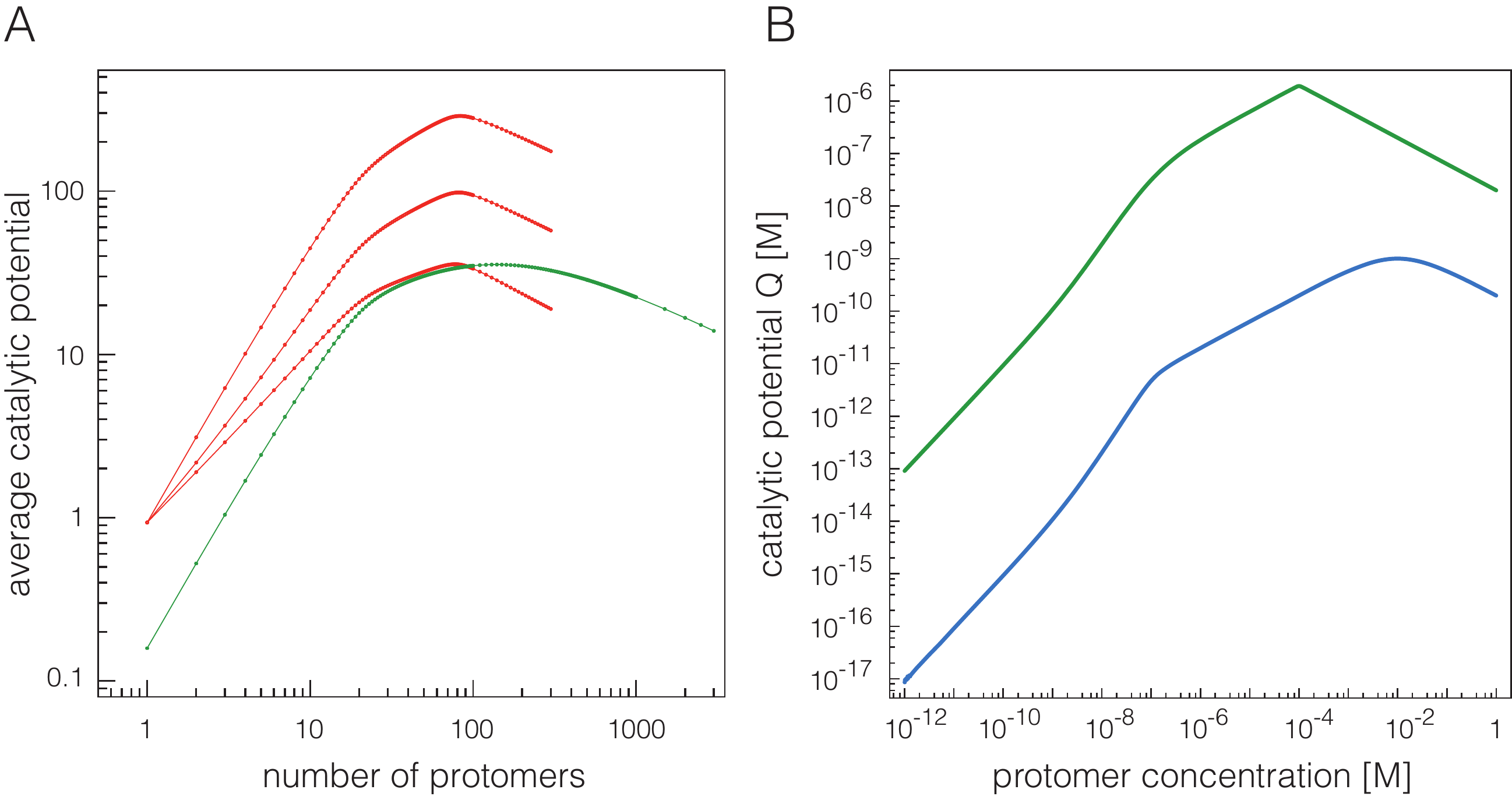}
\caption[Effects in discrete and continuum polymerizing scaffold systems]{Effects in discrete and continuum polymerizing scaffold systems. {\bf A:} The panel illustrates the effects of the polymerization constant $\sigma$, of ligand imbalance, and of unequal ligand affinities on discrete polymerization. Red, ligand imbalance: $t_A=20$, $t_B=80$, $\alpha=\beta=0.9$ molecules$^{-1}$, $\sigma=0.01$ (lower), $\sigma=0.1$ (middle), $\sigma=1$ (upper). Green, unequal ligand affinities: $t_A=t_B=20$, $\alpha=0.01$, $\beta=1$ molecules$^{-1}$, $\sigma=1$ molecules $^{-1}$. $t_S$ on the abscissa. {\bf B:} This panel illustrates the effects of ligand imbalance and of unequal ligand binding constants on continuum polymerization. Blue, unequal binding constants: $\alpha=10^2$ M$^{-1}$, $\beta=10^9$ M$^{-1}$, $t_A=t_B=10^{-7}$ M, $\sigma=10^8$ M$^{-1}$. Green, ligand imbalance: $t_A=10^{-8}$ M, $t_B=10^{-4}$ M, $\alpha=\beta=10^7$ M$^{-1}$, $\sigma=10^8$ M$^{-1}$.}
\label{fig:diffs}
\end{figure}
\ \\

In the continuum case, unlike the discrete case, the initial slope is independent of the polymerization constant $\sigma$ until a level of protomer abundance is reached sufficient for making polymerization effective, as discussed in section \ref{sec:multivspoly} (equation \ref{eq:initialslope_cont}. The inflection point at which the slope changes from $1$ to $2$ (in a log-log plot) will shift accordingly. After that slope change, the responses to ligand imbalance and to differences between ligand binding constants are analogous to the discrete case, as seen in
Figure \ref{fig:diffs}B. 

Neither ligand imbalance or differences in binding constants appear to affect the downward slope at large $t_S$ in the continuum or the discrete case.

\subsection{Multivalent scaffold system}

\begin{figure}[!h]
\centering
\includegraphics[width=0.9\linewidth]{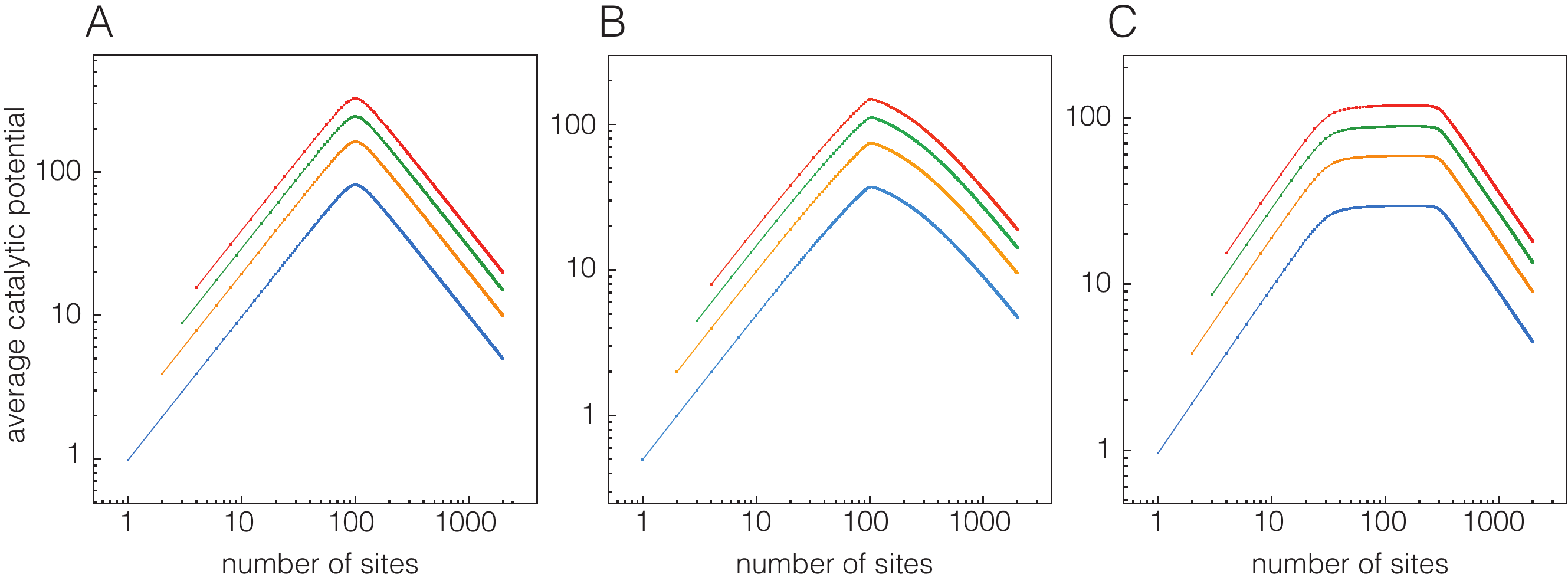}
\caption[Catalytic potential of multivalent scaffolds (discrete case)]{Catalytic potential of multivalent scaffolds (discrete case). {\bf A:} $\langle Q_{\text{multi}}\rangle$, equation \eqref{eq:Qmultidiscrete}, when particle numbers and binding affinities are the same for both ligand types: $A$ and $B$ are $100$ particles each, binding affinities are $0.9$ molecules$^{-1}$. Valencies: $1$ (blue), $2$ (orange), $3$ (green), $4$ (red). The abscissa shows the total number of sites, but $\langle Q_{\text{multi}}\rangle$ is calculated for site increments that reflect the valency of each scaffold. {\bf B:} Like panel A, but unequal ligand binding affinities: $\alpha=0.01$ and $\beta=9$ molecules$^{-1}$. {\bf C:} Like panel A, but unequal numbers of ligand particles: $A=30$ and $B=300$, binding affinities for both are $0.9$ molecules$^{-1}$. Colors indicate valencies as in panel A.}
\label{fig:diffsmv}
\end{figure}
\begin{figure}[!h]
\centering
\includegraphics[width=0.6\linewidth]{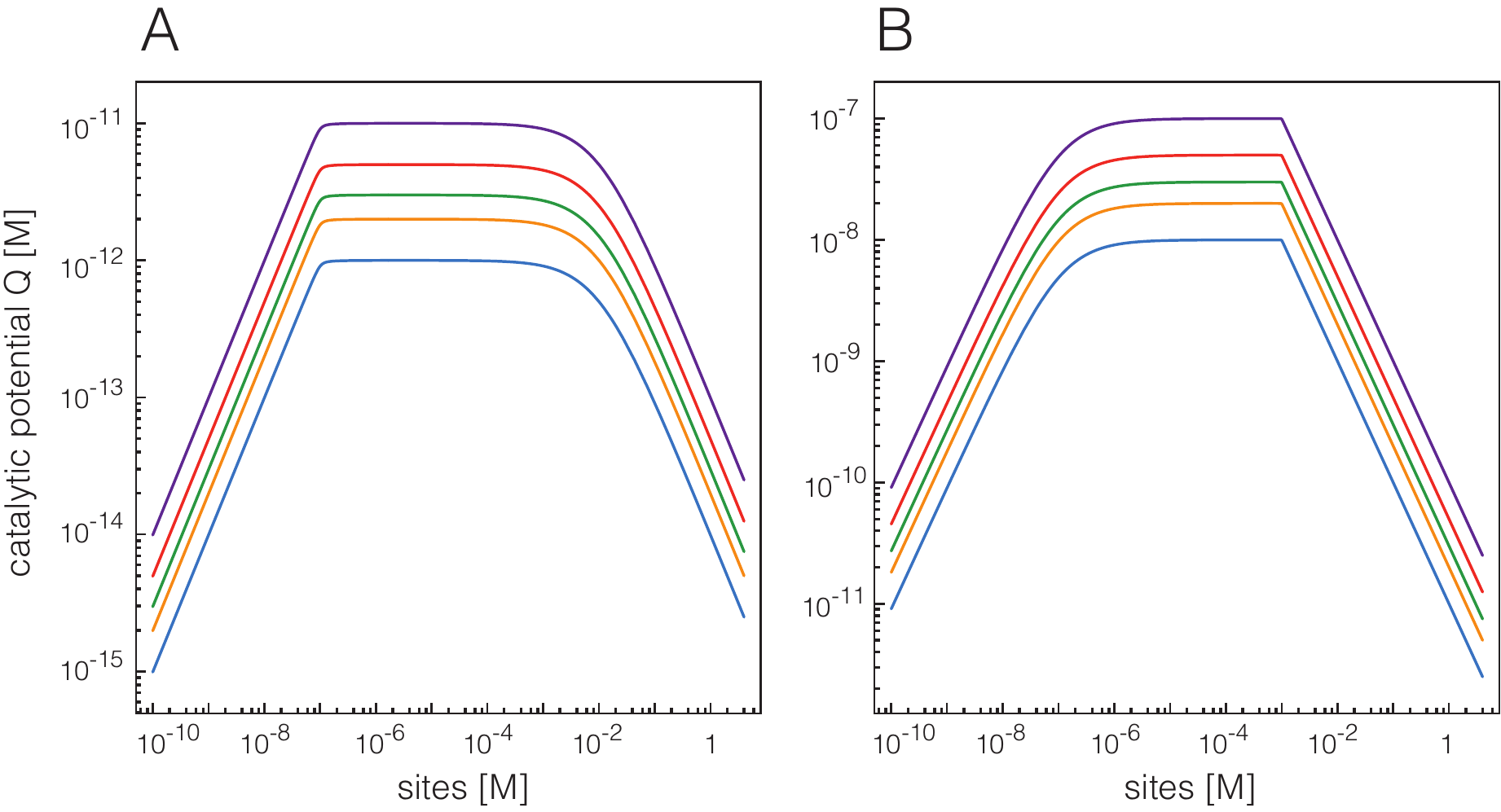}
\caption[Catalytic potential of multivalent scaffolds (continuum case)]{Catalytic potential of multivalent scaffolds (continuum case). {\bf A:} The panel provides an example for the effect of unequal ligand binding affinity. $t_A=t_B=10^{-7}$ M, $\alpha=10^2$ M$^{-1}$, $\beta=10^9$ M$^{-1}$, valencies: $1$, $2$, $3$, $4$. {\bf B:}  The panel illustrates the effect of ligand concentration imbalance. $t_A=10^{-8}$ M, $t_B=10^{-3}$ M, $\alpha=\beta=10^7$ M$^{-1}$, valencies: $1$, $2$, $3$, $4$.}
\label{fig:diffsmvcont}
\end{figure}
\ \\

The responses to ligand and affinity imbalances in a multivalent scaffold system follow similar lines as in the polymerizing case. When both ligand types are present with the same number of particles, the ligand with higher affinity experiences the prozone later, since the amount of scaffold-bound ligand is higher compared to the other type. This is seen in Figure \ref{fig:diffsmv}B with the steepening of the downward slope associated with the stronger binding ligand. The situation with ligand imbalance is analogous. The ligand with higher abundance keeps binding while the scarcer ligand is undergoing its prozone; thus the subdued effect on catalytic potential, which, in the example of Figure \ref{fig:diffsmv}C is mainly holding a constant level until the prozone for the more abundant ligand sets in. Although affinity and number imbalance mimic each other, the affinity imbalance exhibits a much less pronounced plateau around the prozone peak and consequently the drop-off is less sharp than in the case of number imbalance. Extremely high affinity differences would be required to generate a plateau similar to number imbalance. This is seen in the continuum case, shown in Figure \ref{fig:diffsmvcont}A, where affinities differ by 7 orders of magnitude. The concentration imbalance in the continuum case yields a similar picture as in the discrete case (Figure \ref{fig:diffsmvcont}B).

\vspace*{0.5cm}

\section{Stochastic simulations}

\begin{figure}[!h]
\centering
\includegraphics[width=\linewidth]{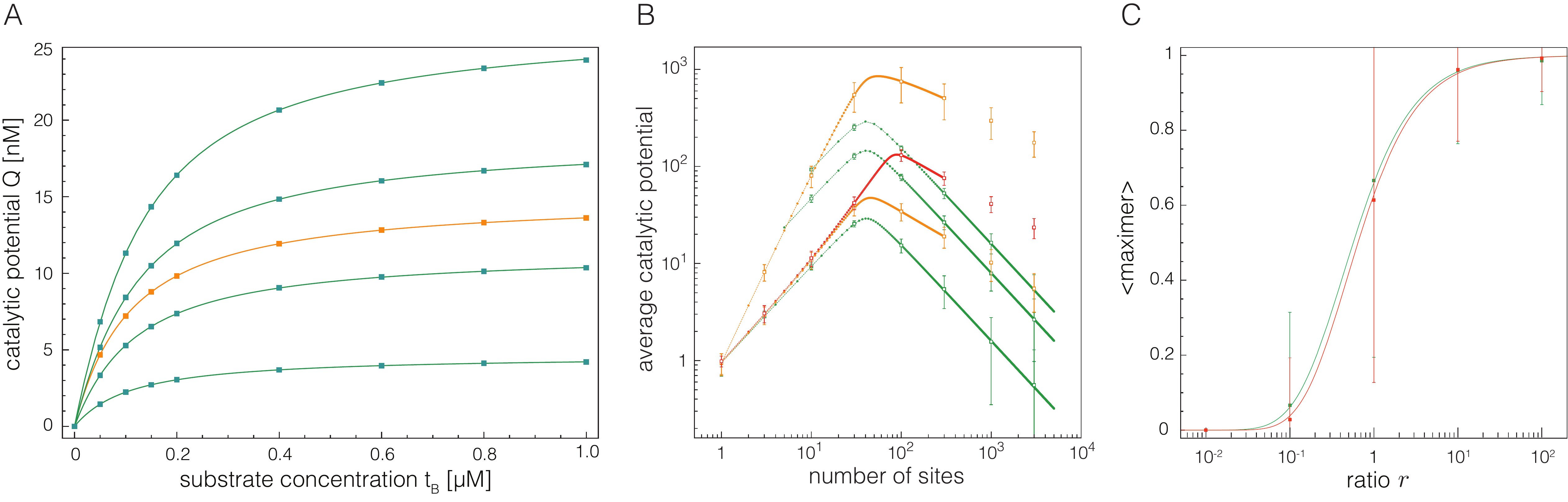}
\caption[Stochastic simulations]{Stochastic simulations. For all stochastic simulations, we used a volume on the order of an human erithrocyte, $V=10^{-12}$ L. All summary statistics were computed with 500 samples, each an independent and equilibrated state. {\bf A:} The solid curves in this panel are identical to those in Figure 3A of the main text. Stochastic simulations were performed by converting deterministic affinities into stochastic affinities as described in the main text (section \enquote{The discrete case in equilibrium}) and by converting concentrations into particle numbers at the given volume $V$. Averages of catalytic potnetial are indicated by filled squares. Green: polymerizing system at various protomer numbers, descending from top: $36120$ molecules ($60$ nM), $27090$ molecules ($45$ nM), $18060$ molecules ($30$ nM), $9030$ molecules ($15$ nM). Orange: reference Michaelian system with $60200$ (100nM) enzymes. Because of the large numbers of particles, the standard deviation is smaller than the squares at the chosen scale. This panel is meant as a sanity check that simulations at large particle numbers indeed reproduce the continuum picture as we derived it analytically. {\bf B:} The curves in this panel are identical to those in Figure 6A of the main text and refer to discrete scaffolding systems. Stochastic simulations were performed using the same parameters listed in that Figure. The squares mark the average catalytic potential, which coincides with the theoretical calculations; the error bars mark one standard deviation. In the polymerizing scaffold case, the simulation allowed us to extend the range of the rather time-consuming calculations using the analytical expression \ref{eq:Qpolydiscrete}. Note the log-log scale of the axes distorting the error bars; for a linear-log scale see Figure \ref{fig:varlin}. Green: multivalent scaffolds of valencies $n=10$ (upper), $n=5$ (middle), and $n=1$ (lower). Orange: polymerizing scaffold system with polymerization affinities $\sigma=10$ (upper) and $\sigma=0.01$ (lower). Red: polymerizing scaffold system at the same affinity as the lower orange curve, but with twice the number of ligand particles. {\bf C:} The curves are identical to those in Figure \ref{fig:scaling}B. As in that Figure, $r$ is the ratio of affinity to the number of protomers. Squares mark the average number of maximers and error bars mark one standard deviation. Green: system with 10 protomers. Red: system with 1000 protomers.}
\label{fig:variance}
\end{figure}
\ \\

Our analysis of the discrete case focuses on average behavior. Analytic techniques for higher moments are beyond the scope of this contribution and will be presented elsewhere. In lieu of an analytic treatment, we performed several stochastic simulations using the Kappa platform \cite{boutillier2018,kappaman} and GNU Parallel \cite{tange_ole_2018_1146014}. Figure \ref{fig:variance} displays the essential observations in the context of Figures 3A and 6A of the main text and \ref{fig:scaling}B of this Supplement.

\begin{figure}[!h]
\centering
\includegraphics[width=0.7\linewidth]{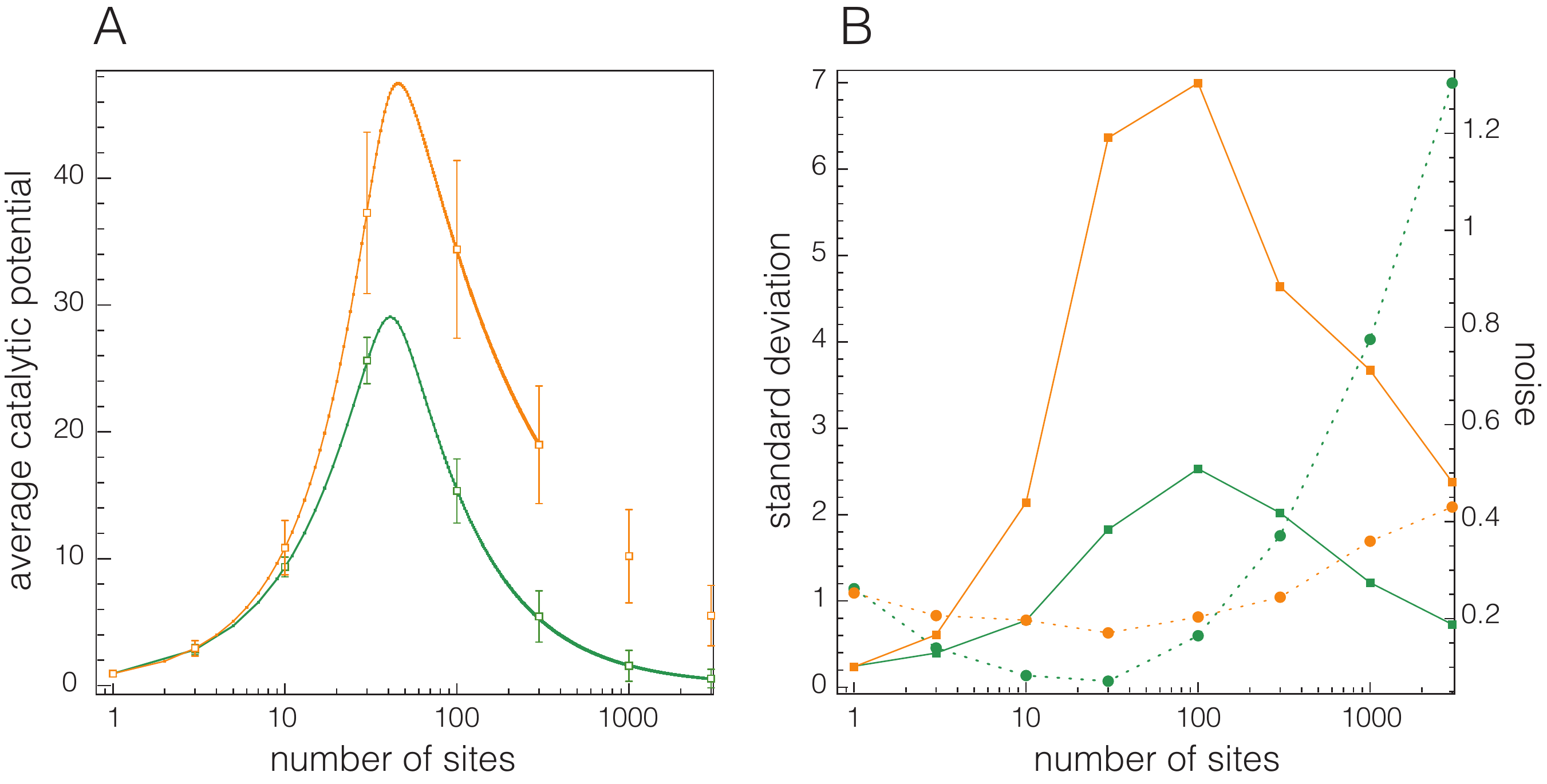}
\caption[Variance and noise]{Variance and noise. {\bf A:} This panel reproduces a subset of data from Figure \ref{fig:variance}B on a linear-log scale to enable a more direct visual interpretation of fluctuations. The green curve in this panel corresponds to the lowest green curve in Figure \ref{fig:variance}B. It belongs to a system of multivalent scaffolds with valency $1$. The orange curve belongs to the polymerizing scaffold system and corresponds to the lowest orange curve in Figure \ref{fig:variance}B. Because the valency of individual scaffolds in both systems is $1$, the number of sites on the abscissa corresponds to the number of scaffold agents, polymerizing or not. The main observation is that for the same average catalytic potential $\langle Q\rangle$ the standard deviation is larger after the prozone peak than prior to it. {\bf B:} This panel recasts the information in panel A by directly displaying the standard deviation (solid curves). The dashed curves (right ordinate) depict the noise, i.e.\@ the ratio of standard deviation to the mean. The main observation here is that the polymerizing system (orange) is significantly less noisy than the monovalent scaffold system (green).} 
\label{fig:varlin}
\end{figure}
\ \\

\begin{figure}[!h]
\vspace*{0.4cm}

\centering
\includegraphics[width=0.7\linewidth]{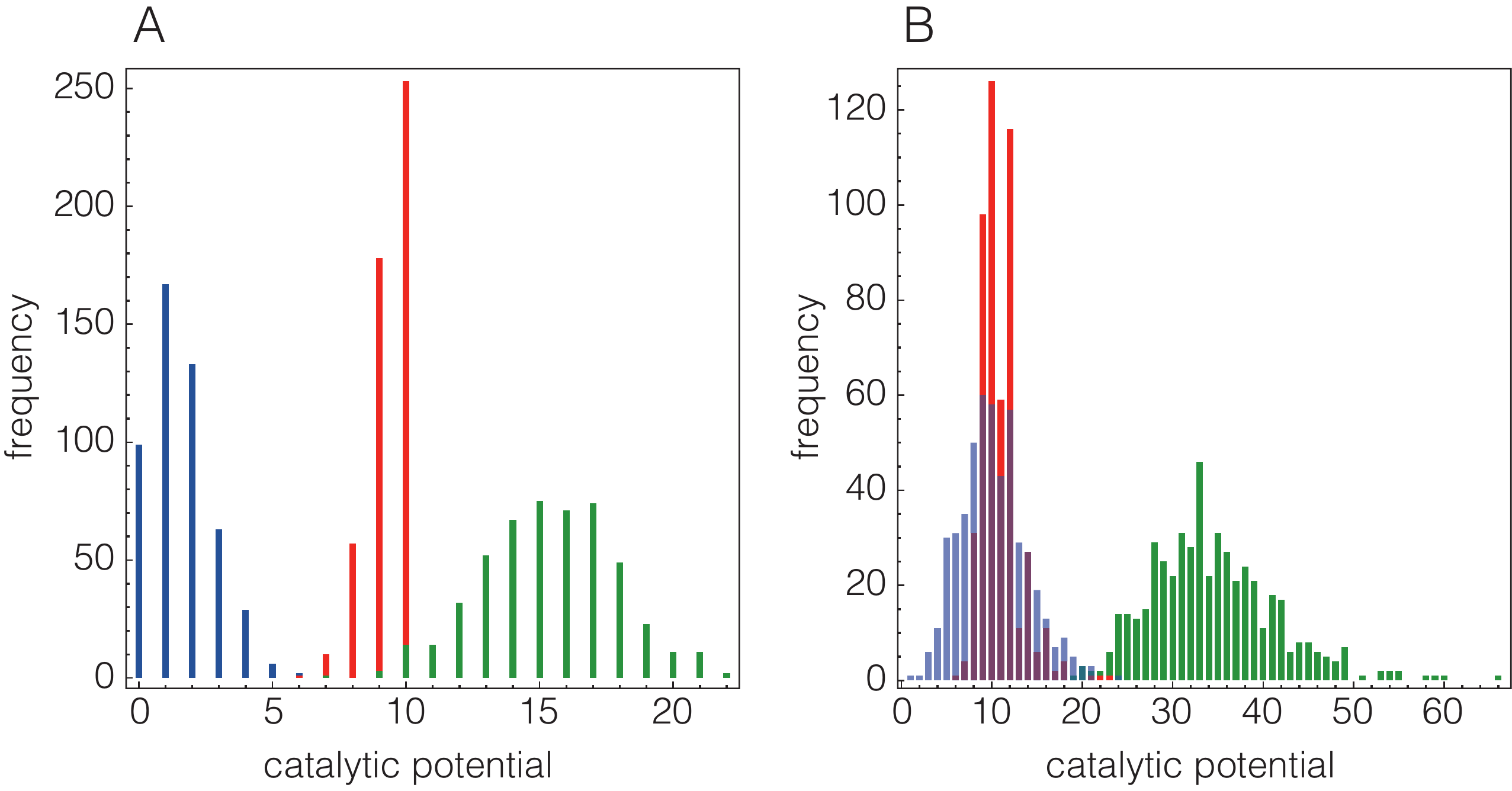}
\caption[Distributions of catalytic potential]{Distributions of catalytic potential. Panels {\bf A} (monovalent scaffold system) and {\bf B} (polymerizing scaffold system) depict the distribution of catalytic potential for a state sampled prior to the prozone peak ($10$ scaffold particles, red), just past the peak ($100$ particles, green) and well past the peak ($1000$ particles, blue). Other parameters as in Figure 6A of the main text.}
\label{fig:Qdist}
\end{figure}
\ \\

Fluctuations in the binding of ligands translate into $Q$-fluctuations on the basis of how sites are partitioned into agents. There are three regimes, which we describe in the case of a monovalent scaffold system for simplicity (lowest green curve in Figure \ref{fig:variance}; green curve in Figure \ref{fig:varlin}; and Figure \ref{fig:Qdist}): (i) At low scaffold numbers, prior to the prozone peak, most scaffolds are fully occupied by both ligands. Fluctuations cause transitions between system states with similar $Q$ and variance is therefore low (see red distributions in Figure \ref{fig:Qdist}). (ii) Just past the prozone peak, many scaffolds are still occupied by both ligands, but there is an increasing number of singly bound and some empty scaffolds. Unbinding from a fully occupied scaffold is statistically offset by re-binding to the pool of singly-bound scaffolds, which yields a net effect similar to situation (i). However, in addition, singly-bound scaffolds may also lose their ligand. This event is neutral in $Q$, but free ligands may re-bind an already singly-bound scaffold, thereby increasing $Q$. Likewise, dissociation from a fully occupied scaffold an re-association with an empty one will decrease $Q$. As a result of this expanded $Q$-range, the variance has increased compared to a situation with similar average $Q$ prior to the prozone peak (see green distributions in Figure \ref{fig:Qdist}). (iii) Well past the prozone peak, a number of scaffolds are bound by one ligand and many have no ligands at all. Ligand binding fluctuations will mainly shift ligands from singly-bound scaffolds to empty scaffolds with no effect on $Q$. As a result, $Q$-variance is now decreasing again (see blue distributions in Figure \ref{fig:Qdist}).

\afterpage{\clearpage}
\addcontentsline{toc}{section}{Supplementary references}
\renewcommand\refname{Supplementary references}

\putbib
\end{bibunit}

\end{raggedy}
\end{document}